\documentclass[twocolumn]{aastex701}
\usepackage{xspace}
\usepackage{amsmath}
\usepackage{framed} 
\usepackage{txfonts}
\usepackage{epstopdf}
\usepackage{subcaption}
\usepackage{xcolor}
\usepackage{float}
\definecolor{maroon}{RGB}{160, 0, 0}
\definecolor{olivegreen}{RGB}{85, 107, 47}
\definecolor{lnavy}{RGB}{25, 25, 160}

\begin{document}

\title{Towards robust simulations of the galactic dynamo: a cross-code comparison between Eulerian and Lagrangian schemes}

\author[0000-0003-2021-8536]{Mark Ivan Ugalino}
\affiliation{Department of Astronomy, University of Maryland, College Park, MD 20742, USA}
\email[show]{ugalino@umd.edu}  

\author[0000-0001-9568-7287]{Benedikt Diemer}
\affiliation{Department of Astronomy, University of Maryland, College Park, MD 20742, USA}
\email{diemer@umd.edu}

\author[0000-0002-6648-7136]{Vadim Semenov}
\affiliation{Center for Astrophysics $|$ Harvard \& Smithsonian, 60 Garden St, Cambridge, MA 02138, USA}
\email{vadim.semenov@cfa.harvard.edu}

\author[0009-0007-6250-8925]{Jin Young Kim}
\affiliation{Department of Astronomy, University of Maryland, College Park, MD 20742, USA}
\affiliation{Department of Physics and Astronomy, University of North Carolina at Chapel Hill, Chapel Hill, NC 27599, USA}
\email{kjinyoung02@gmail.com}

\author[0000-0003-3816-7028]{Federico Marinacci}
\affiliation{Department of Physics and Astronomy ``Augusto Righi'', University of Bologna, Via P. Gobetti 93/2, I-40129 Bologna, Italy}
\affiliation{INAF, Astrophysics and Space Science Observatory Bologna, Via P. Gobetti 93/3, I-40129 Bologna, Italy}
\email{federico.marinacci2@unibo.it}

\author[0000-0002-7275-3998]{Christoph Pfrommer}
\affiliation{Leibniz-Institut für Astrophysik Potsdam (AIP), An der Sternwarte 16, 14482 Potsdam, Germany}
\email{cpfrommer@aip.de}

\author[0000-0002-2395-4902]{Matthieu Schaller}
\affiliation{Lorentz Institute for Theoretical Physics, Leiden University, PO Box 9506, NL-2300 RA Leiden, the Netherlands}
\affiliation{Leiden Observatory, Leiden University, PO Box 9513, NL-2300 RA Leiden, the Netherlands}
\email{schaller@strw.leidenuniv.nl}

\author[0009-0005-5547-7580]{Nikyta Shchutskyi}
\affiliation{Lorentz Institute for Theoretical Physics, Leiden University, PO Box 9506, NL-2300 RA Leiden, the Netherlands}
\affiliation{Leiden Observatory, Leiden University, PO Box 9513, NL-2300 RA Leiden, the Netherlands}
\email{shchutskyi@lorentz.leidenuniv.nl}

\author[0000-0001-5976-4599]{Volker Springel}
\affiliation{Max-Planck-Institut für Astrophysik, Karl-Schwarzschild-Str. 1, D-85748, Garching, Germany}
\email{vspringel@mpa-garching.mpg.de}

\author[0000-0001-7689-0933]{Romain Teyssier}
\affiliation{Department of Astrophysical Sciences, Princeton University, Peyton Hall, Princeton, NJ 08544, USA}
\email{teyssier@princeton.edu}

\author[0000-0003-1811-8915]{Oliver Zier}
\affiliation{Center for Astrophysics $|$ Harvard \& Smithsonian, 60 Garden St, Cambridge, MA 02138, USA}
\email{oliver.zier@cfa.harvard.edu}

\begin{abstract}
Galactic magnetic fields are thought to be seeded by microscopic processes such as the Biermann battery and are primarily amplified to observed microgauss strengths by dynamo processes. Much of this amplification likely occurs within proto-galaxies, where turbulent and rotational kinetic energy is converted into magnetic energy via dynamo action. However, two aspects of this process remain poorly constrained in simulations: the saturation level of the amplified field and the timescale on which it is reached. In particular, different numerical methods, resolution strategies, and subgrid prescriptions have not converged to similar results, making it difficult to establish robust, code-independent conclusions about dynamo amplification in galaxies. In this work, we present the first systematic comparison of both Eulerian and Lagrangian codes that employ diverse MHD schemes, including \textsc{ART}, \textsc{Ramses}, \textsc{Arepo}, and \textsc{Swift}. Using identical initial conditions for Milky Way-type galaxies and without including star formation and stellar feedback, we find that the amplification history of the magnetic field within the galactic disk depends strongly on the choice of hydrodynamical scheme, with orders-of-magnitude differences between Eulerian and Lagrangian approaches. We further demonstrate that reducing the mesh velocity in the moving-mesh code \textsc{Arepo} leads to results consistent with those obtained using the static grid codes \textsc{ART} and \textsc{Ramses}. These results highlight the importance of carefully assessing numerical effects when interpreting magnetic field amplification in galaxy simulations.
\end{abstract}


\section{Introduction}
Magnetic fields can be found everywhere in the Universe. They play an essential role in astrophysical systems of varying length and energy scales, from planets \citep{1983Icar...54..466S, 2026arXiv260916720O}, to black hole accretion disks \citep{2023Sci...381..961Y}, and to the cosmic web \citep{2021A&A...652A..80L, 2022MNRAS.512..945C}. Primordial magnetic fields have been proposed as a remedy for the cosmological $H_0$ tension by introducing inhomogeneities in the density distribution of the Universe at recombination \citep{2020PhRvL.125r1302J}. Magnetic fields also affect thermal conduction in the intergalactic medium (IGM), which influences the evolution of galaxy clusters as observed in X-ray \citep{1989ApJ...338..761R, 2001ApJ...554..561M, 2002MNRAS.335L..71F, 2023MNRAS.526.6205R}. The propagation of cosmic rays that drive feedback in galaxies can also be influenced by complex magnetic field structures \citep{2018MNRAS.473.4544S, 2019MNRAS.488.3716C,2021MNRAS.501.3663H, 2023A&ARv..31....4R}. It has also been shown that these structures can affect star formation in strongly magnetized molecular clouds \citep{2015MNRAS.450.4035F, 2019FrASS...6....7K, 2023MNRAS.522.6203H}. 

Multi-wavelength observations, including radio synchrotron emission, Faraday rotation, and polarized dust measurements, demonstrate that magnetic fields in nearby spiral galaxies are pervasive, exhibiting ordered and turbulent components of a few to several tens of $\mu$G across disks and halos \citep{2013A&A...556A.142S, 2013MNRAS.433.1675B, 2017ApJ...836..185T, 2023A&A...670L..23H, 2023A&A...669A...8H, 2023A&A...678A..56B}. 

The nearby Local Group spiral M33 provides one of the best-resolved cases for connecting magnetic structure to galactic dynamics, exhibiting a coherent spiral field with a total strength of $\sim6$--$7\,\mu$G and magnetic energy in equipartition with the disk's turbulent kinetic energy \citep{2008A&A...490.1005T, 2009IAUS..259..545T, 2026A&A...712A.120S}, while the actively star-forming spiral NGC 6946 hosts magnetic fields up to $20\,\mu$G and distinct magnetic arms offset from the optical ones, tracing an ordered large-scale field despite vigorous feedback activity \citep{1991A&A...251...15B, 2007A&A...470..539B, 2013MNRAS.433.1675B, 2013A&A...556A.142S}. An extreme case is the nearby starburst galaxy M82, which exhibits some of the strongest measured galactic magnetic fields, with radio observations indicating central field strengths of $\sim100\,\mu$G in the core and $\sim25\,\mu$G in the halo \citep{2013A&A...555A..23A, 2017A&A...608A..29A}, while far-infrared dust polarization measurements reveal ordered and turbulent components of $\sim300\,\mu$G and $\sim220\,\mu$G, respectively, within its central kiloparsec, illustrating how intense star formation and feedback amplify and restructure galactic magnetic fields \citep{2021ApJ...914...24L}. These observations provide critical empirical constraints on the amplification and structure of galactic magnetic fields, motivating theoretical and numerical efforts to uncover the physical mechanisms responsible for generating seed magnetic fields and for their subsequent amplification.

Weak seed fields throughout the cosmic web may have been generated first through particle-level processes that precede galaxy formation, such as the Biermann battery \citep{1950ZNatA...5...65B, 1997ApJ...480..481K} and Weibel instability \citep{1959PhRvL...2...83W,1959PhFl....2..337F, 2015NatPh..11..173H}, with initial field strengths around $\sim 10^{-20}-10^{-18}$ Gauss \citep{2000ApJ...539..505G, 2016PhPl...23e6304S, 2021MNRAS.504.2346A}. Competing models instead attribute the origin of seed fields to primordial magnetogenesis during inflation or cosmological phase transitions \citep{2005MNRAS.356..778S, 2013A&ARv..21...62D, 2016RPPh...79g6901S, 2023ApJ...944..100M}. Such fields are many orders of magnitude weaker than the $\mu$G strengths observed in nearby galaxies, implying that efficient amplification mechanisms must have operated during galaxy formation and evolution. 

A popular theory posits that field amplification can be driven by the so-called $\alpha-\Omega$ dynamo \citep{1975A&A....42...85S}, which is the combined effect of turbulence and differential rotation acting on large scales in the galactic disk. Alternatively, the small-scale fluctuation dynamo can amplify seed fields on much shorter timescales by stretching, twisting, and folding field lines in turbulent flows, independent of large-scale rotation \citep{2004ApJ...612..276S, 2014ApJ...783L..20P, 2016MNRAS.457.1722R, 2017MNRAS.471.2674R, 2017MNRAS.472.4368R, 2022MNRAS.513.3326M}. Other proposed mechanisms include the magnetorotational instability (MRI), which amplifies fields through the same shear that drives angular momentum transport in differentially rotating disks \citep{2004A&A...424..565K, 2006ApJ...641..862N, 2012MNRAS.425...74R, 2013ApJ...764...81M, 2015MNRAS.454.3049S, 2018MNRAS.474.2212R}, and cosmic-ray-driven dynamos, in which cosmic-ray pressure gradients help drive the turbulence and outflows that can drive field growth (\citealp{2004ApJ...605L..33H, 2009ApJ...706L.155H, 2010A&A...510A..97S, 2011IAUS..274..398K, 2022MNRAS.515.4229P}, see extensive review by \citealp{2023A&ARv..31....4R}).

Over the past two decades, significant progress has been made in modeling the evolution of magnetic fields in galaxies. Global and local simulations of isolated disk galaxies show exponential amplification of initially weak seed fields to near-equipartition strengths, with growth rates that are highly sensitive to the driving mechanism \citep{2009ApJ...696...96W, 2013MNRAS.432..176P, 2016MNRAS.457.1722R}. These results are consistent with observed field strengths in the Milky Way \citep{2015ASSL..407..483H} and in galaxies at higher redshifts \citep{2017NatAs...1..621M, 2023Natur.621..483G}. Such field growth has been shown to be driven either by turbulence generated by stellar feedback from supernovae and/or radiation, or by accretion shocks arising from gravitational collapse. Stratified shearing-box simulations show that magnetic fields grow exponentially, with growth rates that scale with the rotation period and depend only weakly on the supernova rate \citep{2008A&A...486L..35G}. Similar behavior is also reported in the Milky Way-like global disk simulations of \citet{2016MNRAS.457.1722R} (hereafter, RT16), where comparable magnetic field amplification is obtained in simulations both with and without supernova feedback, suggesting that differential rotation and shear dominate the dynamo growth in those models. It is further shown in RT16 that a small-scale dynamo sets in only once additional infrared radiative feedback is included, since magnetic energy fails to grow appreciably beyond the level reached during the initial collapse when only supernova feedback is present, whereas increasing the strength of the radiative feedback boosts the exponential growth rate (see Figure 11 of RT16). On the other hand, \citet{2022MNRAS.515.4229P} finds that a small-scale dynamo can be driven in two stages, first by turbulence generated by the accretion shock during gravitational collapse, and then by turbulence sustained through disk–halo shear, without invoking supernova or radiative feedback in Milky Way-mass galaxies. These varying conclusions likely stem from differences in numerical scheme and effective resolution across the studies, which can affect how well small-scale turbulent motions and shocks are captured, and consequently how efficiently the dynamo amplifies the magnetic field.

Cosmological simulations further indicate that small-scale dynamos operate efficiently at early times, rapidly amplifying primordial seed fields before large-scale ordering occurs during disk assembly \citep{2014ApJ...783L..20P, 2017MNRAS.472.4368R, 2018MNRAS.480.5113M, 2024MNRAS.528.2308P}. Before disk assembly, field amplification occurs through gravitational compression during halo collapse, with field strengths that increase as $B \propto \rho^{2/3}$, and small-scale dynamo action driven by accretion-shock turbulence, which amplifies the field exponentially \citep{2010A&A...522A.115S, 2010ApJ...721L.134S}. Together, these amplify seed fields to $\sim\mu$G strengths within a few $\times 10^8$\,yr, saturating at roughly 10\% of the kinetic energy. Once a rotationally supported disk forms, the $\alpha$--$\Omega$ dynamo takes over, transforming the turbulent small-scale field into an ordered large-scale field coherent on scales comparable to the disk radius and saturating at $\sim 10\ \mu$G in the disk midplane \citep{2014ApJ...783L..20P}.

However, the saturation level and topology of the magnetic field remain sensitive to feedback prescription, numerical resolution, diffusivity and resistivity, and initial conditions. For instance, Milky Way-like disk simulations by \citet{2009ApJ...696...96W} yield saturated magnetic field energies that only reach $\sim 10^{-3}$ of the total kinetic energy, suggesting highly sub-equipartition saturation. In contrast, a number of other numerical studies find saturation energies of at least 10\% equipartition, and in some cases approaching full equipartition between magnetic and kinetic energy densities \citep{2013MNRAS.432..176P, 2014ApJ...783L..20P, 2017MNRAS.471.2674R, 2017MNRAS.472.4368R, 2024MNRAS.528.2308P,2026arXiv260806978K}. The order-of-magnitude spread in magnetic field saturation across galaxy simulations shows that results are far from converged, making it difficult to draw robust, code-independent conclusions about how cosmic magnetic fields are amplified.

Recent studies have examined the numerical dependencies of the galactic dynamo in detail, including systematic surveys across different numerical parameters. \citet{2023A&A...673A..47W} find that the amplification rate and saturation level are highly sensitive to resolution, numerical dissipation, and feedback prescription, with low-resolution simulations failing to reproduce the converged behavior of higher-resolution runs. 
On the other hand, \citet{2026MNRAS.550g1121S} (hereafter, SP26) show that magnetic field amplification is sensitive to the choice of divergence-control scheme, with improved divergence cleaning substantially reducing numerical errors and enhancing magnetic field growth in galaxy cluster outskirts (see Figure 4 of SP26). \citet{2026arXiv260706660P} (hereafter, RP+26) compares the small-scale dynamo across several numerical schemes in idealized, \textit{subsonic} MHD turbulence and finds that the field structure agrees well across codes, while observed residual differences in the amplification rate in the kinematic phase and saturation strength at fixed resolution are due to varying levels of numerical diffusion between schemes (see Figures 3 and 4 of RP+26). While these studies and other similar work \citep{2009MNRAS.397..733K, 2011ApJ...737...13K, 2016MNRAS.463..477M, 2016MNRAS.461.1260T, 2026ApJS..285...64T} have examined dynamo behavior within a single code or similar schemes and in idealized settings, a direct comparison between Eulerian and Lagrangian MHD schemes in a galactic setting has not yet been performed.

We address this gap with \textit{the first systematic comparison of four cosmological MHD codes with different numerical schemes} --- \textsc{ART}, \textsc{Ramses}, \textsc{Arepo}, and \textsc{SWIFT}. We perform this comparison by simulating the co-evolution of an isolated Milky Way-type galaxy and its magnetic field, using initial conditions consistent with those of the AGORA project \citep{2016ApJ...833..202K}. For this paper, we approximate the integral effect of star formation and feedback via a pressure floor (see \S\ref{method}) which keeps the galactic disk thick. Ensuring a consistent treatment of star formation and feedback between different codes is a highly nontrivial task, which we circumvent by adopting this simplification. We find that Eulerian and Lagrangian codes differ by orders of magnitude in their predicted magnetic amplification, and we show that numerical diffusivity is a likely cause.

This paper is structured as follows. In \S\ref{theory}, we provide a theoretical overview of the galactic dynamo and numerical challenges associated with satisfying the solenoidal condition $\nabla\cdot\mathbf{B} = 0$ in ideal MHD galaxy simulations, discussing its potential impact on simulation accuracy and the methods employed to mitigate these errors. In \S\ref{codes}, we briefly describe the features of each participating code that are most relevant to our code comparison. For the busy reader, one may directly jump to \S\ref{method} where we describe our simulation setup and the framework we implement to compare simulations across different codes. In \S\ref{results}, we present our results, focusing on the magnetic field amplification histories across codes and the possible role of $\nabla\cdot\mathbf{B}$ errors and spatial discretization in driving these differences. In \S\ref{discussion}, we discuss the implications of our findings for the interpretation of galactic dynamo simulations. Finally, in \S\ref{conclusion}, we summarize our conclusions and offer recommendations for future numerical studies.

\section{Theory}
\label{theory}
\subsection{Magnetic field amplification across different scales}
As discussed in the previous section, a variety of mechanisms can amplify and organize magnetic fields in galaxies, from small-scale turbulent dynamos to the large-scale $\alpha$-$\Omega$ dynamo. In this section, we outline the theoretical framework underlying each. The dynamics of the gas and magnetic field are governed by the ideal MHD equations, which in conservative form read
\begin{align}
\frac{\partial \rho}{\partial t} + \nabla \cdot (\rho \mathbf{u}) &= 0, \\
\frac{\partial (\rho \mathbf{u})}{\partial t} + \nabla \cdot \left( \rho \mathbf{u} \mathbf{u}^T - \mathbf{B} \mathbf{B}^T + P_{\text{tot}} \right) &= 0, \\
\frac{\partial E}{\partial t} + \nabla \cdot \left( (E + P_{\text{tot}})\mathbf{u} - (\mathbf{u} \cdot \mathbf{B}) \mathbf{B} \right) &= 0, \\
\frac{\partial \mathbf{B}}{\partial t} - \nabla \cdot (\mathbf{B}\mathbf{u}^T - \mathbf{u}\mathbf{B}^T) &= 0, \label{eq:mhd} \\
\nabla\cdot \mathbf{B} &= 0,
\end{align}
where $\langle\cdot\rangle\langle\cdot\rangle^T$ is the outer product of two vectors, ${\bf u}$ is the velocity field, ${\bf B}$ is the magnetic field (with $B = |{\bf B}|)$, $\rho$ is the local gas density, and $P_{\rm tot} = P_{\rm therm} + B^2/2$ is the total gas pressure. The dynamo process proceeds in two stages: a \textit{kinematic phase}, where fluid motion amplifies a weak seed field, and a \textit{saturation phase}, where the resulting magnetic Lorentz force saturates the growth by backreacting on the flow. 

In the \textit{kinematic phase}, turbulent eddies driven by galaxy-scale processes, e.g. supernova feedback and rotational instabilities,  stretch and fold the magnetic field, amplifying its strength over several turnover times. The rate at which this amplification proceeds is set by the eddy turnover time, $\tau_{\rm eddy}\sim \ell / u_\ell$, where $u_\ell$ is the characteristic velocity at scale $\ell$. For Kolmogorov turbulence, in which the velocity field is self-similar, the characteristic velocity scales as $u_\ell\propto \ell^{1/3}$, which implies that the eddy turnover time scales as $\tau_{\rm eddy}\propto \ell^{2/3}$, such that smaller eddies turn over faster.  Since the induction equation, as shown in Equation \eqref{eq:mhd}, is linear in $\mathbf{B}$ during the kinematic phase, the instantaneous growth rate is proportional to the current field strength, $\partial B/\partial t \sim \Gamma B$, where $\Gamma$ is the exponential growth rate of the field strength \citep{1968JETP...26.1031K, 1992ApJ...396..606K}. The linearity of the induction equation thus drives exponential growth $B \sim e^{\Gamma t}$, at a rate $\Gamma \sim \tau_{\rm eddy}^{-1} \sim \ell^{-2/3}$, which is maximized at the dissipation scale \citep{2004ApJ...612..276S}. The growth rate can depend on either the viscous dissipation scale, $\ell_\nu$, or the resistive dissipation scale, $\ell_\eta$, depending on the magnetic Prandtl number, i.e. ${\rm Pr_m} \sim (\ell_\nu/\ell_\eta)^{2}$ for ${\rm Pr_m} \gg 1$ \citep{1959JFM.....5..113B, 2002ApJ...567..828S}, and ${\rm Pr_m} \sim (\ell_\nu/\ell_\eta)^{4/3}$ for ${\rm Pr_m} \ll 1$ \citep{2007NJPh....9..300S, 2007PhRvL..98t8501I}.

At low Pr$_{\rm m}$, the resistive scale is larger than the viscous scale ($\ell_\eta \gg \ell_\nu$), so magnetic diffusion acts before the smallest turbulent eddies can efficiently stretch the field. As a result, the dynamo is controlled by the resistive scale. The corresponding growth rate can be expressed as
\begin{equation}
\Gamma \sim \tau_{\rm eddy}^{-1}(\ell_\eta) \sim \frac{u_{\rm SN}}{\ell_{\rm SN}}{\rm Re}_{\rm m}^{1/2},
\end{equation}
where $u_{\rm SN}$ and $\ell_{\rm SN}$ denote the characteristic velocity and driving scale of turbulence powered by stellar feedback, and the magnetic Reynolds number ${\rm Re}_{\rm m} = u_\ell \ell/\eta \sim (\ell_{\rm SN}/\ell_\eta)^{4/3}$. However, in the warm ISM where $\mathrm{Pr_m} \gg 1$, the viscous scale exceeds the resistive scale, $\ell_\nu \gg \ell_\eta$. The velocity field is therefore damped at $\ell_\nu$, while magnetic fluctuations persist to much smaller scales. Dynamo stretching is thus controlled by shear at the viscous scale, where the smallest active eddies reside. The growth rate is consequently set by the viscous-scale turnover time,
\begin{equation}
\Gamma \sim \tau_{\rm eddy}^{-1}(\ell_\nu) \sim \frac{u_{\rm SN}}{\ell_{\rm SN}}{\rm Re}^{1/2},
\label{eq:highprandtl}
\end{equation}
where ${\rm Re} = u_\ell \ell / \nu$ is the viscous Reynolds number. Strictly, this scaling assumes that the magnetic Reynolds number ${\rm Re}_{\rm m} \gg 1$. Even when $\mathrm{Pr_m} \gg 1$, this scaling only holds once ${\rm Re}_{\rm m}$ is well above its critical value for dynamo action. At moderate ${\rm Re}_{\rm m}$, resistive losses are not negligible compared to the stretching rate, so the net growth rate is reduced below the pure stretching estimate by an amount that depends on ${\rm Re}_{\rm m}$, and $\Gamma$ cannot be reduced to a function of ${\rm Re}$ alone (see Equation 33-35 in \citealp{2025MNRAS.542.2669B}). This trend was observed by \citet{2022MNRAS.513.2457K} in their isothermal periodic-box MHD turbulence simulations (see Table 1), where at fixed ${\rm Re}$, $\Gamma$ increases systematically with ${\rm Re}_{\rm m}$, while at fixed ${\rm Re}_{\rm m}$ and varying ${\rm Re}$, $\Gamma$ remains approximately constant. Equation \ref{eq:highprandtl} should therefore be understood as the asymptotic scaling in the limit $\mathrm{Re_m} \to \infty$.

The \textit{saturation phase} starts when the Lorentz force $({\bf B}\cdot\nabla){\bf B}$ becomes comparable to the inertial term $\rho({\bf u}\cdot\nabla){\bf u}$, at which point backreaction on the flow causes $\Gamma$ to become field-dependent and growth saturates. At the scale $\ell$ where magnetic field amplification is sustained, saturation corresponds to the equipartition between the kinetic and magnetic energy, $\rho u_\ell^2 \sim B_\ell^2$, so that the Alfvén speed $u_A = B_\ell/\sqrt{\rho}$ becomes comparable to the characteristic velocity $u_\ell$. In this regime, the flow can no longer efficiently bend magnetic field lines since any further magnetic field perturbations excite Alfvén waves that propagate away faster than the timescale required for the flow to overcome magnetic tension. For the warm ISM of the Milky Way, adopting a characteristic gas density 
$n \sim 1\,{\rm cm}^{-3}$ (so that $\rho \sim m_p n \sim 
1.7\times10^{-24}\,{\rm g\,cm}^{-3}$) and a turbulent velocity 
$u_\ell \sim 10\,{\rm km\,s}^{-1}$ on scales $\ell \sim 100\,{\rm pc}$ 
driven by supernova feedback, one obtains
\begin{equation}
    B_{\rm sat} \sim \sqrt{1.7\times10^{-24} {\rm g/cm}^3}
    \times 10^6\,{\rm cm\,s}^{-1} \sim 1\,\mu{\rm G},
\end{equation}
in good agreement with the Milky Way field strength inferred from Faraday rotation measurements \citep{2010ApJ...714.1170M, 2025MNRAS.544.2698D}. 

Meanwhile, on galaxy scales ($\gtrsim 1\,\rm kpc$), the $\alpha$-$\Omega$ dynamo complements the small-scale dynamo by organizing the magnetic field into large-scale coherent structures. Differential rotation winds the poloidal (radial) component into azimuthal ones ($\Omega$-effect). Meanwhile, helical turbulence induced by the Coriolis force in supernova-driven outflows regenerates the poloidal component ($\alpha$-effect). The growth rate due to the $\Omega$-effect follows from the shear term in the induction equation, which for the azimuthal component $B_\phi$ yields,
\begin{equation}
    \frac{\partial B_\phi}{\partial t} =rB_r \frac{\partial\Omega}{\partial r} \,.
    \label{eq:shear}
\end{equation}
A full derivation of this expression is provided in Appendix \ref{sec:shear}. Hence, the \textit{linear} growth timescale is set by the shear rate, $\tau_{\rm shear} \sim (q\Omega)^{-1}$, where $\Omega$ is the angular velocity and $q = -\partial\ln\Omega/\partial\ln r$ is the shear parameter. For Milky Way-like galaxies where $q\approx 1$, the dynamical time is about
\begin{equation}
\tau_{\rm shear} \sim \Omega^{-1}\sim 30-40\ {\rm Myr},
\end{equation}
in the outer region of the disk ($r \gtrsim 8$ kpc), and about 10 Myr within the inner, strongly-shearing regions ($r \lesssim 2$--3 kpc). 

On the other hand, the $\alpha$-effect contributes to the growth of magnetic fields as long as the velocity field is strongly correlated with the magnetic field. This correlation is sustained by the turbulence driven by supernova feedback, which preferentially regenerates poloidal field components regardless of the shear configuration. Hence, the timescale at which the \mbox{$\alpha$-effect} acts to amplify the galactic magnetic field is set by the turnover time of turbulent eddies driven by supernova feedback, $\tau_{\alpha} \sim \ell_{\rm SN}/u_{\rm SN}$. For typical galactic outflows due to supernova feedback, the timescale for the $\alpha$-effect is about
\begin{equation}
\tau_{\alpha} \sim \frac{\ell_{\rm SN}}{u_{\rm SN}}\sim 10\ {\rm Myr}.
\end{equation}
These timescales are much longer than the growth time of the small-scale dynamo in the interstellar medium. Under typical conditions in the warm ISM, where the Reynolds number is about $\mathrm{Re} \sim 10^{7}-10^{8}$,
\begin{equation}
\tau_{\rm eddy}(\eta) \sim \frac{\ell_{\rm SN}}{u_{\rm SN}}{\rm Re}^{-1/2} \sim 
\frac{100\ {\rm pc}}{10\ {\rm km}\ {\rm s}^{-1}} \times (10^{8})^{-1/2} \sim 10^3\,{\rm yr}\,.\end{equation}
Despite its slower growth, the large-scale dynamo produces the coherent, ordered spiral magnetic field structures observed in nearby galaxies. How the rapid small-scale amplification couples to and sustains the large-scale galactic magnetic field remains an open problem.
 
\subsection{Numerical effects}
\label{sec:numerical_effects}
Numerical simulations are indispensable tools for studying the growth of galactic magnetic fields. However, violations of the solenoidal condition ($\nabla \cdot {\bf B} = 0$) can drive unphysical forces and introduce spurious sources or sinks of magnetic energy, while excessive numerical resistivity can artificially damp magnetic structures and suppress field amplification. Conversely, low numerical diffusion can lead to unstable simulations, as low-dissipation MHD runs are prone to grid-seeded instabilities \citep{2012ApJS..201...18M, 2016MNRAS.455.4274L}. The reliability of simulated field growth therefore critically depends on how well the numerical scheme controls both \textbf{divergence errors} and the \textbf{diffusive effects} inherent to the discretization scheme. 

Several studies have shown that inadequate control of $\nabla\cdot\mathbf{B}$ errors can artificially enhance or suppress magnetic field growth (e.g. \citealp{2026MNRAS.550g1121S, 2026ApJS..285...64T}, see also Figure \ref{fig:glm_scan1}), underscoring the need for robust divergence-cleaning or constrained transport schemes when interpreting simulation results in the context of galactic dynamo theory. The impact of these errors is made explicit through the magnetic energy density evolution equation, which in the Lagrangian frame is given by
\begin{align}
\label{eq:induction}
\frac{\mathrm{D}}{\mathrm{D}t}\left(\frac{B^2}{2}\right) &= {\bf B}\cdot[\nabla\times({\bf u}\times{\bf B}) + ({\bf u}\cdot\nabla){\bf B}]\nonumber\\
&=  - B^2(\nabla\cdot\mathbf{u}) + \mathbf{B}\cdot[(\mathbf{B}\cdot\nabla)\mathbf{u}] + (\mathbf{u}\cdot\mathbf{B})(\nabla\cdot\mathbf{B})\nonumber\\
&=\underbrace{- \frac{2}{3}B^2(\nabla\cdot\mathbf{u})}_{\text{compression}}
\ \underbrace{+ \mathbf{B}^{\rm T}\mathcal{S}\ {\bf B}}_{\text{traceless shear}} 
\ \underbrace{+ (\mathbf{u}\cdot\mathbf{B})(\nabla\cdot\mathbf{B})}_{\text{spurious divergence}} \,,
\end{align}
where ${\rm D}/{\rm D}t = [\partial/\partial t + ({\bf u}\cdot \nabla)]$ and $\mathcal{S}$ is the traceless strain tensor defined as,
\begin{equation}
    \mathcal{S}_{ij} = \frac{1}{2}\left(\partial_i u_j + \partial_j u_i\right) - \frac{1}{3}(\nabla\cdot\mathbf{u})\,\delta_{ij} \,.
\end{equation}
The final term proportional to $\nabla \cdot \mathbf{B}$ vanishes for a physically solenoidal magnetic field and is dropped from the induction equation in analytical calculations. In smoothed particle magnetohydrodynamics (SPMHD) schemes, this divergence term is likewise omitted from the induction equation \citep{2026arXiv260806978K, 2023FrASS..1088219T}; the divergence errors are instead advected with the flow, similar to the Powell eight-wave scheme (see \S\ref{sec:divergenceschemes}), and require additional cleaning schemes to control their growth. Nevertheless, in simulations, non-zero divergence errors can act as a spurious source or sink of magnetic energy. If left unmitigated, these errors may accumulate over time and contaminate the magnetic energy evolution (see \S\ref{sec:divergence_art} for an illustration of this effect). However, to our knowledge, the numerical effects of local divergence build-up have not yet been systematically quantified in the literature.

For Eulerian codes, if we let $u_0$, $B_0$, and $\Delta x$ denote characteristic scales for velocity, magnetic field strength, and length, respectively, we can estimate the spurious energy injection rate as
$\Gamma_{\nabla\cdot\mathbf{B}} \sim \epsilon\,u_0 B_0^2/\Delta x$, where $\epsilon$ is the dimensionless divergence error, $\epsilon \equiv |\nabla\cdot\mathbf{B}| \Delta x/B_0$. Normalizing by the magnetic energy density $B_0^2/2$ gives the fractional growth rate
\begin{align}
\left.\frac{{\rm D}\ln(B^2/2)}{{\rm D}t}\right|_{\nabla\cdot\mathbf{B}} &= \left(\frac{2}{B_0^2}\right)[(\mathbf{u}\cdot\mathbf{B})(\nabla\cdot\mathbf{B})]\\
&\sim \frac{2 u_0 B_0 (\pm \epsilon B_0 /\Delta x)}{B_0^2} \nonumber \\
&= \pm\frac{2\epsilon}{\tau} = \pm \frac{2\epsilon}{\tau_{\rm sound}}\mathcal{M}\,,
\end{align}
where $\tau = \Delta x/u_0$ is the crossing time across one cell, $\tau_{\rm sound} = \Delta x/c_s$ is the sound crossing time across one cell, and $\mathcal{M} = u_0 /c_s$ is the local Mach number. The spurious growth rate is therefore $2\epsilon$ times the advection rate, $\tau^{-1}$. Hence, over many crossing times, even modest departures from $\epsilon\ll 1$ can accumulate depending on the sign of $\nabla\cdot{\bf B}$. The total fractional energy error grows or decays as $\mathcal{O}(2n\epsilon)$ after $n$ crossing times, i.e. a simulation with $\epsilon\sim 10^{-2}$ could have accrued a $\sim 20\%$ spurious energy error after 10 crossing times. For $\epsilon\sim 1$, the exponential growth renders the solution unphysical within a single crossing time. 

In a similar way, the local Mach number could also boost the influence of divergence errors on field amplification, i.e., while divergence error effects may not be as consequential in subsonic flows, they could be important in supersonic flows. Under typical conditions within the Milky Way galaxy, where the galactic rotation speed is $\sim 200$ km/s and the sound speed ranges from $10-12$ km/s, the bulk Mach number sits between $\mathcal{M}\sim 16-20$, which is strongly supersonic. To our knowledge, how the distribution of dimensionless divergence errors $\epsilon$ varies with bulk Mach number $\mathcal{M}$, from subsonic to supersonic regimes, and across different simulation setups in different codes has not been reported in the literature.

A similar argument can be made for effects of numerical resistivity. The local effective resistivity, $\eta_{\rm eff}$, scales with the grid spacing and flow velocity as
\begin{equation}
\eta_{\rm eff} \sim \alpha_{\rm eff} u_0 \Delta x,
\end{equation}
where $\alpha_{\rm eff}$ is a dimensionless coefficient depending on the reconstruction scheme and flux computation method. The magnetic energy dissipation rate due to numerical resistivity can be estimated from the energy equation,
\begin{align}
\frac{\mathrm{D}\ln(B^2/2)}{\mathrm{D}t}\Bigg|_{\eta_{\rm eff}} &= -\left(\frac{2}{B_0^2}\right)\mathbf{B}\cdot\nabla\times(\eta_{\rm eff} \nabla \times \mathbf{B})  \\
&\sim -\frac{2\eta_{\rm eff} (B_0/\Delta x)^2}{B_0^2} \nonumber \\
&= -\frac{2\alpha_{\rm eff}}{\tau} = -\frac{2\alpha_{\rm eff}}{\tau_{\rm sound}}\mathcal{M}\,.
\end{align} 
The dimensionless coefficient $\alpha_{\rm eff}$ has been measured using test problems with known analytical solutions, such as the damping of magnetosonic and Alfv\'{e}n waves and the growth rate of the tearing-mode instability (\citealp{2017ApJS..230...18R}). They show that $\alpha_{\rm eff}$ is not a universal constant for a given scheme but depends on spatial resolution, the specific limiter and Riemann solver, the timestep size (or the Courant-Friedrichs-Lewy number), and whether the flow is smooth or contains strong gradients that activate slope limiting.

The argument is slightly different in (quasi-)Lagrangian codes in which the magnetic divergence is ``carried'' with the particles or cells. Instead of a local characteristic velocity $u_0$ within a grid cell, the spurious energy injection depends on velocity gradients between neighboring fluid parcels (particles or cells), $\delta u_0 = |\nabla {\bf u}|(\Delta x)$, integrated along the path of the fluid parcel. Here, the gradient tensor $|\nabla {\bf u}|$ can be thought of as a shearing timescale $\tau_{\rm shear}\equiv |\nabla {\bf u}|^{-1}$, that is the amount of time for two neighboring parcels to move away from or towards each other by their initial separation distance, $\Delta x$, via shearing or compression. Here, the velocity gradient can be decomposed as $|\nabla{\bf u}| = (\theta^2/3 + 2\sigma^2)^{1/2}$, where $\theta = \nabla\cdot\mathbf{u}$ is the local velocity divergence and $\sigma_{ij} = (\partial_j u_i + \partial_i u_j)/2 - \theta\,\delta_{ij}/3$ is the traceless shear tensor. We exclude the anti-symmetric rotational contribution to $|\nabla\mathbf{u}|$ here because the rigid rotation of two neighboring parcels preserves their separation distance $\Delta x$ and therefore does not change the field gradient across which the divergence error is sourced.

A reasonable estimate for the spurious injection rate is therefore given by $\Gamma_{\nabla\cdot\mathbf{B}} \sim \epsilon\,\delta u_0 B_0^2/\Delta x$, which if normalized like we did before yields
\begin{align}
\left.\frac{{\rm D}\ln(B^2/2)}{{\rm D}t}\right|_{\nabla\cdot\mathbf{B}} &\sim \frac{2\,\delta u_0\, B_0\,(\pm \epsilon B_0/\Delta x)}{B_0^2} \nonumber \\
&= \pm\frac{2\epsilon\,\delta u_0}{\Delta x} = \pm\frac{2\epsilon}{\tau_{\rm shear}} = \pm\,2\epsilon\,|\nabla\mathbf{u}|\,.
\label{eq:div_lagrangian}
\end{align}
This expression can be recasted in terms of a local shear Mach number,
\begin{equation}
\mathcal{M}_{\rm shear} \equiv \frac{\delta u_0}{c_s} = \frac{|\nabla\mathbf{u}|\,\Delta x}{c_s} = \frac{\tau_{\rm sound}}{\tau_{\rm shear}}\,,
\end{equation}
such that Equation \ref{eq:div_lagrangian} is equivalent to 
\begin{equation}
\left.\frac{{\rm D}\ln(B^2/2)}{{\rm D}t}\right|_{\nabla\cdot\mathbf{B}} = \pm \frac{2\epsilon}{\tau_{\rm sound}}\mathcal{M}_{\rm shear}\,.
\end{equation}
This is directly analogous to the Eulerian case, but with the bulk Mach number $\mathcal{M}$ replaced by $\mathcal{M}_{\rm shear}$. This shear Mach number could be thought of in terms of the local velocity gradient's ability to distort the flow before thermal pressure ``smoothens'' it out. When $\mathcal{M}_{\rm shear}\gtrsim 1$, neighboring fluid parcels are being sheared or compressed faster than sound waves can communicate across the separation $\Delta x$, so any \mbox{divergence} error carried by the field is amplified on a timescale shorter than the local sound crossing time. On the other hand, when $\mathcal{M}_{\rm shear}\lesssim 1$, pressure forces have time to smooth out the relative motions between neighbors before the divergence errors can appreciably act on the magnetic field $\mathbf{B}$. Therefore, $\mathcal{M}_{\rm shear}$ plays the same role for Lagrangian codes that the bulk Mach number $\mathcal{M}$ plays for Eulerian codes. One can show that a similar result can be obtained for effects due to numerical dissipation, but with $\eta_{\rm eff} \sim \alpha_{\rm eff}|\nabla{\bf u}|(\Delta x)^2$ such that,
\begin{equation}
\frac{\mathrm{D}\ln(B^2/2)}{\mathrm{D}t}\Bigg|_{\eta_{\rm eff}} = -\frac{2\alpha_{\rm eff}}{\tau} = -\frac{2\alpha_{\rm eff}}{\tau_{\rm sound}}\mathcal{M}_{\rm shear}\,.
\end{equation} 

In this framework, neighboring parcels that are rotating in a disk at the same angular velocity would experience no relative shearing motion and hence would not induce numerical field growth due to divergence errors or numerical dissipation, i.e. $\tau_{\rm shear}\to \infty$ or $\mathcal{M}_{\rm shear} \to 0$. However, while field amplification is governed by differential rotation ($\propto r\,\partial\Omega/\partial r$), numerical effects that may seed spurious field growth can depend on the full velocity gradient, 
\begin{equation}
    |\nabla{\bf u}| =\left[\Omega^2+\left(\Omega+r\frac{\partial\Omega}{\partial r}\right)^2\right]^{1/2}\, ,
\end{equation}
and therefore potentially on both the local angular velocity $\Omega$, and its radial variation, $\partial\Omega/\partial r$. Consequently, the effect of divergence errors and numerical dissipation is thus expected to vary sensitively with the local flow morphology and particle arrangement, potentially becoming more significant in regions undergoing strong shear, compression, or repeated shock interactions where neighboring particle trajectories diverge rapidly. In (quasi-)Lagrangian methods, the resulting numerical field amplification or dissipation can therefore depend sensitively on the specific treatment of interactions between neighboring fluid parcels.

\begin{deluxetable*}{lll}[htbp!]
\tablecaption{Summary of the simulation codes used in our code comparison study.\label{table:codes}}
\tablehead{
\colhead{Simulation code} & \colhead{Spatial discretization} & \colhead{Divergence constraint scheme}
}
\startdata
\textsc{Ramses} & Cartesian oct-tree AMR \citep{2002AandA...385..337T} & Constrained Transport (CT)  \\
& &\citep{1988ApJ...332..659E, 2006AandA...457..371F}\\
\textsc{ART} & Cartesian oct-tree AMR  & Dedner  \\
&\citep{1997ApJS..111...73K, 2008ApJ...672...19R} & \citep{2002JCoPh.175..645D, 2010JCoPh.229.2117M}\\
\textsc{Arepo} & Unstructured moving-mesh \citep{2010MNRAS.401..791S} & Powell   \\
&& \citep{1999JCoPh.154..284P, 2013MNRAS.432..176P}\\
\textsc{Swift} & Smoothed-particle hydrodynamics (SPH) & Dedner + Powell (\citealp{1999JCoPh.154..284P}; \\
&\citep{2024MNRAS.530.2378S} & \citealp{2002JCoPh.175..645D, 2026arXiv260806978K})
\enddata
\end{deluxetable*}

\subsection{Mitigating divergence errors in simulations}
\label{sec:divergenceschemes}
Modern MHD solvers employ a variety of divergence-control algorithms each with different trade-offs between accuracy, conservation, and computational cost. 

The \textit{Powell scheme} \citep{1999JCoPh.154..284P} treats the divergence of the magnetic field as a passively advected scalar field $\phi=(\nabla\cdot \textbf{B})/\rho$ that evolves according to the equation of the form,
\begin{equation}
    \frac{\partial(\rho\phi)}{\partial t} + \nabla\cdot(\rho\textbf{u}\phi) = 0 \,.
\end{equation}
This form is derived by adding a source term that is proportional to $\nabla\cdot\textbf{B}$ to the momentum, energy and induction equations (see Equations 20-22 of \citealp{2013MNRAS.432..176P}). In this framework, the divergence of \textbf{B} is a conserved fluid variable that correspond to an eighth MHD wave, in addition to seven waves that are already baked into the ideal MHD equations. However, since the monopole term is just advected away from the source, the scheme is not conservative and can produce unphysical results (see Figures 5 and 8 in \citealp{2016MNRAS.462..576H}, and Figures 5 and 7 in \citealp{2016MNRAS.455...51H}).

In the \textit{Generalized Lagrange Multiplier (GLM) / Dedner scheme} \citep{2002JCoPh.175..645D}, divergence errors are both propagated away from their source at a finite cleaning speed $c_h$ and dissipated by a parabolic correction with artificial conductivity $c_p^2$. The scalar field $\Psi$ acts as a divergence charge density-- wherever $\nabla\cdot\mathbf{B}\neq 0$, $\Psi$ is sourced proportionally and then either carried away as a wave or locally damped, preventing the error from accumulating in the magnetic field. The damping rate per timestep is
\begin{equation}
    \Psi^{n+1} = \Psi^n \exp\left(-\gamma c_h \frac{\Delta t}{\Delta x}\right),
\end{equation}
where $\gamma$ is the damping coefficient ($\alpha$ in \citealp{2010JCoPh.229.2117M}), while $c_h$ itself is set as a fraction $f_{\rm ch}$ of the maximum CFL-limited signal speed. Large $\gamma$ dissipates divergence errors rapidly before $\Psi$ can propagate across a cell, while small $f_{\rm ch}$ allows errors to accumulate
locally before being propagated away (see \S\ref{sec:divergence_art}).

Lastly, the \textit{Constrained Transport (CT) scheme} \citep{1988ApJ...332..659E, 2005JCoPh.205..509G} evolves the magnetic flux through cell faces rather than the cell-centered field components such that the discrete divergence across a cell is identically zero at all times. The magnetic field is staggered with respect to the fluid variables, i.e. electromotive forces along cell edges are used to update the face-centered magnetic flux. Because the divergence-free condition is satisfied by construction at the cell interfaces rather than enforced as a post-processing correction, CT does not introduce free parameters and does not incur additional wave modes. However, its reliance on a staggered mesh representation makes it difficult to implement on unstructured, adaptive \citep{2006AandA...457..371F, 2009JCoPh.228..952L} or moving-mesh codes \citep{2016MNRAS.463..477M}. This difficulty was recently discussed by \citet{2026arXiv260706660P}, who showed that the constrained transport scheme in \textsc{Arepo} is numerically unstable for complex 3D flows such as isothermal, subsonic MHD turbulence, despite illustrating good performance in simpler test problems \citep{2016MNRAS.463..477M}. Among the divergence schemes listed above, CT remains the benchmark, maintaining $\nabla\cdot\mathbf{B}$ at machine precision by construction.

\section{Codes and Methodology}
\subsection{Codes in this comparison}
\label{codes}
The codes in this ideal MHD simulation comparison are the Eulerian adaptive mesh refinement (AMR) codes ART and \textsc{Ramses}, the moving-mesh code \textsc{Arepo}, and the SPMHD code \textsc{Swift}. A summary of the most relevant features of each code is presented in Table \ref{table:codes}.

\subsubsection{\textsc{Ramses} + CT}
\textsc{Ramses} is a grid-based AMR code for astrophysics that implements an unsplit second-order Godunov method \citep{2002AandA...385..337T} and enforces the divergence-free condition of magnetic fields using constrained transport \citep{2006AandA...457..371F}. The code supports different Riemann solvers, such as HLL and HLLD, to compute fluxes, a multigrid solver for self-gravity, and a particle-mesh method for modeling collisionless dynamics. \textsc{Ramses} has been extensively used in major simulation campaigns such as \textsc{Horizon-AGN} \citep{2014MNRAS.444.1453D}, \textsc{Horizon-MareNostrum} \citep{2008MNRAS.390.1326O}, \textsc{NewHorizon} \citep{2021A&A...651A.109D} and in previous studies of the galactic dynamo \citep{2016MNRAS.457.1722R, 2017MNRAS.471.2674R, 2017MNRAS.472.4368R}.

\subsubsection{ART + Dedner}
The Adaptive Refinement Tree (ART) code is a cosmological N-body \citep{1997ApJS..111...73K} and hydrodynamics \citep{2008ApJ...672...19R} code designed for high-resolution simulations of structure formation in the universe. This project is the first application of the most recent version of \textsc{ART}  (Diemer, Semenov, Ugalino et al, \textit{in preparation}) which includes an ideal MHD solver based on the MHD-HLLC solver by \citet{2005JCoPh.203..344L}. In this version, we implement a directionally split, cell-centered MHD solver based on the CTU-GLM scheme by \citet{2010JCoPh.229.2117M} to satisfy the divergence constraint. \textsc{ART}  had been used in studies of star formation, feedback, ISM dynamics and chemistry, and reionization physics in both idealized and cosmological galaxy simulations \citep{2014ApJ...793...29G, 2016ApJ...833..202K, 2017ApJ...845..133S, 2017ApJ...834...69L, 2018ApJ...861....4S, 2024ApJ...966..172P, 2025ApJ...989..219S}.

\begin{figure*}[ht!]
\centering
\includegraphics[width=0.95\linewidth]{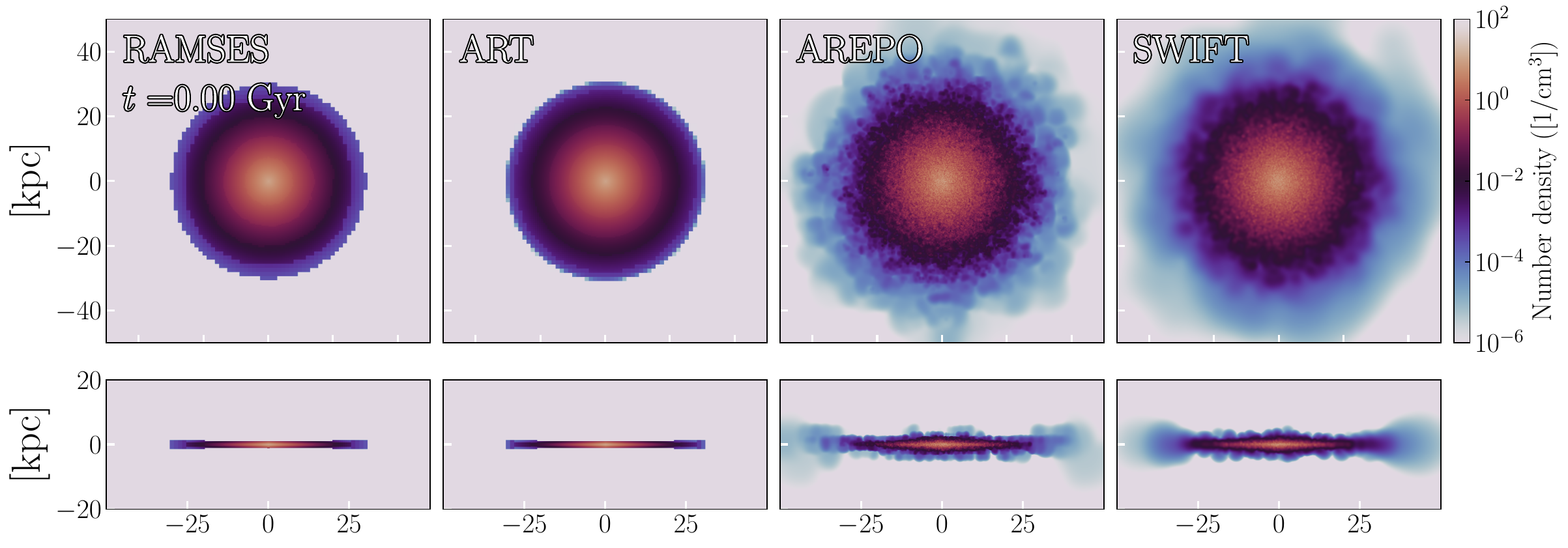}
\caption{\textbf{Initial number density slices for each code}. The gaseous disk is initialized with a double-exponential density profile characterized by a radial scale length of $r_d = 3.432$ kpc and a vertical scale height of $z_d = 0.343$ kpc. Minor differences in the resulting density distributions reflect the different spatial discretization and refinement schemes adopted by each code.}
\label{fig:init_cond}
\end{figure*}

\subsubsection{\textsc{Arepo} + Powell}
\textsc{Arepo} is a moving-mesh astrophysical simulation code that solves the equations of ideal MHD using a Godunov scheme on an unstructured Voronoi mesh that moves with the local fluid velocity \citep{2010MNRAS.401..791S}. Unlike AMR codes such as \textsc{Ramses} and \textsc{ART}, which employ a hierarchical grid structure for refinement, \textsc{Arepo} uses a quasi-Lagrangian approach where the mesh is regenerated every timestep as gas moves in the domain. \textsc{Arepo} employs divergence control for magnetic fields using Powell \citep{2013MNRAS.432..176P} and Dedner \citep{2011MNRAS.418.1392P} on a moving mesh, with support for Riemann solvers such as HLL and HLLD. Most applications of \textsc{Arepo} in the literature use the Powell method to impose the divergence-free constraint. \textsc{Arepo} includes a tree-based gravity solver, radiative cooling, star formation, and feedback modules. This code had been used in large cosmological simulations to study galaxy formation like IllustrisTNG \citep{2019ComAC...6....2N}, in zoom-in simulation suites like Auriga \citep{2017MNRAS.467..179G} and in previous studies of the galactic dynamo \citep{2013MNRAS.432..176P, 2016MNRAS.463..477M, 2017MNRAS.469.3185P, 2018MNRAS.481.4410P, 2024MNRAS.528.2308P, 2020A&A...641A.165N, 2022MNRAS.515.4229P, 2025A&A...701A.114T}.

\subsubsection{\textsc{Swift} + Dedner and Powell}
\textsc{Swift} is a task-based highly-parallel code for cosmological simulations \citep{2024MNRAS.530.2378S} that solves the equations of ideal MHD using an SPMHD scheme \citep{2026arXiv260806978K} on top of the \textsc{Sphenix} flavor of SPH \citep{2022MNRAS.511.2367B}. The divergence-free 
condition of the magnetic field is enforced using the Dedner scheme with an additional Powell term embedded in the equations of motion \citep[see also][]{2018PASA...35...31P}. The code includes support for self-gravity via a fast multipole method, star formation, stellar and AGN feedback modules. \textsc{Swift} 
has been used (without MHD) in studies of cosmology and galaxy formation as part of the FLAMINGO and COLIBRE projects \citep{2023MNRAS.526.4978S, 2025arXiv250821126S}.

\subsection{Methodology}
\label{method}
We run isolated Milky Way-mass galaxy simulations with a common set of initial conditions. We perform these simulations without star formation and feedback to isolate the effects of the numerical methods implemented in each code. In the following sections, we discuss the initial conditions, the AMR criteria applied for (de-)refinement, and the cooling prescription and pressure floor implemented in each code.

\subsubsection{Common initial conditions}
We use the isolated disk initial conditions from the AGORA Project \citep{2016ApJ...833..202K} in this work. These initial conditions consist of a dark matter halo, a stellar bulge, and a disk of gas and stars. The gaseous disk has a total mass of $8.593\times 10^9\ M_\odot$ with a double exponential profile given by,
\begin{equation}
\rho(r,z) = \rho_0 \exp\left(-\frac{r}{r_d}\right)\exp\left(-\frac{|z|}{z_d}\right) \,,
\end{equation}
where $\rho_0=M_{\rm gas}/(4\pi r_d^2 z_d)$. We impose a gas density floor $\rho_\mathrm{gas, floor} \approx 10^{-31}\ \mathrm{g}\ \mathrm{cm}^{-3}$ or a number density floor $n_\mathrm{floor} \sim 10^{-7} /\mathrm{cm}^3$ on the entire domain. The stellar component of the disk has a total mass of $3.438\times 10^{10}\ M_\odot$ and $10^5$ particles. We adopted the same values for the characteristic radius and height of the gas and stellar disk as in AGORA, $r_d = 3.432$ kpc and $z_d = 0.1\ r_d = 0.343$ kpc. The circular velocity profile of the disk is set such that all components are in rotational equilibrium. 

The dark matter (DM) halo follows a Navarro-Frenk-White (NFW) profile \citep{1997ApJ...490..493N} with concentration parameter $c=10$ and spin parameter $\lambda = 0.04$. The halo has $10^5$ particles and a total mass of $M_{200c}=1.074\times 10^{12}\ M_\odot$ within a radius $R_{200c}=205.5$ kpc with circular velocity $v_c = 150\ \mathrm{km}\ \mathrm{s}^{-1}$. The stellar bulge follows a Hernquist profile \citep{1990ApJ...356..359H} with a total mass of $4.297\times 10^9\ M_\odot$ and $1.25\times 10^4$ particles, which correspond to a bulge-to-disk ratio $M_\mathrm{bulge}/M_\mathrm{disk} = 0.1$. In our fiducial setup, the DM particles are two orders-of-magnitude more massive than the gas particles. Because massive collisionless particles can source strong local gravitational perturbations that may drive additional small-scale gas motions in (quasi-)Lagrangian codes, we also tested whether increasing the DM and stellar particle resolution at fixed gas resolution significantly affects magnetic field amplification in \textsc{Arepo} (see \S\ref{sec:resolution_effects}).

We set the initial temperature constant across the entire domain at $T = 10^4$ K. The initial magnetic field is uniform and aligned with the $z$-axis, $\mathbf{B} = B_0\ \hat{z}$, with $B_0 = 10^{-12}\,\mathrm{G}$. Although we have tested multiple field orientations and found no appreciable dependence of our results on this choice, we note that such insensitivity may not hold in general. While \citet{2020MNRAS.499.2076S} demonstrated that the initial field geometry does not affect the amplification of small-scale magnetic fields, \citet{2016MNRAS.457.1722R} found differences in large-scale field evolution when comparing dipolar and quadrupolar seed field configurations.

\begin{figure*}[ht!]
\centering
\includegraphics[width=\linewidth]{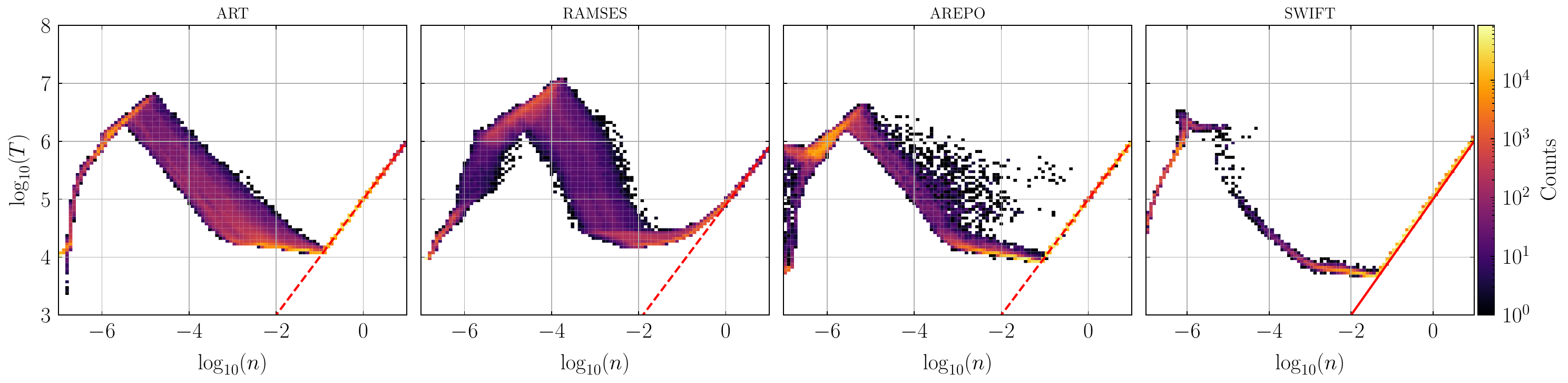}
\caption{\textbf{Number density vs. temperature ($n$–$T$) phase diagrams} for the entire simulation domain at $t = 4.0\,\mathrm{Gyr}$ for each code. The pressure floor adopted in each simulation is shown as a dashed line. Differences in the cooling prescriptions primarily affect gas immediately surrounding the galactic disk and in the circumgalactic medium, where the number density is below $10^{-1}\,\mathrm{cm}^{-3}$.}
\label{fig:nTplot}
\end{figure*}

For the \mbox{(quasi-)Lagrangian} codes \textsc{Arepo} and \textsc{Swift}, we generate the initial distribution of gas particles for our disk galaxy simulations using a Monte Carlo sampling approach \footnote{A random Monte Carlo sampling does not, in general, produce a particle distribution that minimizes the SPH Lagrangian. The subsequent relaxation and rearrangement of particles can therefore introduce additional particle noise.}. The particle and cell positions are sampled from the corresponding probability distribution,
\begin{equation}
\mathcal{P}(r, z) \propto r \exp\left(-a\frac{r}{r_d}\right) \exp\left(-\frac{|z|}{z_d}\right),
\end{equation}
where the factor of $r$ arises from the cylindrical volume element, ensuring that the particle distribution sampled reproduces the desired density profile. The parameter $a$ controls the radial concentration, which we set to 0.9 for our simulations to carefully match their radial and vertical density profiles. We sample particle positions from this distribution using \texttt{emcee} \citep{2019JOSS....4.1864F} by evaluating the log-probability, $\ln[\mathcal{P}(r,z)]$, with 128 walkers and 8,000 steps, discarding the first 500 steps as burn-in. This produces an accurate sampling of the three-dimensional density profile. From the posterior samples, we randomly select particles and convert their cylindrical coordinates $(r, z)$ to Cartesian coordinates $(x, y, z)$ by uniformly distributing the azimuthal angle $\phi \in [0, 2\pi)$. The initial density slices are shown in Figure \ref{fig:init_cond}. We tested the effects of this random sampling by running several realizations and found no appreciable differences in either the field amplification or the dynamics of the galactic disk. Using the circular velocity profile provided by the AGORA collaboration \citep{2016ApJ...833..202K}, we interpolate the circular velocity of the disk at each grid point or particle position. 

In \textsc{Arepo}, we first regrid the initial conditions onto an oct-tree AMR mesh using the \texttt{ADDBACKGROUNDGRID} option before running the full simulation. This step ensures that the ambient medium surrounding the galactic disk remains stable in the early stages of the simulation. Skipping this step leads to the formation of overdensities that later merge into the galactic disk. We also activate the \texttt{NODEREFINE\_BACKGROUND\_GRID} flag in \textsc{Arepo} which prevents cells in the low-resolution background grid from being derefined during the course of the simulation. Similarly, we model the \textit{static} ambient medium in \textsc{Swift} by adding gas particles of the same mass on a uniform Cartesian mesh that does not overlap with the galactic disk, with a prescribed target density and a volume that encloses all dark matter and star particles.

The gaseous disk, with an exponential density profile and total mass of $8.593 \times 10^9\ M_\odot$ (representing 20\% of the stellar disk mass), is sampled using $10^5$ particles. This corresponds to gas particle mass of $8.593\times10^4\ M_\odot$. For grid-based codes, a corresponding spatial resolution of $\Delta x = 80$ pc is used, providing comparable resolution to the Lagrangian particle-based approaches at each refinement level.

In \textsc{ART} and \textsc{Ramses}, refinement is controlled using two independent criteria. Cells with more than 8 dark matter and/or star particles are refined to ensure that there is one particle per cell on average. A gas mass refinement criterion is also applied to ensure that cells at the highest refinement level achieve a gas mass resolution consistent with the corresponding Lagrangian mass resolution. This refinement strategy is effectively quasi-Lagrangian at more refined levels since the refinement criteria follow the local mass distribution, maintaining approximately constant mass per resolution element as the simulation evolves. We also impose the same gas mass refinement criterion for \textsc{Arepo}.

\subsubsection{Cooling and Pressure Floor}
Implementing the same cooling prescription across different simulation codes is non-trivial, since each code implements gas cooling in different ways. The AGORA Collaboration addressed this by integrating the \textsc{Grackle} library \citep{2017MNRAS.466.2217S} into all of the codes included in their comparison. However, this has not been implemented in the code versions used in our work. We therefore use the following cooling prescriptions, calibrated to solar metallicity, together with a pressure floor that is high enough such that differences in low-temperature cooling are irrelevant. Ideally, with this setting, the galactic disk should evolve as a smooth, nearly axisymmetric rotating disk, and by the Cowling anti-dynamo theorem \citep{1933MNRAS..94...39C,1957QJMAM..10..129C}, no self-sustained dynamo is expected to amplify the galactic magnetic field. However, instabilities arising from cooling and the initial collapse of the gas can generate velocity perturbations that violate the theorem's underlying assumptions, potentially allowing dynamo action to occur. 

In our \textsc{ART} simulations, we adopt the cooling rates from \citet{2012ApJS..202...13G}, which provide a self-consistent treatment of metal cooling in the presence of an ionizing ultraviolet background. These rates were computed using \textsc{Cloudy} for a gas in thermal and ionization equilibrium, accounting for both collisional and photoionization processes across a wide range of densities, temperatures, and metallicities. These rates were used to derived approximate functions for cooling and heating that have an explicit dependence on density, metallicity, and the radiation field.

Meanwhile, for \textsc{Ramses}, we utilize the cooling function in \citet{2005A&A...433....1A} (and also \citealp{1995ApJ...443..152W, 2003ApJ...587..278W}), which models the thermal evolution of the neutral atomic interstellar medium (ISM) through a net cooling rate calibrated to reproduce metal-line cooling, primarily from C\,\textsc{ii} and O\,\textsc{i} in the temperature range $10^2 \lesssim T \lesssim 10^4 \ \mathrm{K}$, H cooling due to the Ly$\alpha$ line, and by electron recombination onto positively charged grains. This cooling function yields a thermally bistable ISM with warm and cold neutral phases separated by a thermally unstable regime.

Cooling and heating in \textsc{Arepo} uses tabulated rates that depend on the gas density and temperature within a cell. We employ the cooling and heating prescription of \citet{1996ApJS..105...19K}, which assumes collisional ionization equilibrium for a primordial composition gas consisting of hydrogen and helium. The cooling and photoheating rates are tabulated as a function of temperature and redshift, accounting for the ionizing background radiation from a uniform UV background model. This model includes radiative cooling from collisional excitation, collisional ionization, recombination, free-free emission, and Compton cooling. 

For \textsc{Swift}, we adopt the cooling rates from \citet{2009MNRAS.393...99W}, which tabulate radiative cooling rates for optically thin gas in ionization equilibrium as a function of hydrogen number density, temperature, and redshift, accounting for the effects of photoionization by a UV and X-ray background. These rates are computed using \textsc{Cloudy} and account for the contribution of both primordial and metal-line cooling across a wide range of densities and temperatures.

As we mentioned previously, we approximate the integral effect of star formation and feedback with a pressure floor that keeps the galactic disk thick. Ensuring a consistent treatment of star formation and feedback across different codes is highly nontrivial, so we circumvent it with this simplification. The pressure floor enforces,
\begin{equation}
\frac{P}{k_B n^2} \geq 10^5\ \mathrm{K}\ \mathrm{cm}^3 
\end{equation}
which is equivalent to keeping the Jeans length constant since
\begin{equation}
\lambda_J^2 = \frac{15}{4\pi G\mu^2m_p^2} \left(\frac{P}{n^2}\right)\,.
\end{equation}
\noindent This floor ensures that there is sufficient pressure support to keep the disk thick and smooth. Implementing this high pressure floor removes the sensitivity of our results to differences in the cooling prescriptions within the galactic midplane, where $n \gtrsim 10^{-1}\,\mathrm{cm}^{-3}$. Gas cools rapidly through the atomic cooling regime until it reaches the imposed pressure floor, preventing it from entering the cold, dense ISM regime. In \textsc{Swift}, this is implemented following the EAGLE model, using two polytropic entropy floors (a low-density ``Cool'' floor and a high-density ``Jeans'' floor), each of the form $P = P_c(\rho/\rho_c)^\gamma$, where we set the Jeans floor to $\gamma=2$, while the Cool floor plays no role at these densities.

The treatment of the pressure floor is similar across Eulerian and \mbox{(quasi-)Lagrangian} codes, though the code implementation differs. In \textsc{Ramses}, the pressure floor is imposed as an additional polytropic ($P\propto n^2$) internal energy contribution $e_\mathrm{poly}$ to the thermal energy, such that the cell's effective energy is $e_\mathrm{eff} = e_\mathrm{thermal} + e_\mathrm{poly}$ at all times. In ART, \textsc{Arepo}, and \textsc{Swift}, the floor is imposed as a maximum cut and the thermal energy is replaced by $e_\mathrm{eff} = \max(e_\mathrm{thermal}, e_\mathrm{poly})$, so the polytropic term only takes over when $e_\mathrm{thermal}$ would otherwise fall below it. We caution that this distinction between addition and an actual floor matters primarily at intermediate densities where $e_\mathrm{thermal} \sim e_\mathrm{poly}$; at high densities, where $e_\mathrm{poly} \gg e_\mathrm{thermal}$, the sum and the maximum converge, and the two implementations yield essentially the same effective pressure. As shown in Figure \ref{fig:nTplot}, the $n$-$T$ distribution at $t=4.0$ Gyr for all simulations clearly shows this effect at densities of $n\sim 10^{-1} \,{\rm cm}^{-3}$, corresponding to the outskirts of the galactic disk. 

\section{Results}
\label{results}

\subsection{Field growth in fiducial simulations}
\label{sec:fiducial}

\begin{figure*}[h!]
    \centering
    \includegraphics[width=0.8\linewidth]{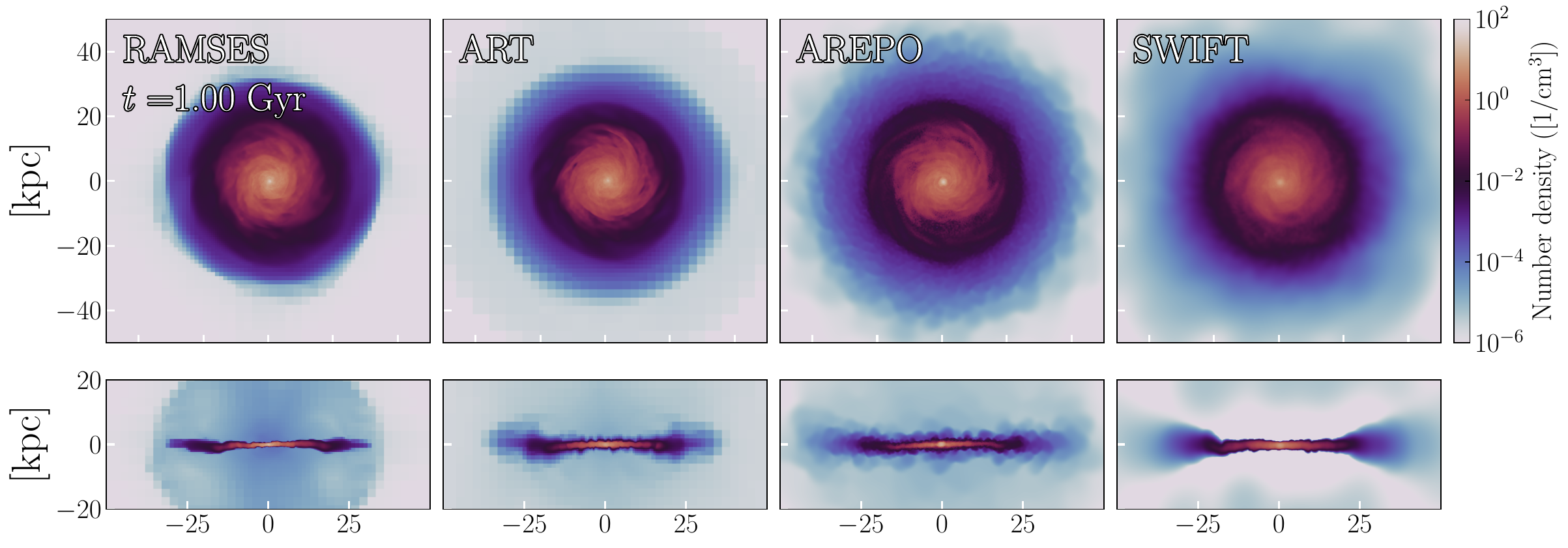}
    \includegraphics[width=0.8\linewidth]{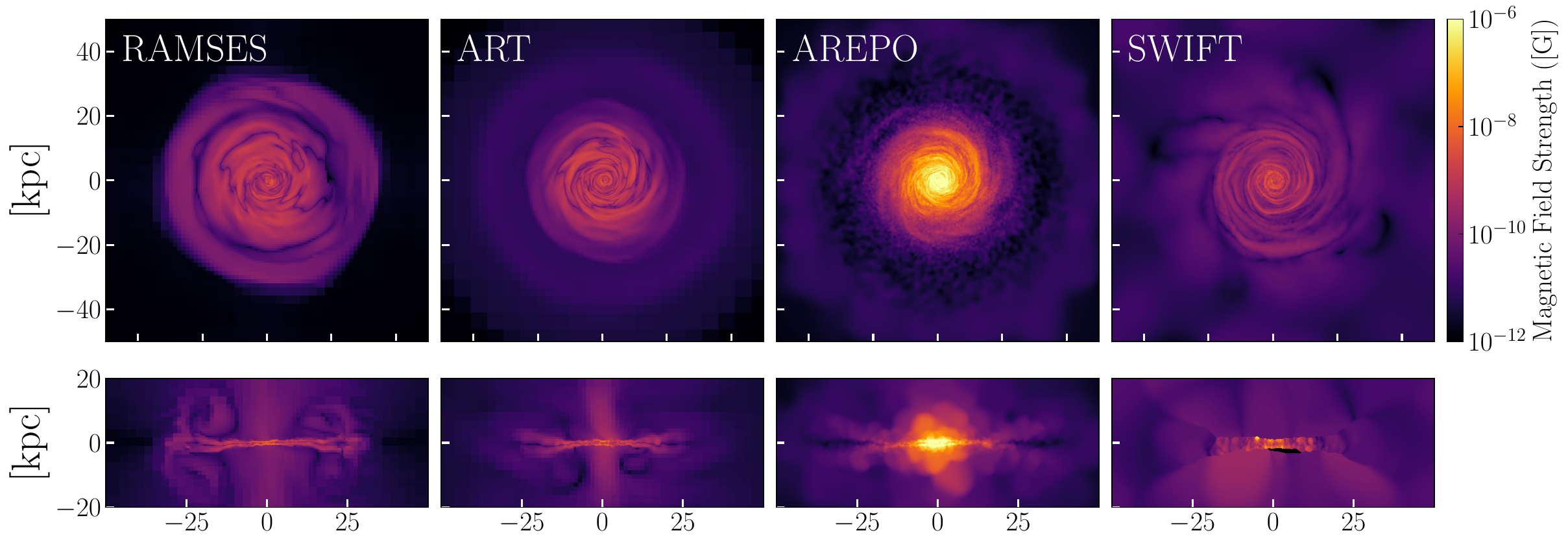}
    \caption{\textbf{Midplane slices of number density (top) and magnetic field strength (bottom)} at $t=1.0$~Gyr for each code within a (100~kpc)$^2$ box centered on the disk's center of mass. The density structure along the midplane is similar across all codes, with spiral arms forming in each case. However, a slightly more dense gaseous halo not observed in the other simulations forms around the disk in our fiducial \textsc{Ramses} run. The magnetic fields in the \textsc{Arepo} simulation are strongest near the central region of the disk, in contrast to the Eulerian and \textsc{Swift} results, where the fields are more uniform in magnitude across the disk.}
    \label{fig:100Gyr}

    \includegraphics[width=0.8\linewidth]{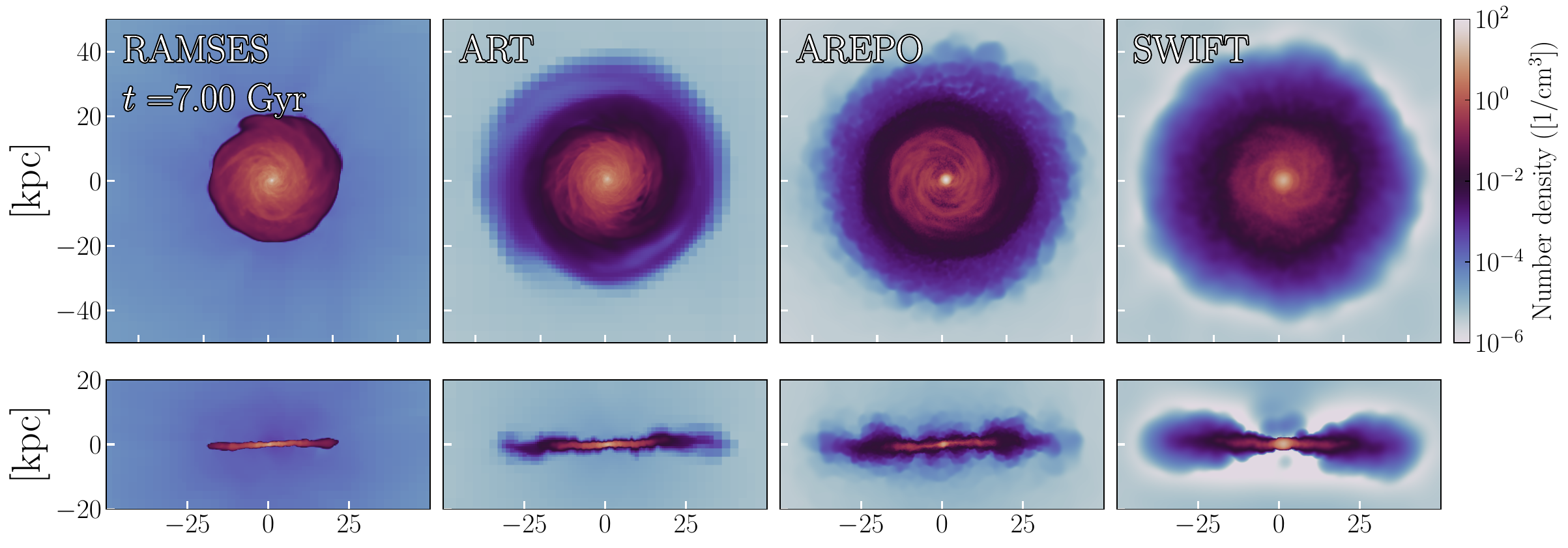}
    \includegraphics[width=0.8\linewidth]{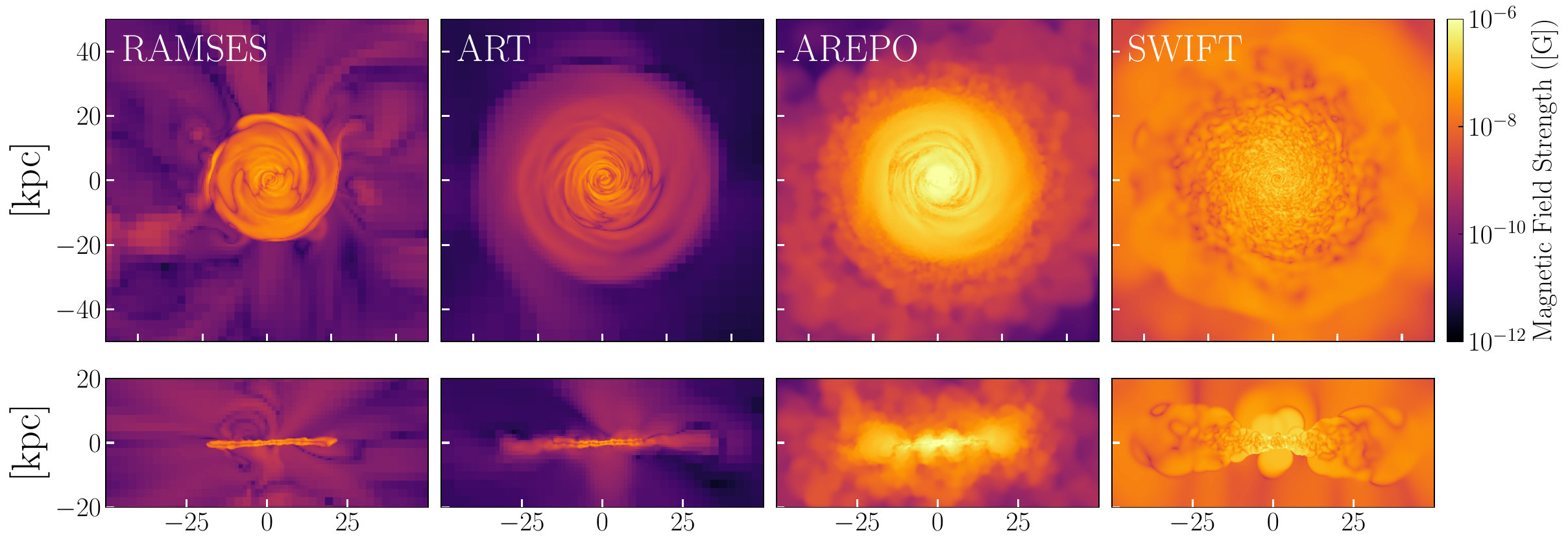}
    \caption{Same as Figure~\ref{fig:100Gyr} but for simulation snapshot at $t=7.0$ Gyr. The galactic magnetic field in the Eulerian codes is weaker and more tightly confined within the disk, whereas the field in the Lagrangian codes are stronger and more radially and vertically extended due to greater magnetic pressure support along the midplane. The galactic disk in the \textsc{Ramses} simulation is more compact than the rest of the codes.}
    \label{fig:700Gyr}
\end{figure*}

\begin{figure*}[ht!]
    \centering
    \includegraphics[width=\linewidth]{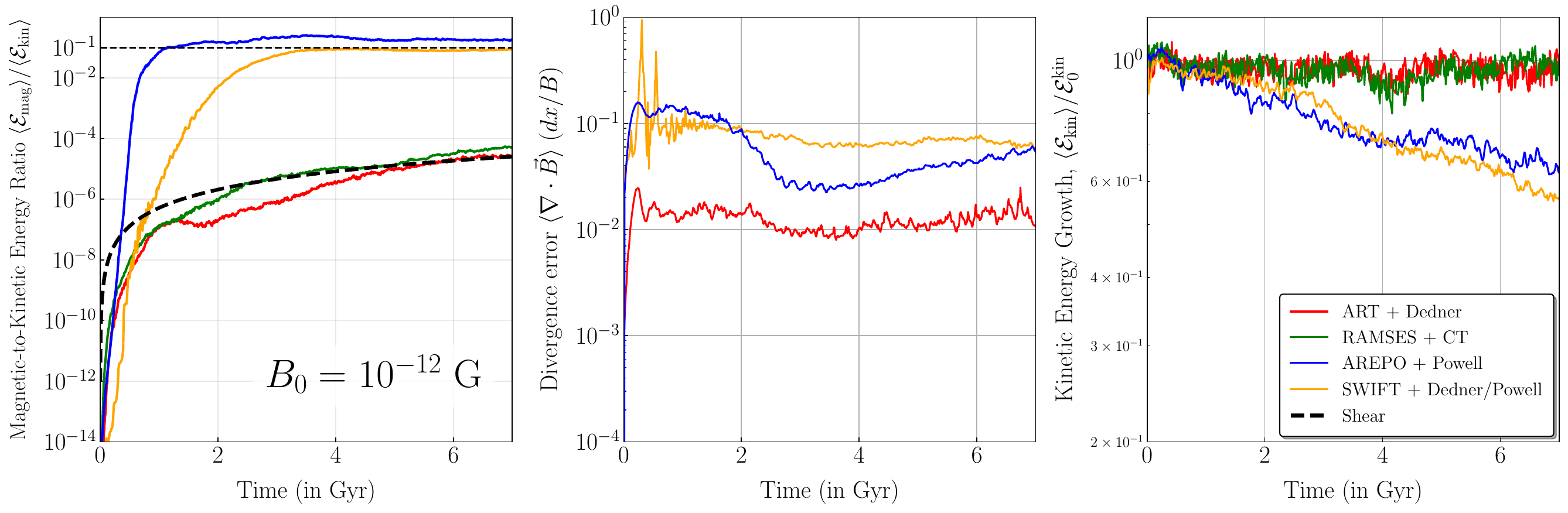}
    \caption{\textbf{Time evolution of the volume-averaged magnetic-to-kinetic energy ratio (left), dimensionless divergence errors (center), $(\nabla\cdot\mathbf{B})(\Delta x/B)$ and volume-averaged kinetic energy (right),} within a cylindrical disk with radius $r_d = 15$ kpc and height $h_d = 0.5$ kpc. The black curve in the left panel represents the expected result for a differentially rotating disk without feedback with shearing rate $q\Omega = 1.0\ {\rm Myr}^{-1}$, following an initial gravitational collapse. Equipartition is reached when the ratio $\langle \mathcal{E}_{\rm mag}\rangle / \langle\mathcal{E}_{\rm kin}\rangle \sim 1.0$. Lagrangian codes yield magnetic-to-kinetic energy ratios that grow exponentially toward near-equipartition (20\% for \textsc{Arepo} and 10\% for \textsc{Swift}), whereas Eulerian codes grow to final values three orders of magnitude lower, in closer agreement with shear-driven amplification. The center plot shows the corresponding volume-weighted average dimensionless divergence error, demonstrating that the Dedner method in \textsc{ART} maintains the lowest divergence among non-CT schemes, while CT in \textsc{Ramses} keeps the divergence at machine precision ($\sim 10^{-16}$). By the end of the simulation, the average kinetic energy within the galactic disk decreases to 60\% of its initial value in \textsc{Arepo} and \textsc{Swift}, whereas in \textsc{Ramses} and \textsc{ART} it fluctuates around the initial value. The decrease in kinetic energy is attributed to an increase in magnetic pressure support within the galactic disk, which is driven by the early growth of the magnetic field to saturation.}
    \label{fig:amplification_divergence_codes}
    \includegraphics[width=\linewidth]{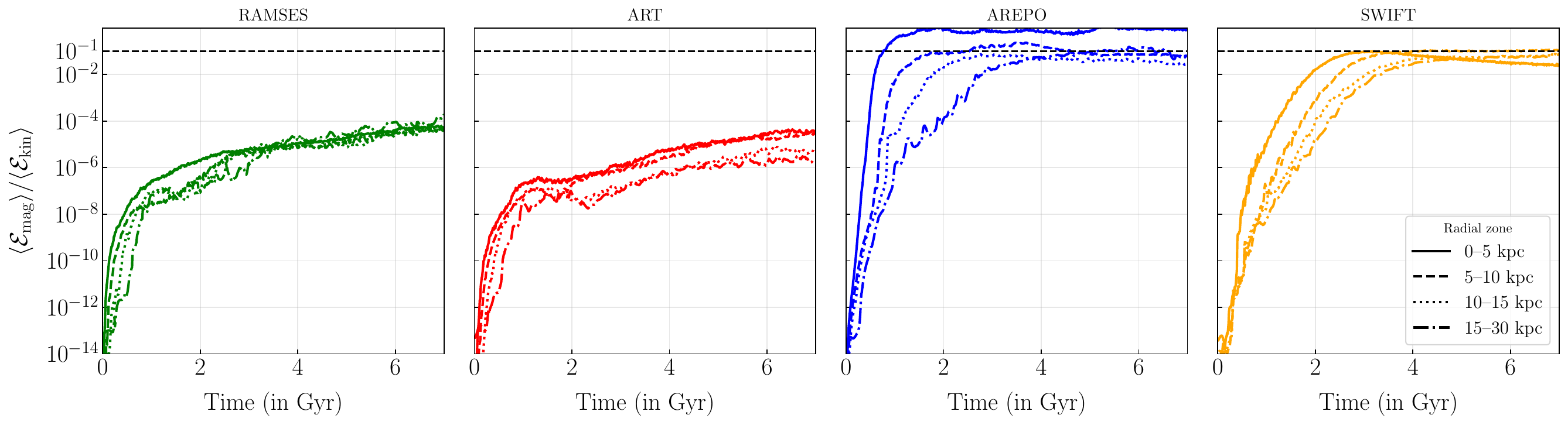}
    \caption{\textbf{Time evolution of the volume-averaged magnetic-to-kinetic energy ratio} within different radial zones away from the galactic disk's center of mass and a fixed height $h_d = 0.5$ kpc. In the \mbox{(quasi-)Lagrangian} codes, the growth of magnetic energy is slower in annuli farther from the rotation axis. This behavior is not observed in the Eulerian codes, where the magnetic energy appears to grow at nearly the same rate across the entire galactic disk. }
    \label{fig:radial1}
\end{figure*}

In this section, we characterize the growth of the magnetic field within the galactic disk and compare the magnetic field structure during the early- and late-time phases in our fiducial simulations.

Figures \ref{fig:100Gyr} and \ref{fig:700Gyr} show slices of the magnetic field strength and number density through the galactic midplane at $t=1.0$ Gyr, and at the end of the simulations, $t=7.0$ Gyr, for each of the four codes. The morphology reflects the substantially different amplification histories. At $t=1.0$~Gyr, \textsc{Arepo} already exhibits a bright, centrally concentrated field with well-developed spiral structure, while the rest of the simulations exhibit a weaker and more uniform profile, remaining well below equipartition, as shown in Figure~\ref{fig:amplification_divergence_codes}. The midplane and vertical density slices show no appreciable difference across codes, except in \textsc{Ramses}, where a denser gaseous halo surrounds the disk.

By $t=7.0$~Gyr, the \textsc{Ramses} and \textsc{ART} slices show a thin, well-confined magnetic field distribution close to the midplane, with field strengths that are still uniform across the disk up to $r\sim 25$ kpc. The disk in \textsc{Ramses} is noticeably more compact than the rest of the codes, with an ambient medium that is an order of magnitude denser. \textsc{Arepo} and \textsc{Swift}, by contrast, produce a more extended vertical profile with significant field strength above and below the disk, together with a thicker gas disk.

Figure~\ref{fig:amplification_divergence_codes} shows the evolution of the volume-weighted mean magnetic energy, the dimensionless divergence error, $(\nabla\cdot{\bf B})(\Delta x/B)$, and the kinetic energy within the inner galactic disk ($r<15$ kpc, $|z|<0.5$ kpc), which are used to track magnetic field amplification and divergence control throughout the simulation. As the galactic disk settles into a quasi-equilibrium state, the gas undergoes adiabatic compression, leading to an initial amplification of the magnetic field. Further amplification is subsequently driven by differential rotation, causing the magnetic energy to grow quadratically with time,
\begin{equation}
    \mathcal{E}_{\rm mag} = \mathcal{E}_{\rm mag, 0} \left[1 + \left(\frac{t}{\tau}\right)^2 \right],
\end{equation}
\citep{2010A&A...523A..72D, 2016MNRAS.457.1722R}, where the characteristic timescale $\tau$ is determined by the shear rate $q\Omega$ (see \S\ref{theory}). We include a reference curve in the evolution plot corresponding to growth driven at a rate $\tau^{-1}=q\Omega = 1.0\ {\rm Myr}^{-1}$ which corresponds to conditions within the inner 1 kpc of a Milky Way-like disk. The left plot shows that the different numerical schemes exhibit substantially different magnetic field amplification histories during the early stages of evolution. The moving-mesh code \textsc{Arepo} amplifies the magnetic energy exponentially and reaches near-equipartition values within roughly 1 Gyr. \textsc{Swift} amplifies the field more slowly but likewise approaches near-equipartition, doing so roughly 2 Gyr after \textsc{Arepo}. By contrast, the grid-based schemes, \textsc{Ramses} and \textsc{ART}, show slower magnetic growth that is bounded by the $\mathcal{E}_{\rm mag}\propto t^2$ reference curve, remaining several orders of magnitude below equipartition throughout the simulation. 

The center panel of Figure~\ref{fig:amplification_divergence_codes} shows the evolution of the dimensionless divergence error $(\nabla\cdot {\bf B})\,(\Delta x/B)$, which normalizes the divergence by the local cell size and field strength, providing a scale-independent measure that is more meaningful for cross-code comparisons where the resolution and field amplitude may differ. We set the resolution scale as the smoothing length $h$ in \textsc{Swift} and $\Delta x \approx 0.62~ V_{\rm cell}^{1/3}$, where $V_{\rm cell}$ is the Voronoi cell volume, in \textsc{Arepo}. The \textsc{Ramses} CT scheme maintains divergence errors at machine-level precision throughout the simulation. Among the divergence-cleaning approaches, the Dedner scheme in \textsc{ART} performs best, with a mean dimensionless divergence error of $\sim 10^{-2}$, while \textsc{Arepo} and \textsc{Swift} exhibit somewhat larger errors of $\sim 10^{-1}$, though in both cases the errors remain bounded over the course of the evolution. Although both \textsc{ART} and \textsc{Swift} employ the Dedner scheme, \textsc{ART} achieves lower dimensionless divergence errors than \textsc{Swift}. The \textsc{Swift} values we observe are consistent with previous \textsc{Swift} MHD simulations, which reported dimensionless divergence errors between $\sim 10^{-2}$ and unity in an adiabatic cosmological box simulation ($R_0$ in Figure~13 of \citealt{2025MNRAS.541.3427S}) and isolated galaxy simulations with star formation and feedback (see Figure~33 of \citealt{2026arXiv260806978K}). Similarly, \citet{2013MNRAS.432..176P} reported dimensionless divergence errors of the same order for \textsc{Arepo} in isolated disk galaxy simulations (see Figure~13 of that work).

\begin{figure*}[ht!]
    \centering
    \begin{minipage}[t]{0.48\textwidth}
        \centering
        \includegraphics[width=0.9\linewidth]{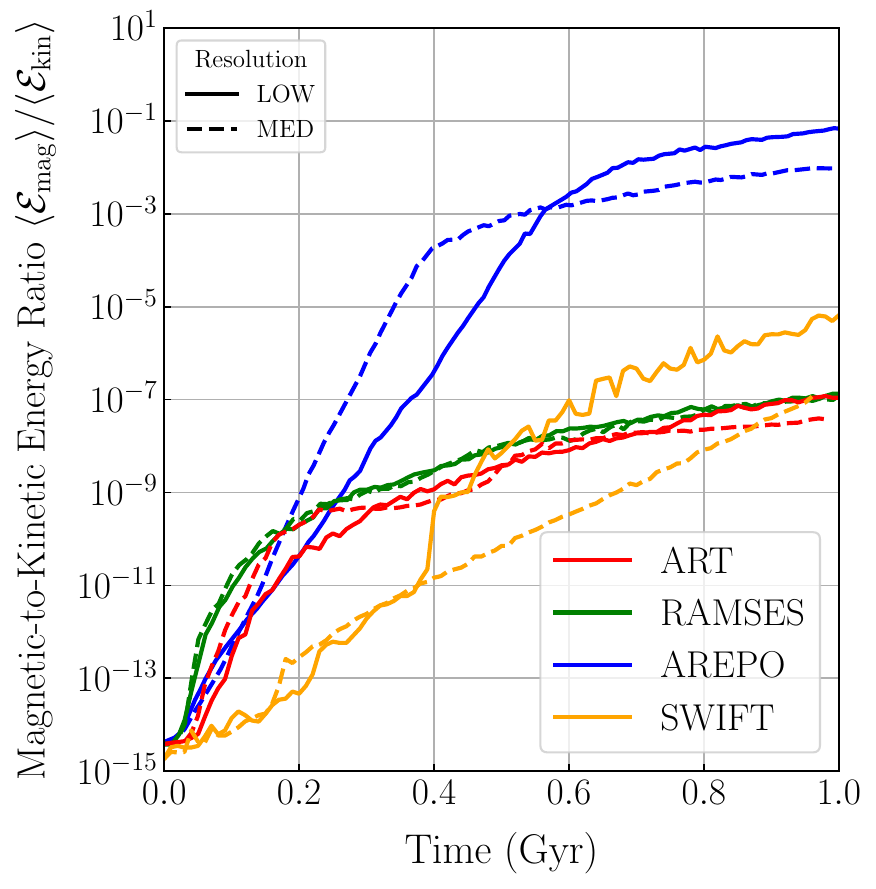}
        \captionof{figure}{Same as the leftmost panel of Figure~\ref{fig:amplification_divergence_codes}, but for simulations run at two mass (and spatial) resolutions, with the higher-resolution (MED) runs having ten times smaller cell (gas) and particle (DM, star) masses (and correspondingly $2\times$ finer spatial resolution) than the fiducial (LOW) runs. Increasing the resolution does not significantly affect the amplification history in the Eulerian codes. In \textsc{Arepo}, by contrast, the higher-resolution run amplifies the magnetic field more rapidly at early times but saturates at a weaker value than the fiducial run. In \textsc{Swift}, increasing the resolution removes the sharp jump seen in the lower-resolution run, though the two resolutions have not yet converged by the end of the simulation.
        }
        \label{fig:resolution}
    \end{minipage}
    \hfill
    \begin{minipage}[t]{0.48\textwidth}
        \centering
        \includegraphics[width=0.9\linewidth]{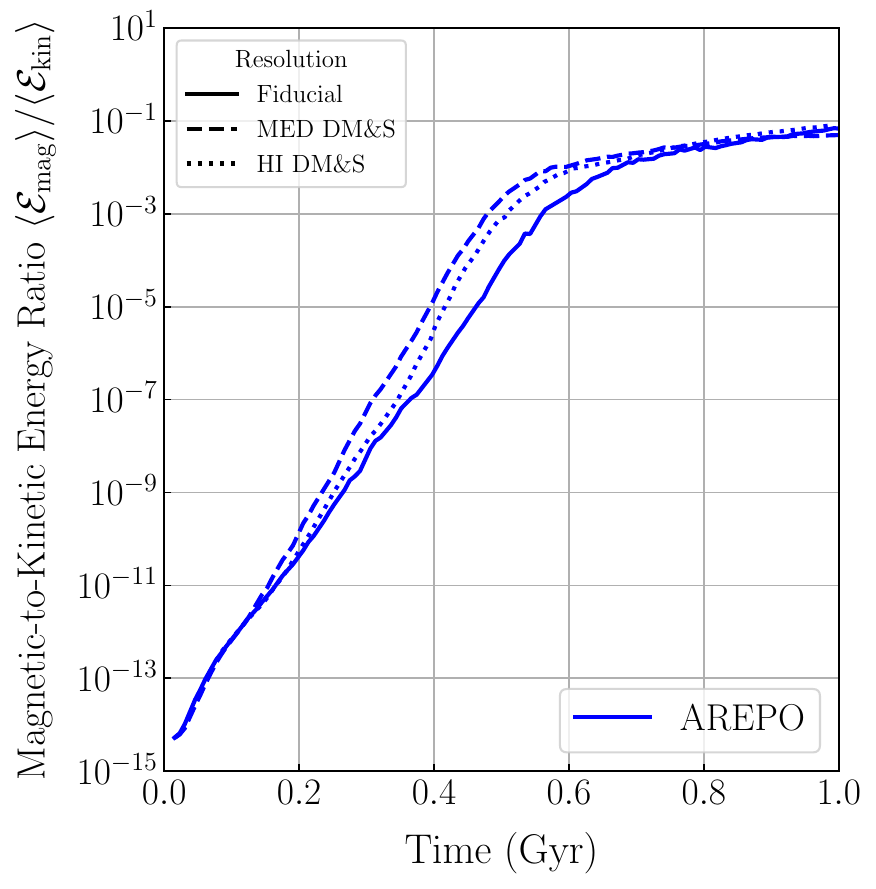}
        \captionof{figure}{Same as the leftmost panel in Figure~\ref{fig:amplification_divergence_codes}, but for \textsc{Arepo} simulations run at three dark matter and star particle mass resolutions: LOW, corresponding to our fiducial run; MED, with 10 times higher mass resolution; and HI, with 100 times higher mass resolution. The gas mass resolution is kept constant in all three runs. There is no significant difference in the growth history across all setups, although the MED and HI runs show closer agreement with each other.}
        \label{fig:arepo_resolution}
    \end{minipage}
\end{figure*}

We next consider the differences in magnetic energy growth rates across the disk for each code (right panel of Figure \ref{fig:amplification_divergence_codes}). We expect this difference to depend on the disk's velocity and cell refinement profiles. The panels in Figure~\ref{fig:radial1} show the time evolution of the magnetic-to-kinetic energy ratio within concentric annuli centered on the disk's center of mass, for each simulation. The \mbox{(quasi-)Lagrangian} codes exhibit slower magnetic energy growth with increasing radius, whereas the Eulerian codes show nearly identical growth rates across the disk. In \textsc{Arepo}, the magnetic energy grows at different rates across the disk's annuli, with the innermost region, where velocity gradients are highest, exhibiting the fastest growth. The field growth in regions beyond $r=5$ kpc saturates at 10\% of equipartition, whereas the innermost region reaches near-equipartition. We observe somewhat different behavior in \textsc{Swift}, where the field energy within a 5 kpc radius grows at a faster rate than the field energy in the annuli beyond 5 kpc. The field energy in these outer annuli initially grows at similar rates, but the growth rates begin to diverge after 1 Gyr. The field growth across the entire disk eventually saturates, reaching a saturation level of approximately 10\% of the equipartition energy. We find similar results in \textsc{ART} and \textsc{Ramses}, though with some differences. In \textsc{Ramses}, the outer annuli initially grow at a slower rate than the inner 5 kpc, but eventually catch up, so that growth across the disk proceeds at the same rate. In \textsc{ART}, the inner two rings (0--5 kpc and 5--10 kpc) grow at the same rate throughout the simulation, whereas the outer rings grow at a lower rate, resulting in a slightly lower final energy than in the inner regions.

These results show that not only does the magnetic energy growth rate differ across codes, but the way that growth rate depends on radius differs as well.

\subsection{Resolution effects}
\label{sec:resolution_effects}

To assess the numerical convergence of our results, we perform additional higher-resolution (MED) simulations with all codes. For \textsc{Arepo} and \textsc{Swift}, we increased the mass resolution of gas cells by a factor of ten, while for \textsc{ART} and \textsc{Ramses} we doubled the spatial resolution, increasing the maximum resolution from $\Delta x = 80$~pc to $\Delta x = 40$~pc. We also increased the number of dark matter and star particles by a factor of 10 for all codes. Figure~\ref{fig:resolution} compares the magnetic energy growth histories of the fiducial and higher-resolution runs. The \textsc{Ramses} and \textsc{ART} simulations exhibit nearly identical magnetic field evolution, indicating that the Eulerian results are largely converged with respect to resolution over the range considered. In contrast, the higher-resolution \textsc{Arepo} simulation shows faster amplification at early times but saturates at a weaker magnetic field amplification than the fiducial run, suggesting a modest resolution dependence. Nevertheless, the MED-resolution \textsc{Arepo} simulation produce substantially stronger magnetic field growth than the other codes. In \textsc{Swift}, we find that the LOW-resolution run exhibits a sharp jump in magnetic energy near $t \approx 0.4$~Gyr, whereas the MED-resolution run amplifies the field more gradually and at a similar rate throughout. By $t = 1$~Gyr, however, the two resolutions remain separated by nearly two orders of magnitude in the magnetic-to-kinetic energy ratio, indicating that the \textsc{Swift} results are not yet converged at the LOW resolution over the time range considered.

We also tested whether increasing the dark matter and star particle resolution, thereby producing a smoother gravitational potential, while keeping the gas resolution fixed at the fiducial level, would significantly affect magnetic field amplification in \textsc{Arepo}. In lower-resolution simulations, collisionless particles are more massive and therefore source stronger local gravitational perturbations. A higher force resolution in \mbox{(quasi-)Lagrangian} codes combined with coarse collisionless mass resolution can make these perturbations dynamically significant, potentially driving additional small-scale gas motions that could result in higher amplification. 

We performed this test using $10^6$ and $10^7$ dark matter particles, $10^6$ and $10^7$ stellar disk particles, and $1.25\times10^5$ and $1.25\times 10^6$ stellar bulge particles for the MED and HI resolution runs, respectively. Figure~\ref{fig:arepo_resolution} shows the magnetic field amplification histories for simulations with varying dark matter and stellar particle mass resolution. We find no significant differences in the magnetic field amplification history within the galactic disk, with the LOW-, MED- and HI-resolution dark matter and star particle mass simulations showing good convergence within a 10\% spread. 

\subsection{Term-by-term contributions to field amplification}
While the dimensionless divergence error provides a useful measure of how well the solenoidal constraint is maintained, it does not directly quantify the potential contribution of non-zero $\nabla\cdot\mathbf{B}$ to the amplification or evolution of the magnetic field. The significance of the divergence error can only be assessed by comparing its contribution to the physical source terms appearing in the induction equation. As shown in Equation \eqref{eq:induction}, the evolution of magnetic energy can be decomposed into three contributions: two physical terms associated with shear and compression, and a third unphysical contribution arising from non-zero magnetic divergence. The relative importance of these terms determines whether the magnetic field amplification is physically meaningful or dominated by numerical artifacts.

To identify the dominant driver of magnetic field growth in our simulations, we directly measure each term in the induction equation and compare their relative amplitudes. For this analysis, all simulation outputs are remapped onto a uniform Cartesian grid covering a $(20\  \mathrm{kpc})^2$ region at the finest available resolution, corresponding to $\Delta x = 80\ \mathrm{pc}$. We select this region because it exhibits the fastest magnetic energy growth at early times in the (quasi-)Lagrangian codes, and because this is where the codes differ most strongly. The values reported in this section are not intended to reproduce the \textit{exact} numerical operations performed internally by each code, which deviate from the induction equation (Equation \ref{eq:induction}) to control divergence errors (e.g., Powell's source-term approach, see \S\ref{sec:divergenceschemes}). Rather, this framework is intended to identify, in an implementation-agnostic manner, the mechanisms that could dominate field amplification within the galactic disk if such numerical remedies are not implemented.

For \textsc{Ramses}, \textsc{ART} and \textsc{Arepo}, we perform the remapping using Monte Carlo volume averaging. We randomly sample points for each cell in the uniform grid. Each sampled point is assigned the value of its nearest cell center or particle using SciPy's cKDTree nearest-neighbor search \citep{2020NaMet..17..261V}. The remapped cell value is then computed as the average of the values assigned to all sampled points. We verified the accuracy of this remapping procedure against slices obtained with the Python package \texttt{yt} \citep{2011ApJS..192....9T}, and found very good agreement for all fields. The accuracy of this procedure is controlled by the number of random samples per cell and, because it relies on nearest-neighbor assignment, it does not explicitly account for the spatial smoothing scale associated with individual particles in SPH. Hence, for our \textsc{Swift} data, we use the Python package \texttt{swiftsimio} \citep{2020JOSS....5.2430B}, which performs grid interpolation by accounting for the smoothing kernel (in this case, the Wendland-C2 kernel; \citealp{1995AdCM....4..389W}) and smoothing length of each particle when mapping the particle data onto a voxel grid. This provides a more physically consistent volume averaging of the SPH data.

We then compute second-order finite-difference gradients using a two-point stencil for the remapped \textsc{Ramses}, \textsc{ART}, and \textsc{Swift} data. We choose this stencil to effectively smoothen out grid-scale noise over two cells. To match the stencil used for the constrained transport scheme in \textsc{Ramses}, we use a stride that is one cell across for our divergence calculations. However, since SPH gradients are kernel-weighted sums over a set of neighboring particles, we expect the gradients we compute for \textsc{Swift} to be slightly different from what is internally used in the code. For \textsc{Arepo}, we instead use the gradients computed directly by the code on its native moving mesh which are better adapted to the underlying Voronoi mesh structure. 

\begin{figure*}[ht!]
    \centering
    \hspace{0.6in}\includegraphics[width=0.8\linewidth]{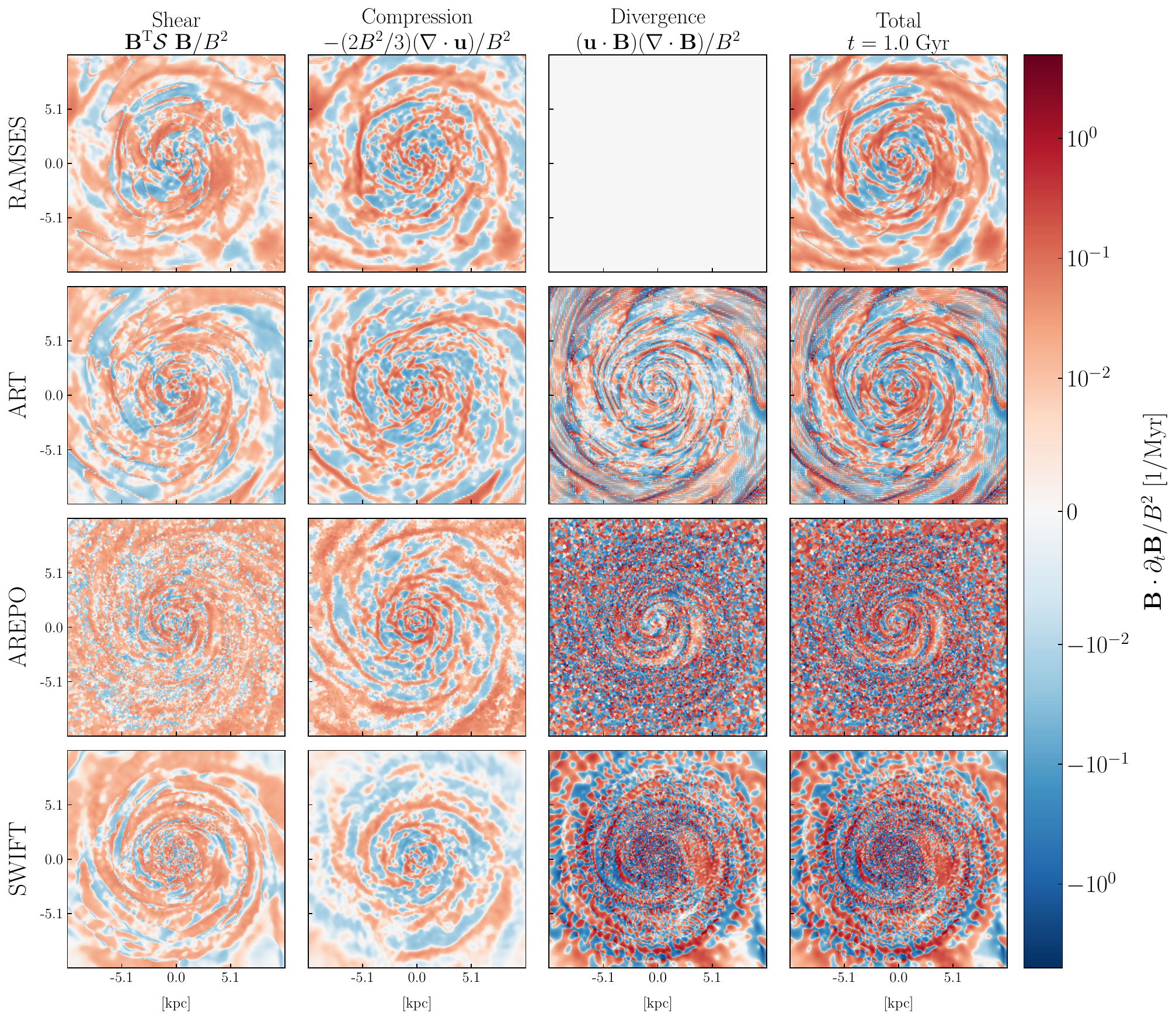}
    \caption{\textbf{Midplane slices of growth rate contributions to the induction equation (Equation \ref{eq:induction}) in the comoving frame} with units of 1/Myr, at $t=1.0$ Gyr for each code. Red and blue regions, shown on a symmetric logarithmic color scale, indicate local growth and decay of the magnetic energy density, respectively. Coherent spiral features are most prominent, across all contributions, in the Eulerian codes and in \textsc{Swift}. Similar features are also present in \textsc{Arepo} but pronounced only in the compression contribution. The shear contribution in \textsc{Arepo} is confined to very small scales. In (quasi-)Lagrangian codes, divergence error contributions to magnetic field growth are highly localized and alternate rapidly between regions of growth and decay.}
    \label{fig:term_codes1a}
    
    \includegraphics[width=0.7\linewidth]{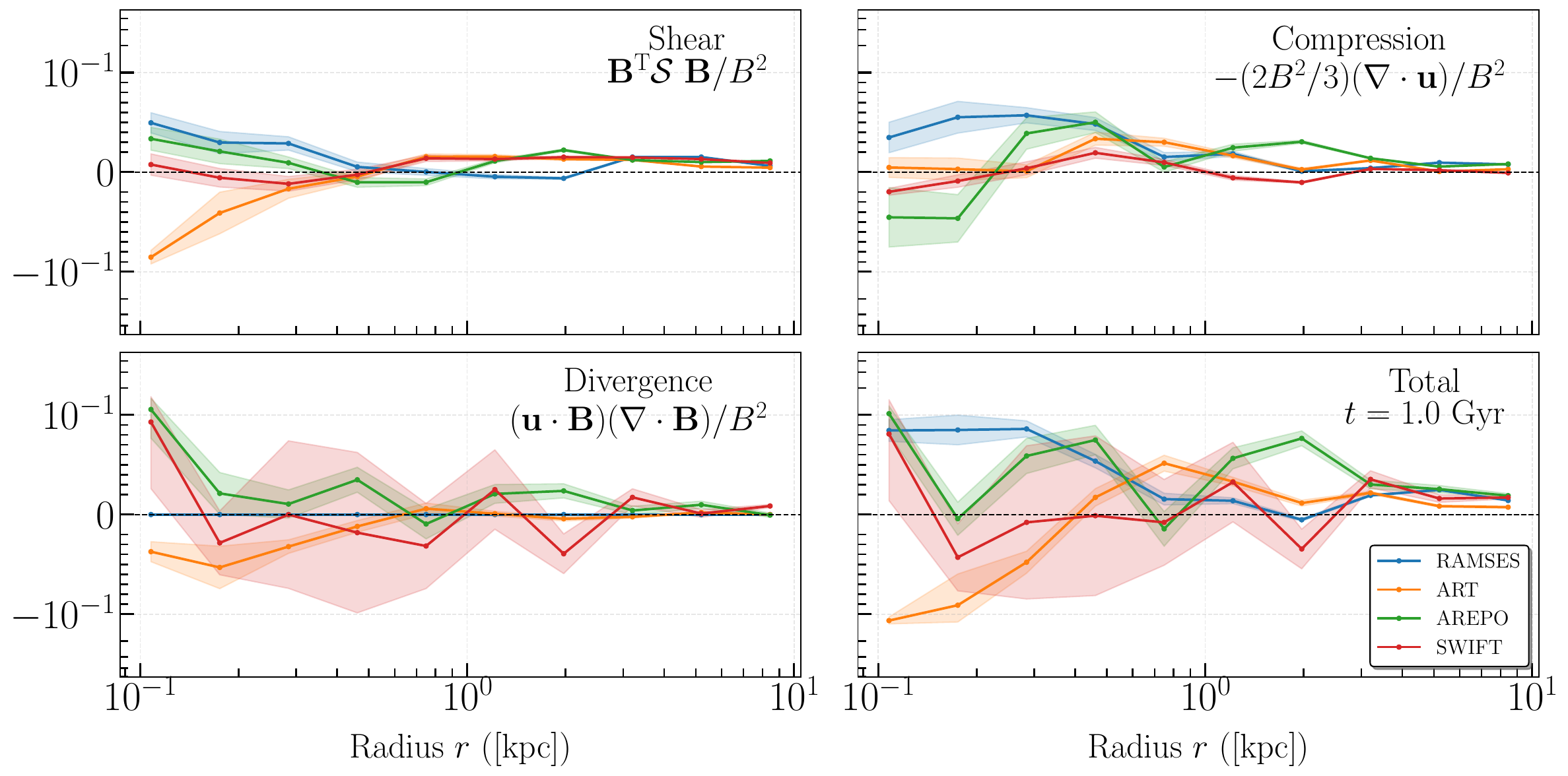}
    \caption{\textbf{Density-weighted radial profiles of growth rate contributions to the induction equation (Equation \ref{eq:induction}) in the comoving frame} at $t=1.0\,\mathrm{Gyr}$. The shaded region indicates the $1\sigma$ interval around the mean value. All terms in the induction equation appear to contribute to the amplification of the magnetic field, with the compression term especially dominating within the inner 1 kpc. The divergence term is subdominant for the Eulerian codes but is comparable to the other physical terms in \textsc{Arepo} and \textsc{Swift}, reaching amplitudes comparable to the shear and compression terms especially within the inner 1 kpc. All simulations exhibit net amplification across the disk, except within the inner few hundred pc in \textsc{ART} and \textsc{Swift}, which show net decay.}
    \label{fig:term_codes2a}
    \vspace{0.5in}
\end{figure*}
\begin{figure*}[htbp!]
    \centering
    \hspace{0.6in}\includegraphics[width=0.8\linewidth]{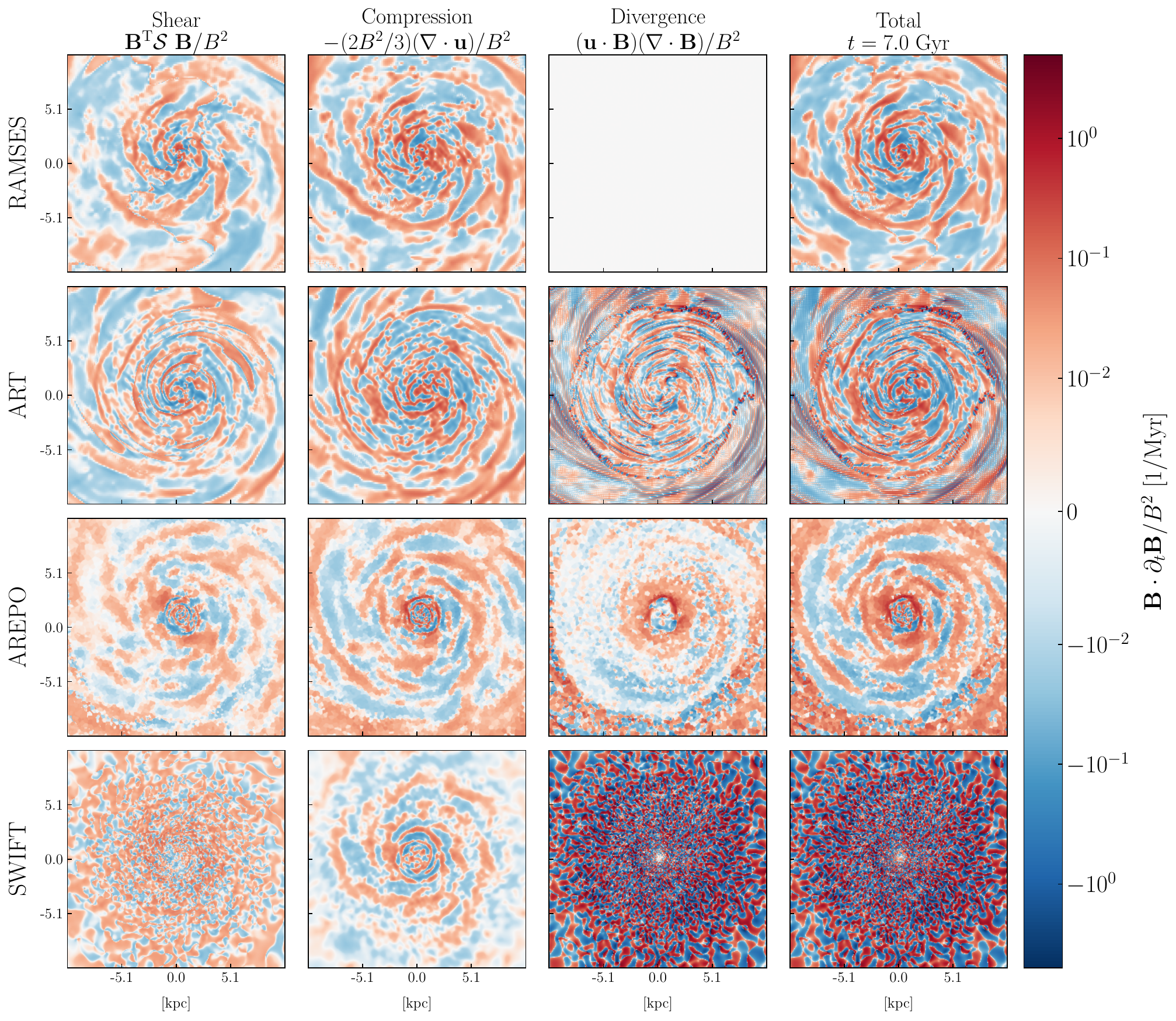}
    \caption{Same as Figure~\ref{fig:term_codes1a} but for simulation snapshot at $t=7.0$ Gyr. Spiral arm features are prominent in the compression contribution, with alternating regions of growth and decay, across all codes. The shear contribution in \textsc{ART} and \textsc{Ramses} traces spiral arms that generally line up with regions of compression. In \textsc{Swift}, by contrast, growth associated with shear is strongly local rather than tracing extended arm-like structure. The divergence term is noticeably lower in \textsc{Arepo}, while the same strongly local features at $t=1.0$ Gyr are observed in \textsc{Swift}.}
    \label{fig:term_codes1b}

    \includegraphics[width=0.7\linewidth]{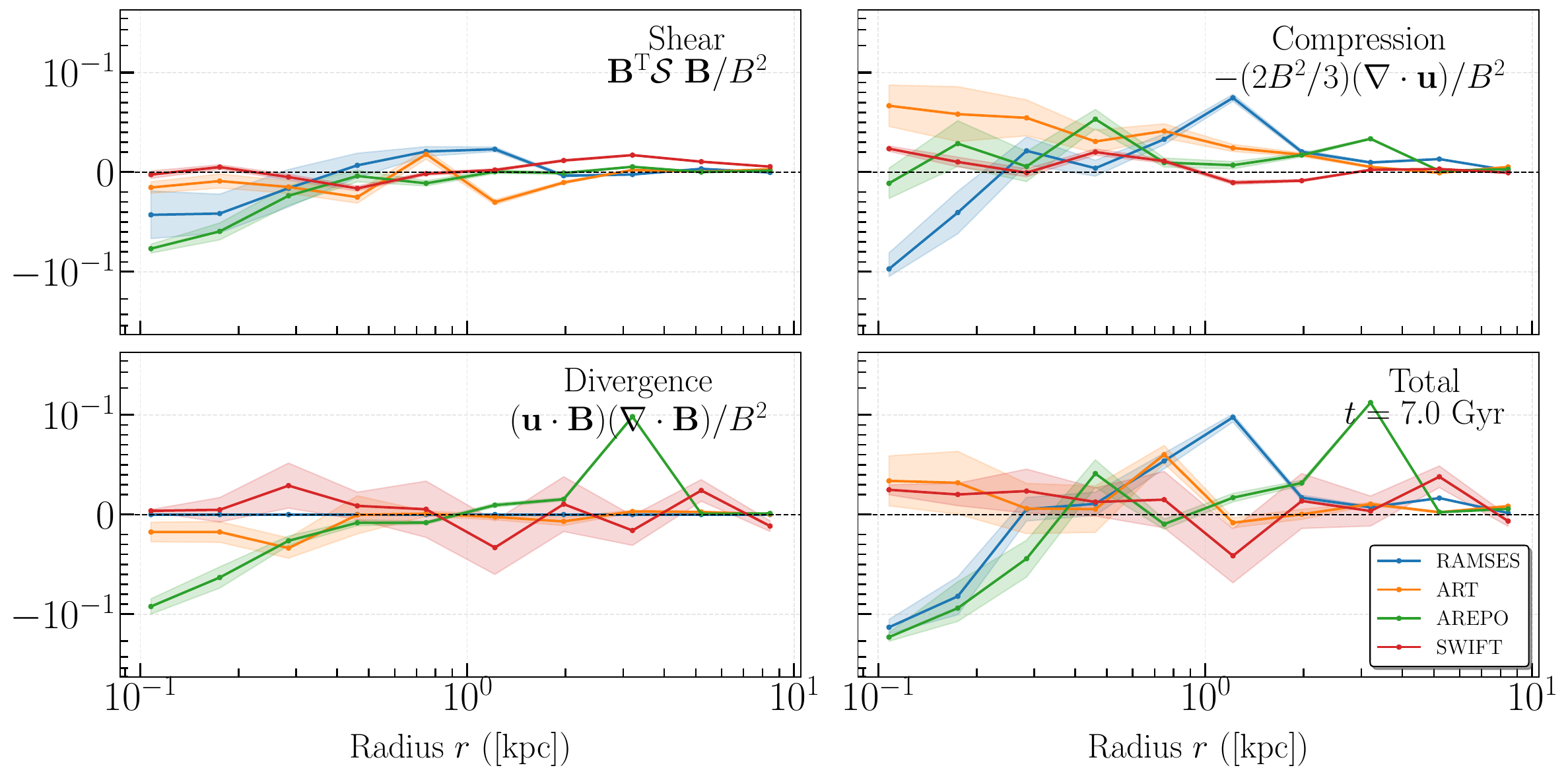}
    \caption{Same as Figure~\ref{fig:term_codes2a}, but for the simulation snapshot at $t = 7.0\,\mathrm{Gyr}$.  Compression in all codes acts to amplify the field. The shear contribution to growth is generally positive across the disk in \textsc{Swift} except within the inner 1 kpc, whereas the average shear contribution in the other codes indicates decay.}
    \label{fig:term_codes2b}
\end{figure*}

We present midplane slices of the individual terms and the total fractional growth rate, ${\bf B}\cdot \partial_t{\bf B} / B^2$, computed per cell at two representative epochs, $t=1.0$ Gyr and $t=7.0$ Gyr, in Figures \ref{fig:term_codes1a} and \ref{fig:term_codes1b}, respectively. To help compare the amplitudes of each term, we also show density-weighted profiles computed within a finite slab with height $h=0.1$ kpc from the midplane, logarithmically binned in radius from 80 pc to 10 kpc, in Figures \ref{fig:term_codes2a} and \ref{fig:term_codes2b}. The density weighting here is important because the magnetic field evolution is coupled to the gas distribution, and an unweighted spatial average would give the tenuous, low-density volume-filling gas the same weight as the denser gas that contains most of the mass and is typically more dynamically relevant. The quantity ${\bf B}\cdot \partial_t{\bf B} / B^2$ has units of inverse time and characterizes the instantaneous fractional rate of change of magnetic energy density $B^2$, up to a factor of two, with positive and negative values corresponding to local amplification and decay, respectively. Note that these snapshots capture \textit{transient} behavior and should \textit{not} be used as a proxy for the cumulative amplification history.

We note that in the case of \textsc{Swift}, the gradient operators were evaluated on quantities remapped from SPH particles onto a Cartesian grid. Since SPH data can exhibit substantial particle-scale noise, part of the observed scatter may arise from the mismatch between the grid-based differential operators and the underlying particle representation.

The induction terms at $t=1.0$ Gyr are shown in Figures \ref{fig:term_codes1a} and \ref{fig:term_codes2a}. The compression term exhibits a large-scale spiral pattern, with alternating bands of field growth (compression, red) and decay (rarefaction, blue) wound around the center. We interpret this pattern as arising from the shearing of an initially large-scale compressible perturbation by differential rotation, analogous to the winding of density perturbations by shear. This feature is not as pronounced in \textsc{Swift}. We find a similar imprint in the shear term for \textsc{Ramses}, \textsc{ART} and \textsc{Swift}, but the alternating growth/decay bands are disrupted by small-scale noise, particularly within the inner regions, suggesting that local velocity shear fluctuations contribute more strongly to this term than to the compression term. Interestingly, the shear contribution in \textsc{Arepo} is highly local, appearing as small disordered patches of growth and decay rather than the large-scale structure seen in the other codes. This is likely a consequence of local shearing motions that may be more resolved in \textsc{Arepo} than in other codes. The divergence term is negligible in \textsc{Ramses}, consistent with its use of a constrained-transport scheme that enforces $\nabla\cdot\mathbf{B} \approx 0$ to machine precision. In \textsc{ART}, the term exhibits a spiral pattern similar to that seen in the compression and shear terms, while in \textsc{Arepo} this structure is obscured by small-scale noise, with only a faint spiral discernible near the center. The divergence contribution in \textsc{Swift} shows a pronounced concentric spiral pattern, overlaid with fine-scale noise that, while substantial, does not obscure the underlying structure to the same degree as in \textsc{Arepo}. The radial profiles show that the divergence term makes a contribution comparable to those of the shear and compression terms in both \textsc{Arepo} and \textsc{Swift}.

At later times ($t = 7.0\ \mathrm{Gyr}$; Figures \ref{fig:term_codes1b} and \ref{fig:term_codes2b}), we find that the Eulerian codes retain broadly similar large-scale structures in the induction terms, whereas the (quasi-)Lagrangian codes develop different morphologies compared to the corresponding $t=1.0\ \mathrm{Gyr}$ snapshot. The shear-term slice for \textsc{Arepo} indicates large-scale structure of growth/decay at this later time, in contrast to the small-scale, disordered patches seen at $t=1.0\ \mathrm{Gyr}$. The opposite trend is seen in \textsc{Swift}, where the shear term becomes more localized and disordered at $t=7.0\ \mathrm{Gyr}$ compared to the more coherent structure present at $t=1.0\ \mathrm{Gyr}$. By late times, compression dominates magnetic energy growth in \textsc{Arepo}, with shear still contributing appreciably and divergence playing a weaker role compared to the earlier snapshot.

\begin{figure*}[ht!]
    \centering
    \includegraphics[width=0.8\linewidth]{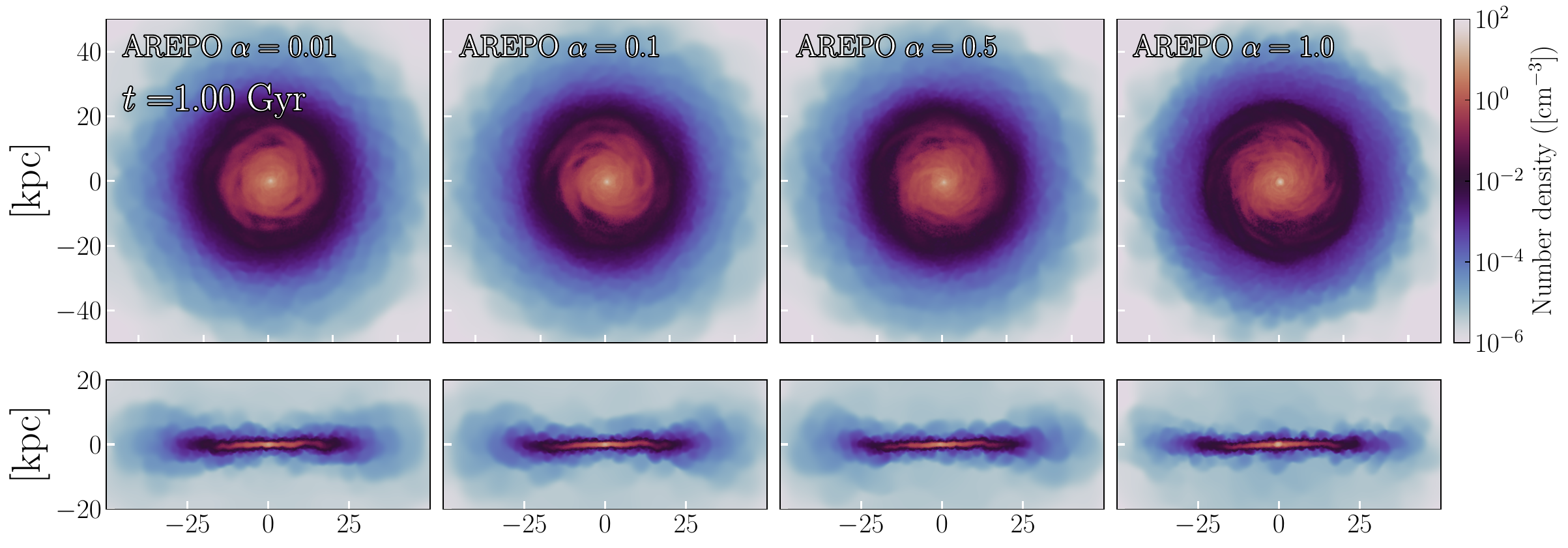}
    \includegraphics[width=0.8\linewidth]{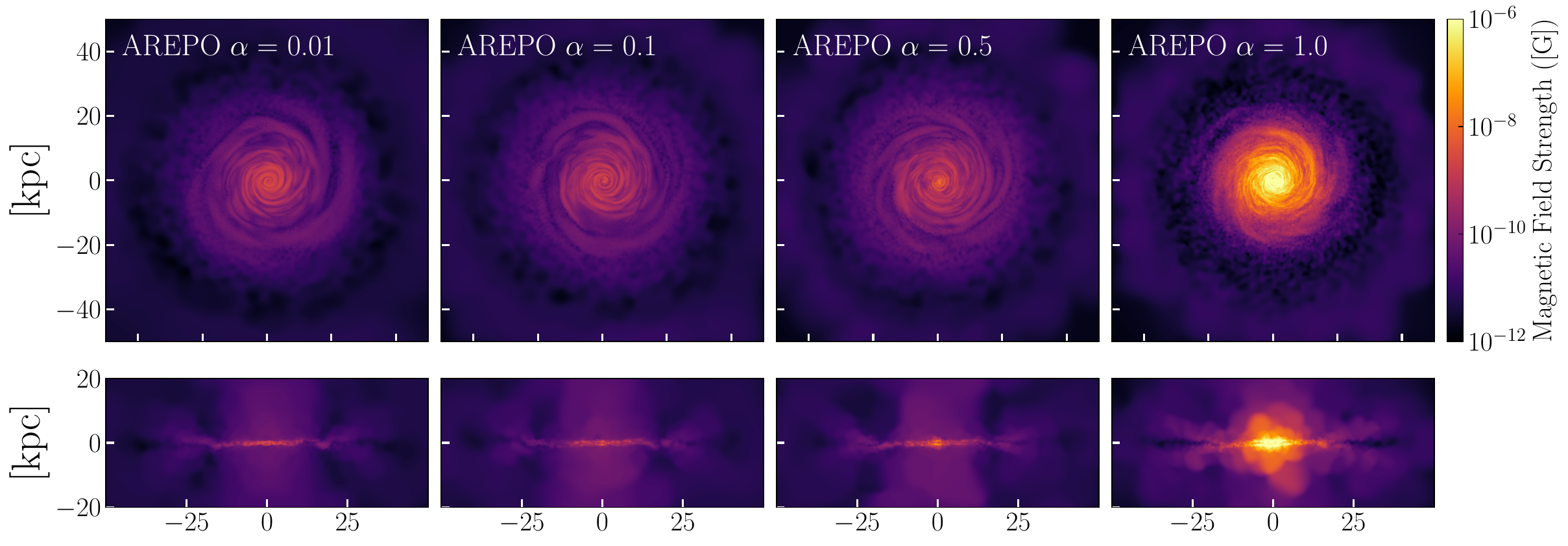}
    \caption{\textbf{Midplane slices of number density (top) and magnetic field strength (bottom)} at the $t=1.0$ Gyr, run using \textsc{Arepo} with different mesh-generating particle velocities, ${\bf u}_{\rm mesh} = \alpha {\bf u}_{\rm gas}$, where the $\alpha=1.0$ corresponds to the fiducial \textsc{Arepo} run. For this numerical test, we reduce the parameter $\alpha$ from 1.0, which corresponds to the fiducial moving-mesh case, to 0.01, which represents the Eulerian-like limit in \textsc{Arepo}. We observe that while the midplane and vertical density slices look similar across all values of $\alpha$, the magnetic field strength is considerably weaker and more confined within the midplane in simulations with $\alpha < 1$.}
    \label{fig:arepo_mesh_slices1}
    
    \includegraphics[width=0.8\linewidth]{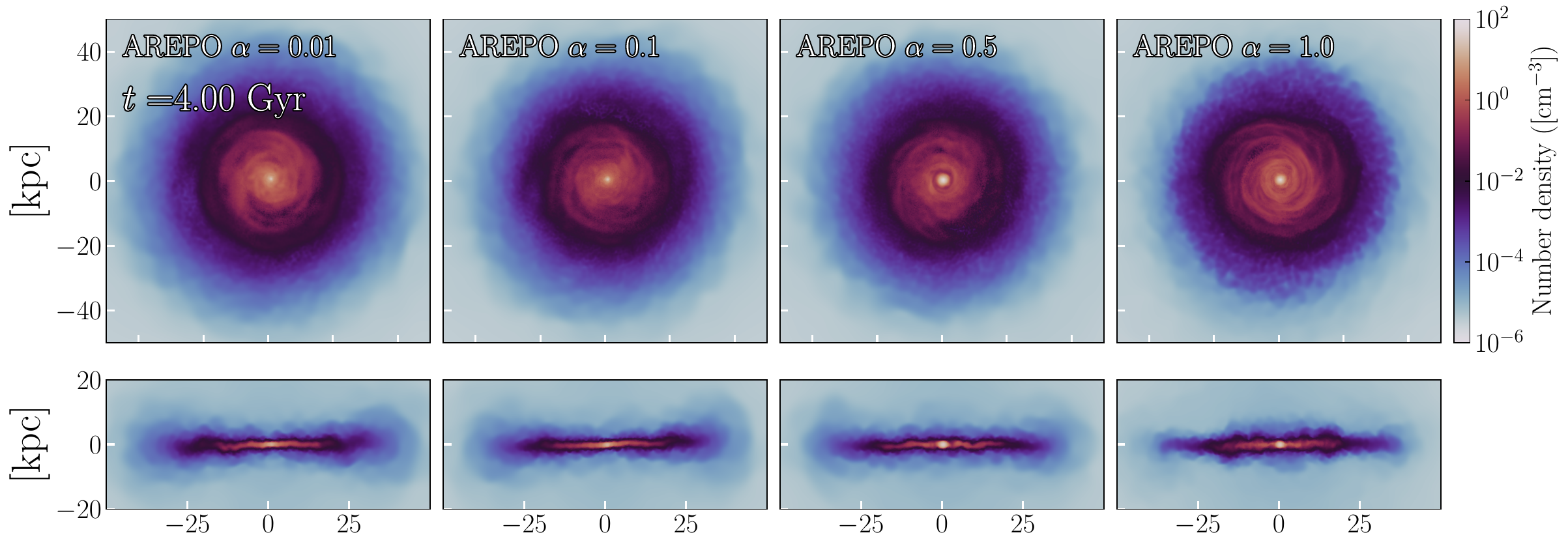}
    \includegraphics[width=0.8\linewidth]{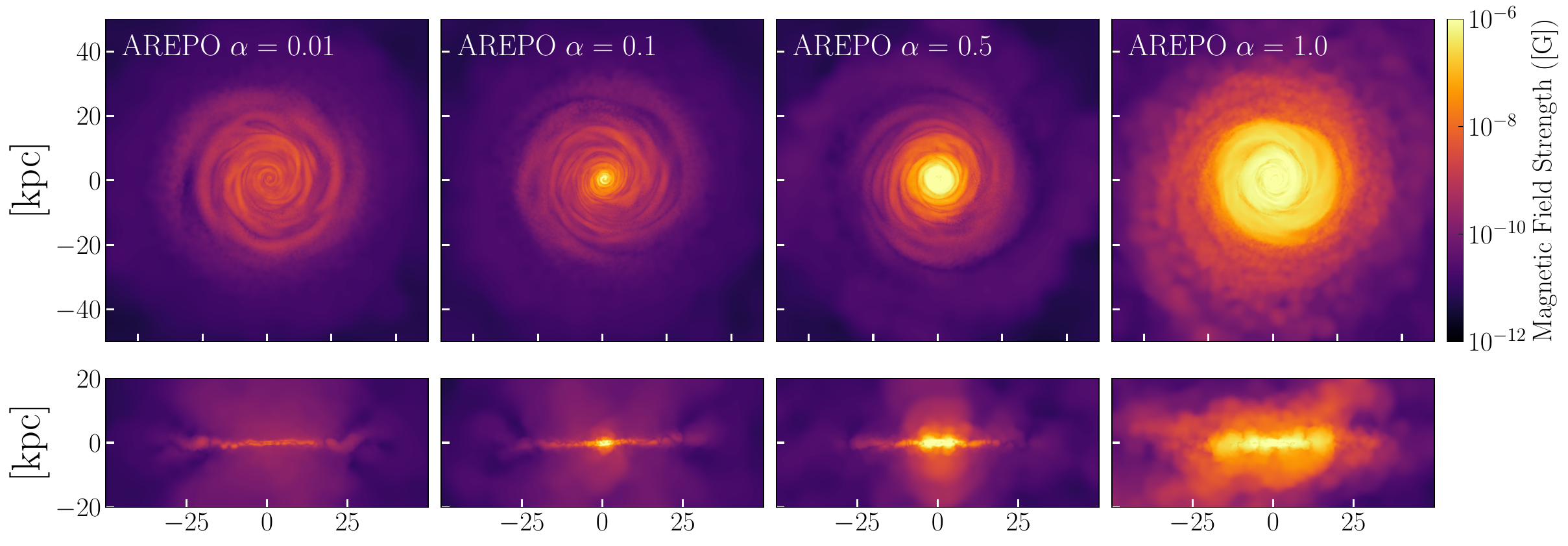}
    \caption{Same as Figure~\ref{fig:arepo_mesh_slices1}, but for the snapshot at $t=4.0$ Gyr. As the simulations approach the Eulerian limit ($\alpha \rightarrow 0$), the magnetic field becomes increasingly confined to the galactic midplane, while its strength decreases throughout the disk. We also find that the field approaches the Eulerian case first in the outer regions of the disk and progressively toward the center. The density profiles likewise vary with $\alpha$ and exhibit structures that closely trace those of the magnetic field.}
    \label{fig:arepo_mesh_slices2}
\end{figure*}

\begin{figure*}[ht!]
    \centering
    \includegraphics[width=\linewidth]{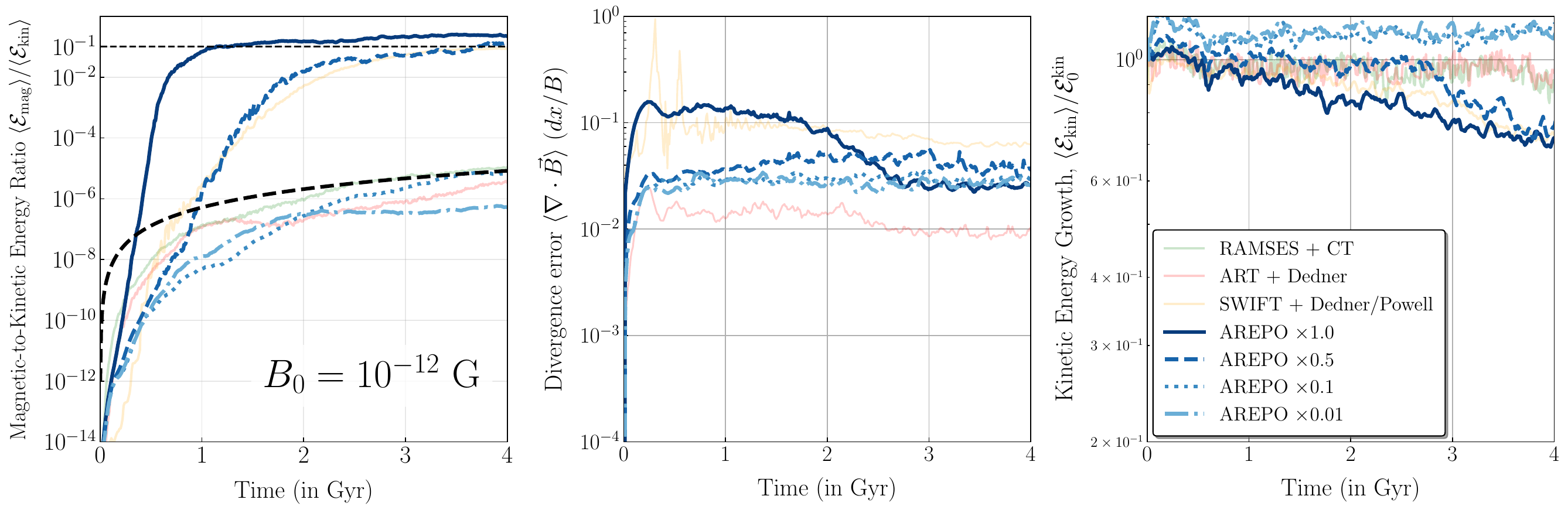}
    \caption{Same as Figure~\ref{fig:amplification_divergence_codes} but for isolated disk galaxy simulations run using \textsc{Arepo} with different mesh-generating particle velocities, ${\bf u}_{\rm mesh} = \alpha {\bf u}_{\rm gas}$ but up to $t=4$ Gyr. The black curve represents the expected result for a differentially rotating disk without feedback with shearing rate $q\Omega = 1.0\ {\rm Myr}^{-1}$. The growth curve of the magnetic energy density within the disk in \textsc{Arepo} approaches the results we obtained using \textsc{Ramses} and \textsc{ART} as we reduce the velocity of the mesh-generating particles with respect to the gas flow. The growth history in \textsc{Arepo} matches that of \textsc{Swift} at $\alpha=0.5$ case.}
    \label{fig:arepo_mesh}
    \vspace{0.1in}
    \includegraphics[width=\linewidth]{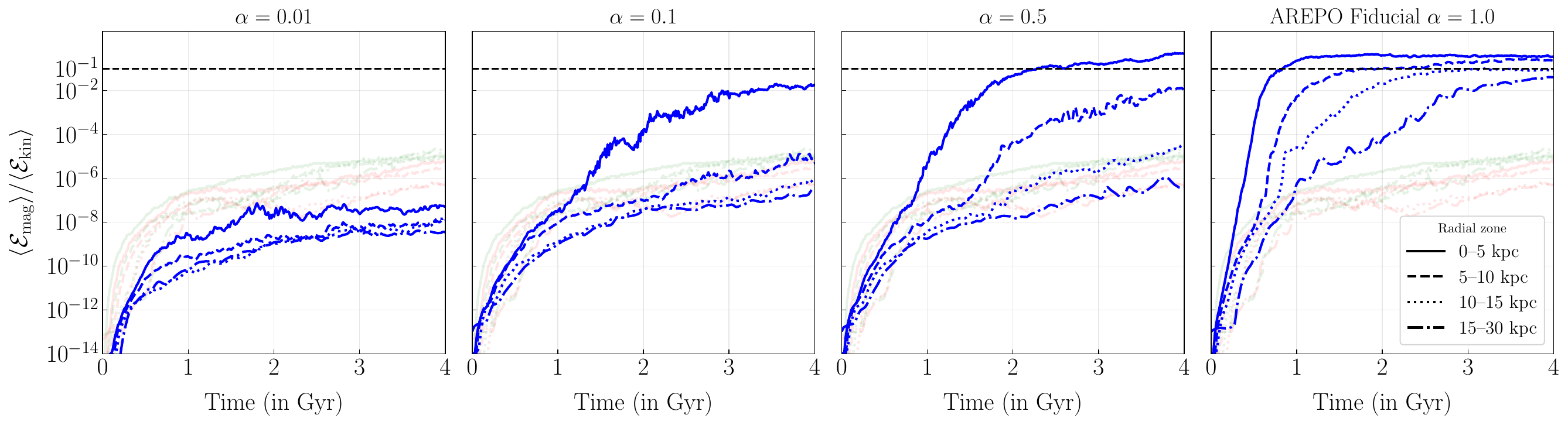}
    \caption{Same as Figure~\ref{fig:radial1} but for \textsc{Arepo} runs with varying mesh velocities ${\bf u}_{\rm mesh}=\alpha{\bf u}_{\rm gas}$ but up to $t=4$ Gyr. As the mesh velocity is reduced from the Lagrangian case ($\alpha = 1.0$) toward the Eulerian limit ($\alpha = 0.01$), the growth of the magnetic energy across the disk progressively approaches the behavior obtained with \textsc{Ramses} and \textsc{ART}, as shown in Figure~\ref{fig:radial1}. The volume-averaged magnetic energy within each annulus grow at the same rate throughout the disk in $\alpha = 0.01$, except within the inner 5 kpc where it is initially faster.}
    \label{fig:arepo_radial}
\end{figure*}

\begin{figure*}[ht!]
    \centering
    \hspace{0.6in}\includegraphics[width=0.8\linewidth]{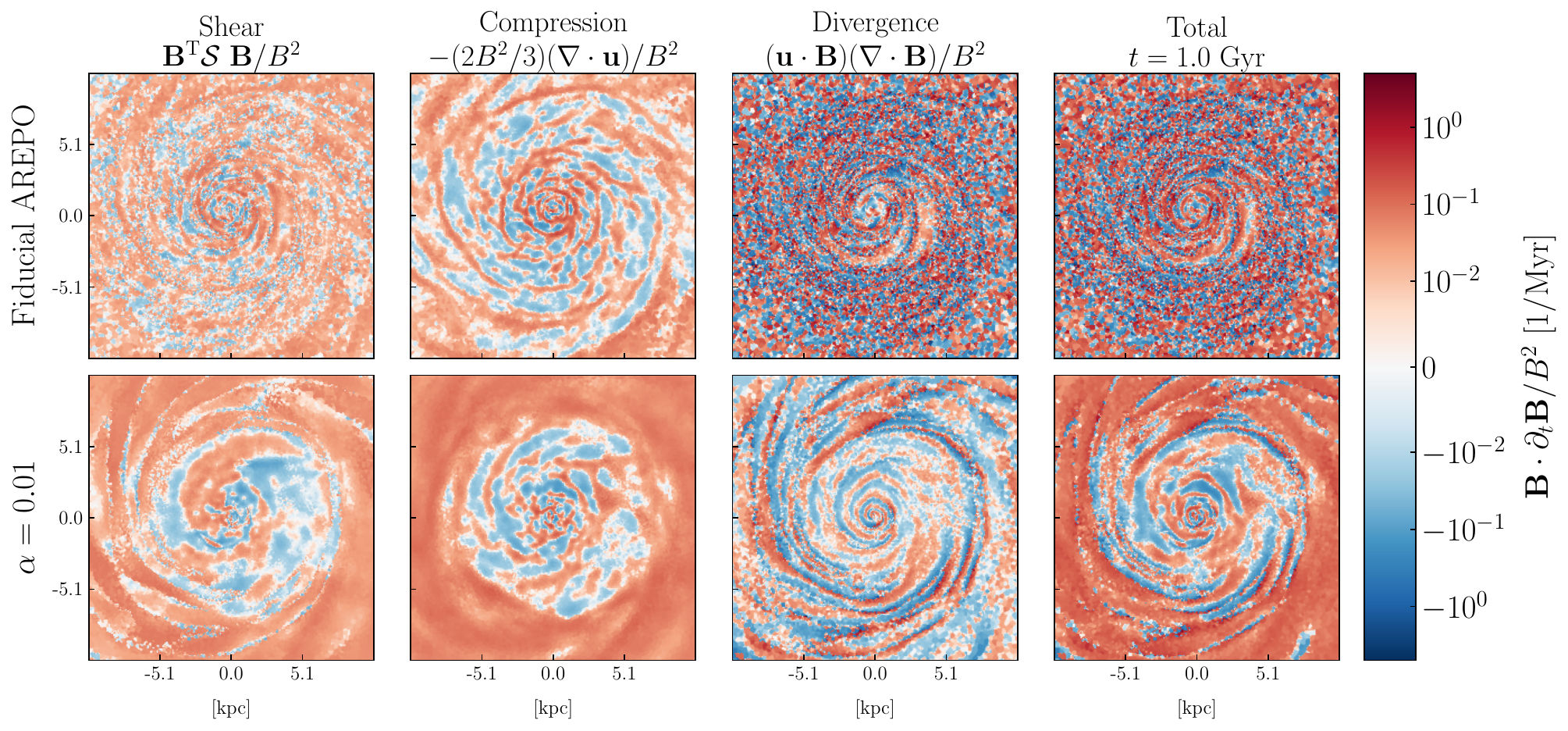}
    \caption{Same as Figure~\ref{fig:term_codes1a}, but for the simulation snapshot at $t = 1.0\,\mathrm{Gyr}$, comparing the \textsc{Arepo} run with a slowed down mesh $\alpha = 0.01$ to the corresponding fiducial run. The slowed-down-mesh run exhibits markedly different features from the fiducial case, with spiral-arm structure considerably more pronounced in the shear and divergence contributions than in the fiducial \textsc{Arepo} run ($\alpha = 1$). The divergence contribution is also lower in magnitude across most regions in the $\alpha=0.01$ run, with the remaining higher values clearly tracing more coherent spiral features.}
    \label{fig:areposlow1}
    
    \includegraphics[width=0.7\linewidth]{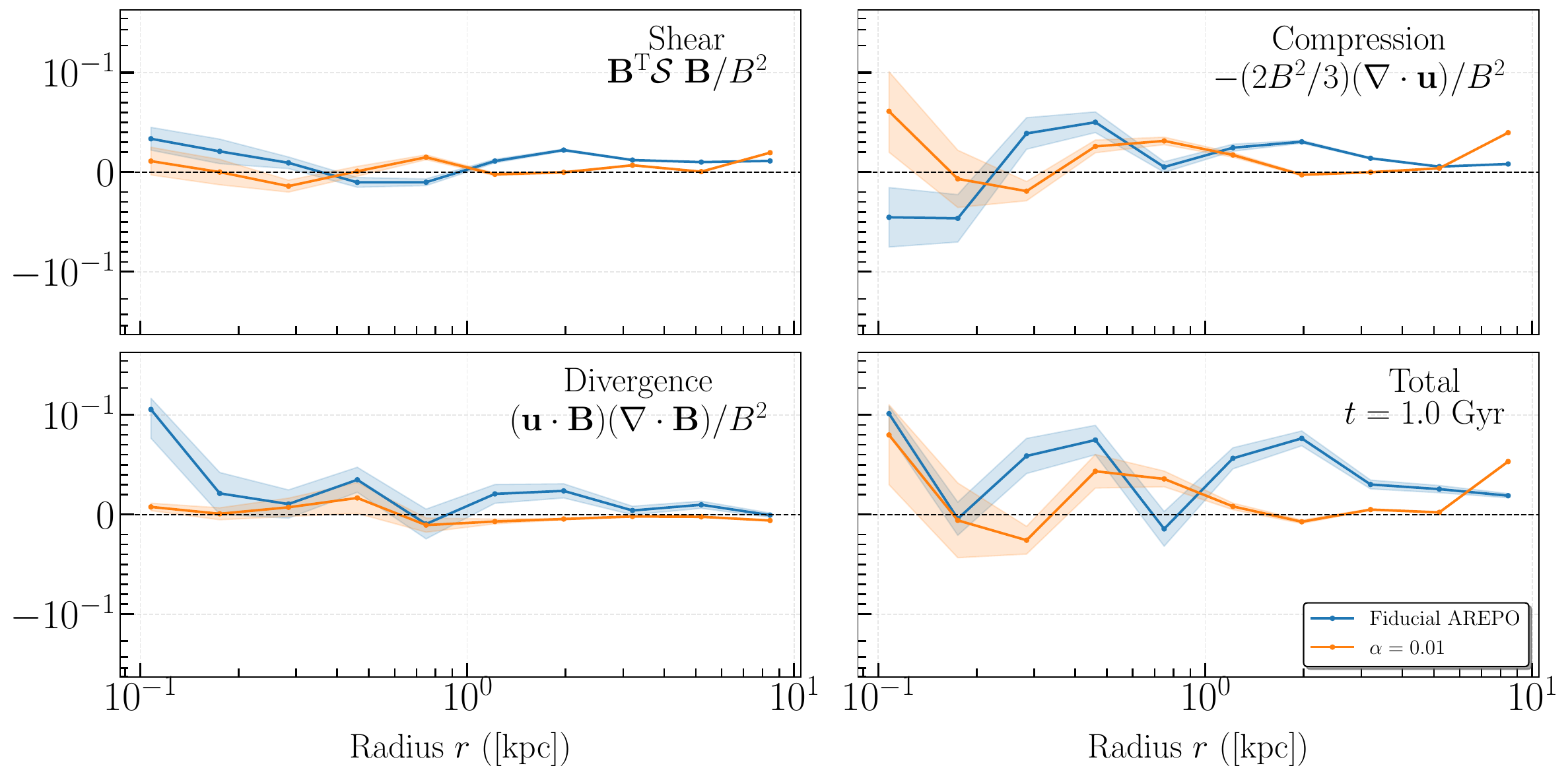}
    \caption{Same as Figure~\ref{fig:term_codes2a} but for simulation snapshot at $t=1.0\,\mathrm{Gyr}$ with \textsc{Arepo} $\alpha = 0.01$ and the corresponding fiducial run. The divergence contribution in the $\alpha =0.01$ run is lower compared to the fiducial run across the entire disk. Compression still dominates the growth/decay throughout the disk.}
    \label{fig:areposlow2}
\end{figure*}

\subsection{Eulerian vs. Lagrangian}
\label{sec:eulerlag}
Our key conclusion so far is that there is a substantial discrepancy between the Eulerian and Lagrangian simulations, both in the growth rate and average magnetic energy within the disk at the end of our simulations. To investigate the origin of this difference, we carry out a numerical experiment using \textsc{Arepo} in which the motion of the mesh-generating points is artificially reduced according to
${\bf u}_{\rm mesh} = \alpha {\bf u}_{\rm gas},$ where $\alpha$ is a constant parameter. For this test, we used $\alpha \in [0.01, 0.1, 0.5, 1.0]$. In \textsc{Arepo}, the computational mesh is generated by a moving Voronoi tessellation. The mesh-generating points follow the local fluid motion, yielding a quasi-Lagrangian discretization. Small correction velocities are applied to the mesh-generating points to keep them close to the centers of mass of their respective cells. Although \textsc{Arepo} is not intended to operate under such conditions, this setup provides a useful diagnostic to identify the mechanism responsible for the discrepancy between Eulerian and Lagrangian methods. Reducing the mesh velocity relative to the gas velocity introduces an effective numerical diffusivity because the mesh no longer co-moves with the fluid. As a result, gas parcels continuously cross cell boundaries and are subject to numerical diffusion. In the limiting case $\alpha \rightarrow 0$, the method approaches the behavior of a purely Eulerian scheme, whereas $\alpha = 1.0$ recovers the standard moving-mesh formulation. 

Shown in Figures~\ref{fig:arepo_mesh_slices1} and \ref{fig:arepo_mesh_slices2} are gas density and magnetic field strength slices for different values of $\alpha$ at $t=1.0$ and 4.0 Gyr. At $t=1.0$ Gyr, the midplane and vertical number density slices show no appreciable difference across different $\alpha$ values. This is not the case for the magnetic field strength slices, where the slices from simulations with $\alpha < 1.0$ show results that resemble the \textsc{ART} snapshot shown in Figure~\ref{fig:100Gyr}. Prior to the amplification of the field to near-equipartition levels at $t=1.0$ Gyr, the pressure, density, and velocity magnitudes within the disk are similar across all values of $\alpha$. Once the field reaches near-equipartition, these quantities begin to diverge, most notably in the distribution of pressure and density within the disk. We find stronger discrepancies at the end of each simulation $t=4.0$ Gyr. In the face-on gas density slices, the $\alpha = 0.01$ run shows well-defined spiral structure, which becomes progressively smoother and more diffuse as $\alpha$ increases. The magnetic field is also more confined within the galactic midplane for low-$\alpha$ runs, whereas high-$\alpha$ runs exhibit strongly magnetized structures above and below the midplane. 

The difference in amplification histories can be further observed in the evolution of magnetic energy within and in different regions of the disk, shown in Figures~\ref{fig:arepo_mesh} and \ref{fig:arepo_radial}. As $\alpha$ is reduced from 1.0 to 0.01, the growth rate and saturation level of $\langle E_\mathrm{mag}\rangle/\langle E_\mathrm{kin}\rangle$ decrease monotonically, with the $\alpha = 0.01$ run approaching the behavior of the Eulerian codes. Interestingly, the $\alpha=0.5$ case follows the same amplification history as the fiducial \textsc{Swift} run, suggesting that differences in diffusivity may also contribute to the discrepancy we observe between the fiducial \textsc{Arepo} and \textsc{Swift} runs. Despite the difference in amplification histories, the average divergence does not evolve differently for different $\alpha$ values except at early times ($t=1-2$ Gyr), suggesting that the energy growth is not driven by divergence errors on the grid. Furthermore, since the magnetic field within the disk remains sub-equipartition ($\langle E_\mathrm{mag}\rangle/\langle E_\mathrm{kin}\rangle \sim 10^{-4}$) in the low-$\alpha$ runs, we also do not observe any significant decrease in volume-averaged kinetic energy within the disk. 

\begin{figure*}[ht!]
\centering
\includegraphics[width=\linewidth]{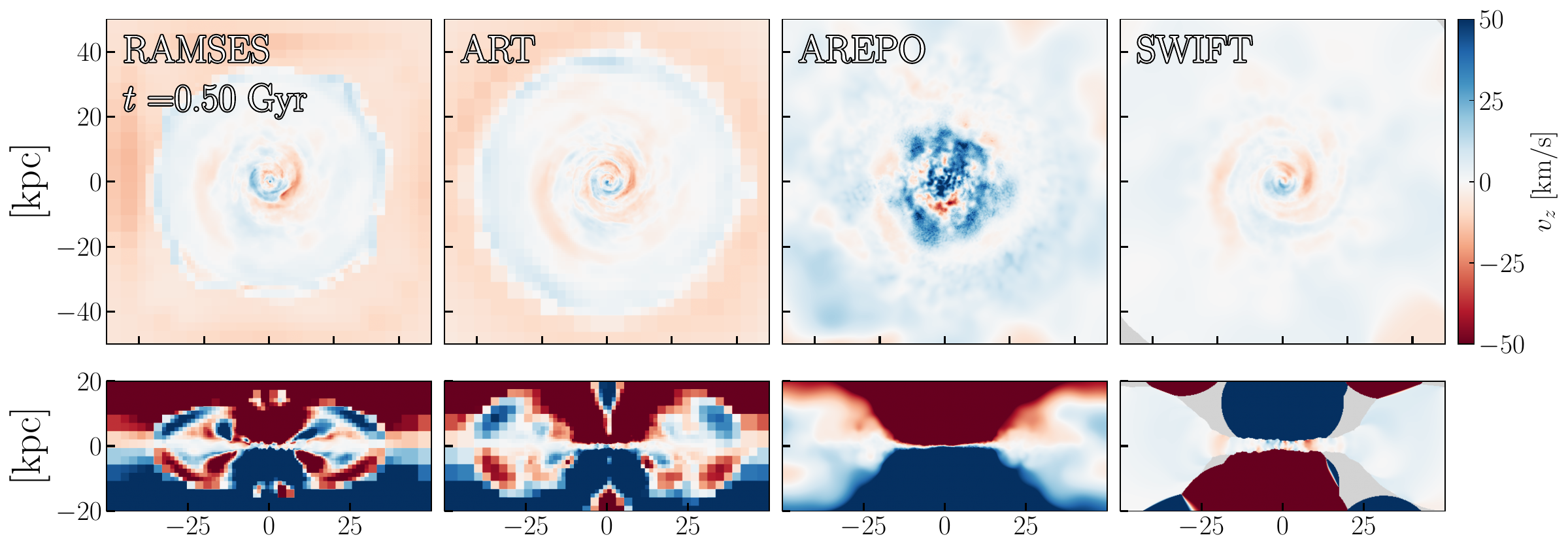}
\caption{Midplane and edge-on slices of the $z$-component of the gas velocity at $t=0.5$ Gyr for each code. Velocity perturbations seeded by the initial conditions and by physical processes such as cooling break the axisymmetry required by Cowling's anti-dynamo theorem, so dynamo amplification remains possible in principle. The blob-like structures in the \textsc{Swift} panel arise because few particles sit above and below the disk, so the slice there is set by the smoothing kernels of individual particles.} 
\label{fig:velz}
\end{figure*}

The evolution of magnetic energy within radial bins across the disk, shown in Figure~\ref{fig:arepo_radial}, further illustrates the transition toward Eulerian-like behavior as $\alpha$ decreases. The rightmost panel, representing our fiducial run, shows that the growth rate slows monotonically with radial distance from the disk's center of mass. As we approach the Eulerian limit ($\alpha = 0.01$), the volume-averaged magnetic energy in different annuli grows at comparable rates throughout the entire disk, consistent with the results obtained with \textsc{Ramses} and \textsc{ART}. A comparison of the amplification histories for $\alpha = 0.5$, $0.1$, and $0.01$ further indicates that the outer annuli, where the effective resolution $\Delta x$ is coarser, are the first to converge toward similar amplification histories.

To complete the analysis of our numerical experiment in \textsc{Arepo}, we compare the term-by-term contributions to the induction equation between the fiducial ($\alpha = 1$) and Eulerian-like ($\alpha = 0.01$) runs in Figures~\ref{fig:areposlow1}--\ref{fig:areposlow2}. At $t = 1.0$~Gyr, the shear term suppresses magnetic field growth across the disk in both runs, but with a substantially higher amplitude within the inner 1~kpc of the fiducial run. Coherent large-scale structures are considerably more pronounced in the shear and divergence contributions of the $\alpha = 0.01$ run than in the fiducial case, while the divergence term is also lower in magnitude across most of the disk, with the remaining higher values clearly tracing these more coherent spiral features. Compression is mostly positive in the outer disk and dominates the field evolution throughout the disk in the $\alpha = 0.01$ run. Overall, field growth in the Eulerian-like run is driven primarily by shear and compression, whereas in the fiducial run all three contributions remain comparable in magnitude.

We find similar behavior at $t = 4.0$~Gyr. The shear and divergence contributions are spatially disordered in both runs, with no clear large-scale morphology, while compression overwhelmingly dominates the growth rate in the outer regions of the disk. Both the slices and the radial profiles indicate that compression drives the field growth across the inner region of the galactic midplane, consistent with the behavior found across all simulations with different codes and setups at late times. We also observe that the divergence term is an order of magnitude lower in the $\alpha = 0.01$ run like in the previous snapshot at $t=1.0$ Gyr.

\section{Discussion}
\label{discussion} 

\subsection{What drives the Eulerian–Lagrangian differences?}

Our fiducial setup for this code comparison is an idealized test, i.e. without star formation, stellar feedback, or cosmological effects, and so it is not meant to reproduce what happens in real galaxies. However, its simplicity makes it a cleaner and more tractable problem than a full galaxy simulation for isolating the effects of the numerical scheme. The comparison presented above demonstrates that the numerical method used to solve the ideal MHD equations has a leading-order effect on the amplification of the galactic magnetic field. The\mbox{(quasi-)Lagrangian} schemes (\textsc{Arepo} and \textsc{Swift}) amplify the field to near-equipartition ($\sim 10-20\%$ the disk's average kinetic energy) strength within a few hundred Myr. On the other hand, results produced from Eulerian schemes (\textsc{Ramses} and \textsc{ART}) remain close to the theoretical prediction for shear-driven amplification, $\mathcal{E}_{\rm mag}\propto t^2$, several orders of magnitude below equipartition for the full duration of the simulation. By the end of the simulations, the magnetic energy in \textsc{Arepo} and \textsc{Swift} exceeds that in \textsc{Ramses} and \textsc{ART} by roughly three orders of magnitude, despite identical seed fields and initial disk conditions across all four codes. This discrepancy is comparable to, or larger than, the spread reported between resolution levels in single-code convergence studies, indicating that the numerical scheme is at least as consequential as commonly-invoked resolution effects in the literature \citep{2013MNRAS.432..176P, 2018A&A...609A.104K, 2018MNRAS.476.2890B, 2022MNRAS.513.3326M, 2023A&A...673A..47W}. Which code is physically correct cannot be determined with certainty from this comparison alone. The Cowling anti-dynamo theorem does not strictly apply here, since velocity perturbations are seeded both by our choice of initial conditions and by physical mechanisms such as cooling, as shown in Figure \ref{fig:velz} --- either of which could drive dynamo amplification, in principle. Even without stellar feedback to drive turbulence on the disk, all four simulations develop coherent, large-scale $v_z$ motions, sourced by early disk instabilities during the cooling-driven relaxation of the gas towards the midplane. As these flows shear against the surrounding medium, they can become Kelvin-Helmholtz unstable and decay into a turbulent cascade, seeding velocity fluctuations that can drive a small-scale dynamo. The lower numerical diffusivity of the \mbox{(quasi-)Lagrangian} schemes lets this cascade develop further than in more diffusive Eulerian schemes, which may help explain the orders-of-magnitude difference in magnetic amplification we report above. 

We also find that while higher divergence errors may contribute to the faster amplification seen in the \mbox{(quasi-)Lagrangian} simulations, the shear and compression contributions to magnetic energy growth also differ between schemes. This decomposition is not necessarily immune to divergence errors either, since spurious ($\nabla\cdot{\bf B}$)-sourced terms can project onto either mechanism. It is also important to note that the scale at which divergence errors are produced matters, in addition to their overall magnitude. If the divergence error is predominantly confined to and fluctuates rapidly on small spatial scales, perturbations sourced by these errors should have a minimal impact on strong, large-scale magnetic fields. We therefore do not claim the discrepancy is purely physical, but our \textsc{Arepo} experiment in \S\ref{sec:eulerlag} suggests it cannot be attributed solely to divergence errors.

\begin{figure*}[htbp!]
    \centering
    \hspace{-0.1in}\includegraphics[width=0.95\linewidth]{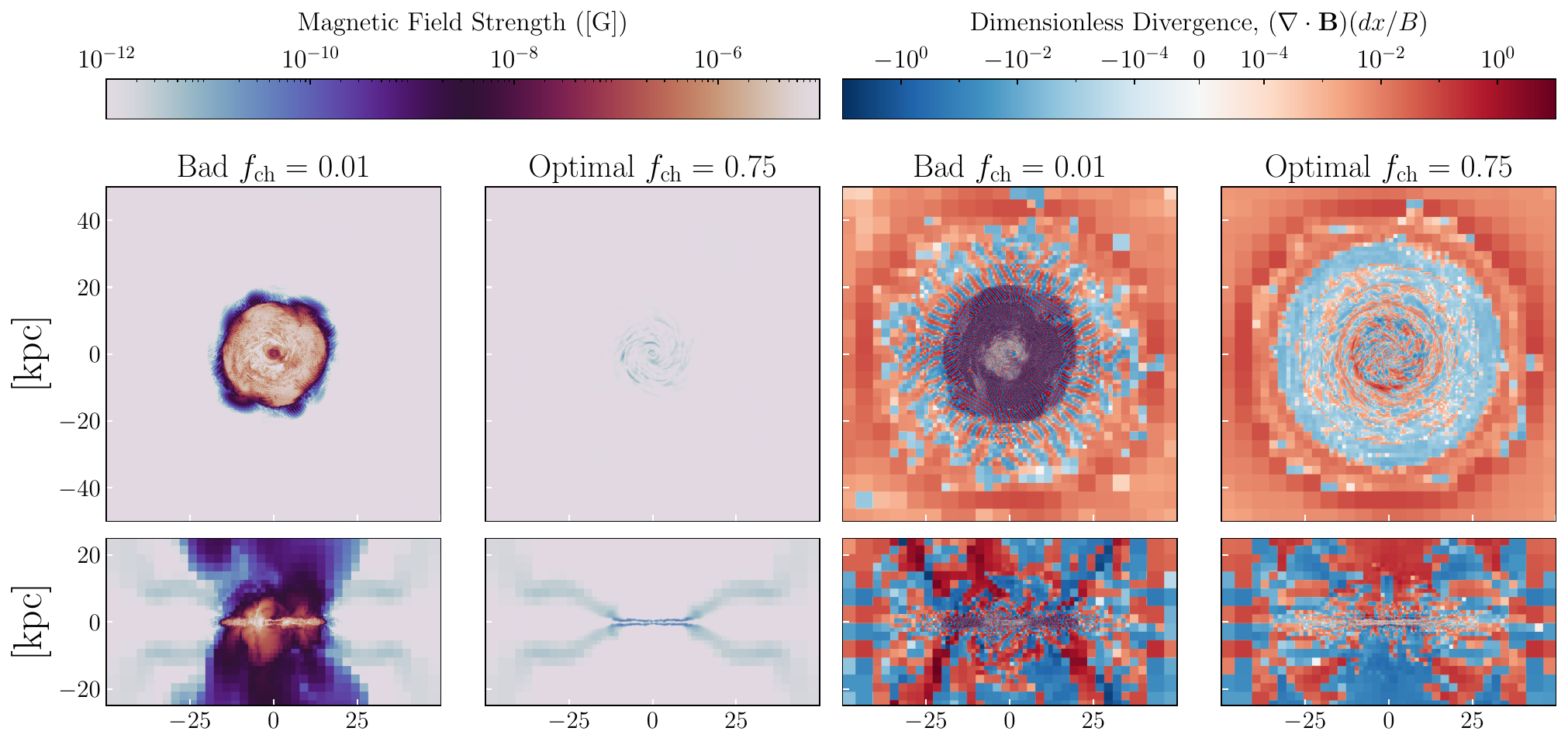}
    \caption{Midplane slices of the magnetic field strength (left) and dimensionless divergence (right) for fiducial \textsc{ART} simulations run with different Dedner cleaning scheme parameters, $\gamma=0.5$ and $f_{\rm ch} \in [0.01, 0.75]$, at $t=200$ Myr. Here, $\gamma$ and $f_{\rm ch}$ control the damping and propagation of the scalar $\Psi$ in the Dedner scheme, respectively. The fiducial \textsc{ART} simulation adopted in this comparison uses $(f_{\rm ch}, \gamma) = (0.75, 0.5)$. This test demonstrates that a low $f_{\rm ch}$, such as $f_{\rm ch}=0.01$, which corresponds to slow propagation of divergence errors away from their source, results in a local build-up of divergence errors and spurious growth of the magnetic field.}
    \label{fig:glm_scan1}
\end{figure*}

Nonetheless, the fast amplification of magnetic energy in the \mbox{(quasi-)Lagrangian} codes has clear dynamical effects on the disk. Once the magnetic field becomes dynamically important, Lorentz forces efficiently suppress gas motions that initially powered the amplification process. The earlier onset and increased strength of magnetic braking in \textsc{Arepo} and \textsc{Swift} therefore arise naturally from their faster magnetic growth, whereas the more gradual kinetic energy evolution in \textsc{Ramses} and \textsc{ART} reflects the weaker magnetic backreaction operating in the shear-dominated, sub-equipartition regime.

We have also examined how the magnetic growth rate varies with radius. Shear-driven field growth (Eq.~\ref{eq:shear}) scales with $B_r \, \partial\Omega/\partial r$, where
\begin{equation}
    \frac{\partial \Omega}{\partial r} = \frac{1}{r}\frac{\partial v_\varphi}{\partial r} - \frac{v_\varphi}{r^2}.
\end{equation}
The AGORA rotation curve $v_\varphi(r)$ climbs gradually from the center, peaks around $r\approx5$ kpc, and declines slowly thereafter, so $\Omega=v_\varphi/r$ falls steeply from the innermost radii outward. Shear, $|\partial\Omega/\partial r|$, is therefore largest at small $r$ and decreases monotonically with radius, dropping by an order of magnitude by the time the rotation curve flattens. Differential rotation should therefore wind radial field into azimuthal field most efficiently at small radii, driving faster amplification there, consistent with the delayed growth seen at larger radii in the \mbox{(quasi-)Lagrangian} simulations. The nearly radius-independent growth rates in the Eulerian codes instead suggest numerical effects, such as enhanced diffusion from coarser mesh refinement, that weaken this dependence on local shear; such advection errors are expected to be larger in supersonic flows like ours, where the typical Mach number $\mathcal{M}\sim10$. We note, however, that the spiral perturbation superposed on the rotation curve in our simulations can make the local shear rate vary non-monotonically in radius and azimuthal angle, complicating this simple picture.

The resolution tests in \S\ref{results} further indicate that the faster growth in \textsc{Arepo} and \textsc{Swift} is not simply a matter of insufficient resolution of either the Eulerian or Lagrangian simulations. Doubling the spatial resolution in \textsc{Ramses} and \textsc{ART} left the amplification history essentially unchanged (Figure~\ref{fig:resolution}), and increasing the mass resolution in \textsc{Arepo} and \textsc{Swift} by an order of magnitude did not suppress, and if anything mildly reduced, its excess growth (Figures~\ref{fig:resolution} and \ref{fig:arepo_resolution}). This differs from the behavior expected if the discrepancy were purely a consequence of under-resolving the gas within the disk, as in the quasi-Lagrangian AMR refinement tests of \citet{2022MNRAS.513.3326M}, and instead suggests that the numerical diffusivity in each code is set by something other than the spatial or mass resolution alone.

Another striking result comes from our numerical experiment in which we artificially slow the \textsc{Arepo} mesh to probe its behavior in the Eulerian or static-grid limit. Reducing the mesh velocity brings \textsc{Arepo}'s amplification history for the average magnetic energy into agreement with \textsc{ART} and \textsc{Ramses}, suggesting that the Eulerian vs. Lagrangian discrepancy is numerical rather than physical in origin, arising from differences in numerical diffusivity. In the fully Lagrangian limit ($\alpha = 1$), low numerical diffusion allows small-scale motions to persist, boosting the growth rate of the magnetic field and driving earlier saturation. As $\alpha$ decreases, the outer disk ($r\in[5,30]$ kpc), where coarser resolution leads to larger advection errors, is the first to recover Eulerian-like behavior, consistent with the picture where lower resolution increases sensitivity to the choice of discretization. This convergence toward Eulerian behavior as $\alpha$ decreases indicates that the enhanced amplification in fully Lagrangian runs is primarily a numerical consequence of lower diffusivity. As noted earlier in \S\ref{sec:numerical_effects}, there is generally a tradeoff between diffusivity and numerical stability, so a lower-diffusivity algorithm should not automatically be regarded as more physically correct.

\subsection{How do $\nabla\cdot\mathbf{B}$ errors affect magnetic field amplification in simulations?}
\label{sec:divergence_art}

To illustrate the effects of divergence errors in MHD simulations of galaxies, we present a pair of \textsc{ART} simulations that sample the parameter space of the Dedner scheme (see \S\ref{sec:divergenceschemes}), varying the damping parameter $\gamma$ and the propagation speed $c_h$ that govern the evolution of the scalar field $\Psi$. Increasing $\gamma$ more aggressively damps the scalar field $\Psi$, suppressing divergence errors at the risk of over-damping $\Psi$ modes, while increasing $f_{\rm ch}$ propagates errors away from their source more rapidly. We show in \S\ref{sec:fiducial} that, with an optimal combination of Dedner parameters, our \textsc{ART} results are comparable to those of a \textsc{Ramses} simulation using constrained transport, despite non-zero divergence errors. In Figure \ref{fig:glm_scan1}, we show the opposite case, in which choosing a small propagation speed $f_{\rm ch}$ prevents divergence errors from being efficiently propagated away or damped, allowing them to accumulate locally and artificially drive magnetic-field amplification in the galactic disk. We find that the propagation speed $c_h$ has a leading-order effect on suppressing divergence errors, while the damping coefficient $\gamma$ plays a secondary role in controlling the divergence errors. More generally, this behavior is not specific to the Dedner scheme. We expect that for any numerical divergence cleaning scheme, a sufficiently large local accumulation of $\nabla\cdot\mathbf{B}$ errors can lead to spurious magnetic-field amplification. Thus, regardless of the specific scheme used, preventing the local build-up of divergence errors is essential for producing physically reliable magnetic field evolution in galaxy simulations.

\begin{figure*}[ht!]
    \centering
    \hspace{-0.15in}\includegraphics[width=\linewidth]{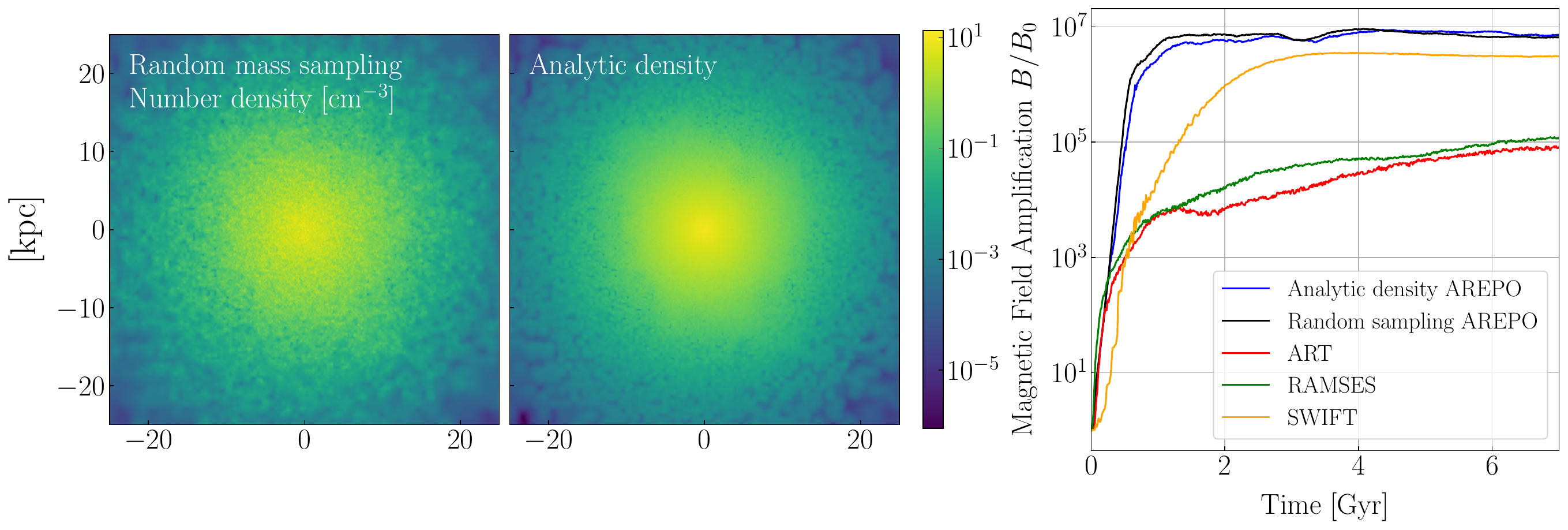}
    \caption{Slices of the initial number density for the simulation setups that sample initial conditions with randomly distributed, fixed-mass particles, compared to the setup that instead imposes the analytic density profile described in \S\ref{method} (left), and corresponding magnetic field amplification histories for these setups alongside the other fiducial runs (right). We find no significant deviation between the simulation using the analytic density profile and our fiducial particle-sampled \textsc{Arepo} run, indicating that the pressure noise introduced by random particle sampling does not significantly affect the magnetic field amplification.}
    \label{fig:ic_comparison}
\end{figure*}

\subsection{Insensitivity of field amplification to initial conditions}

\begin{figure*}[ht!]
    \centering
    \includegraphics[width=0.95\linewidth]{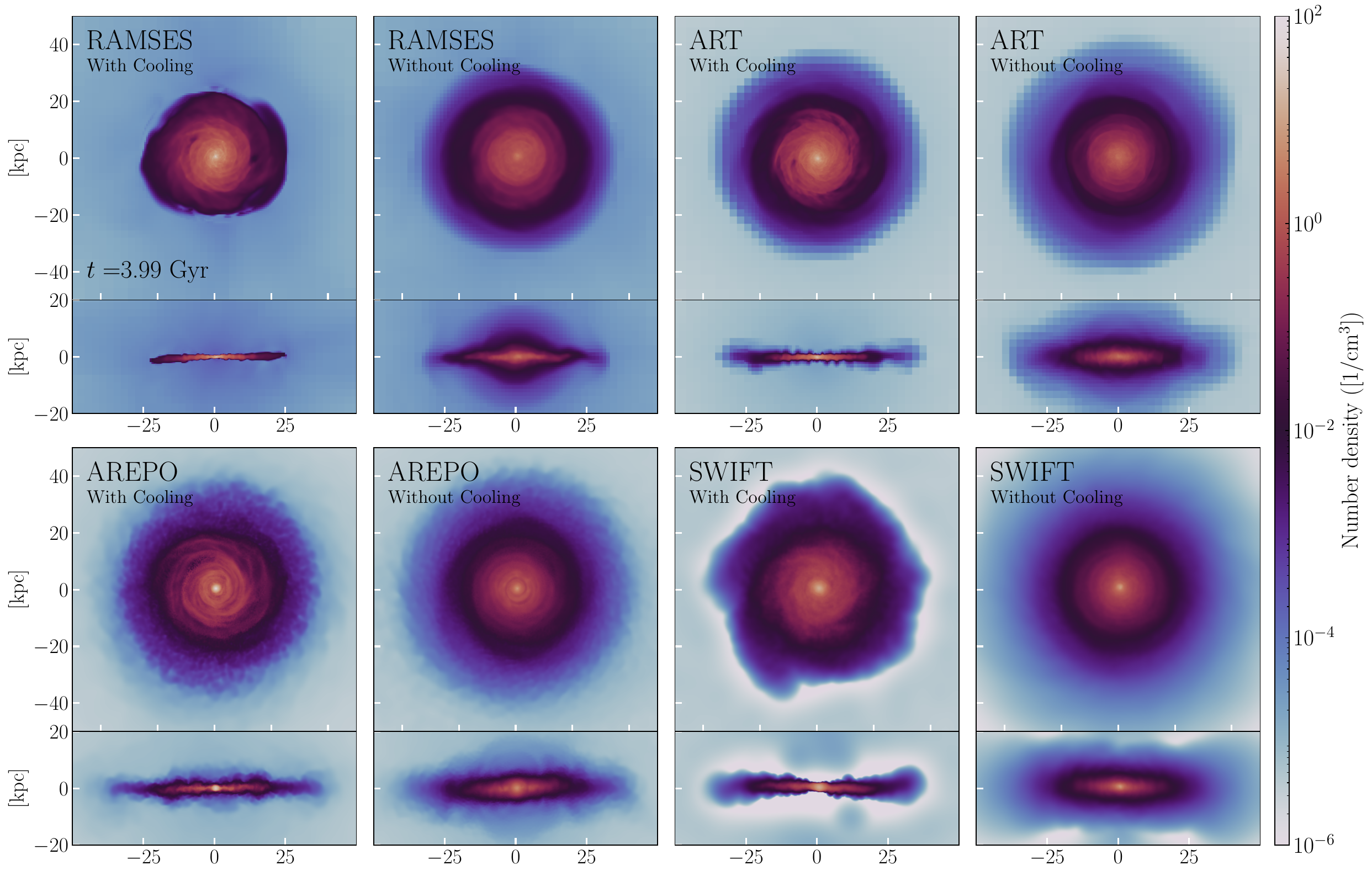}
    \includegraphics[width=0.95\linewidth]{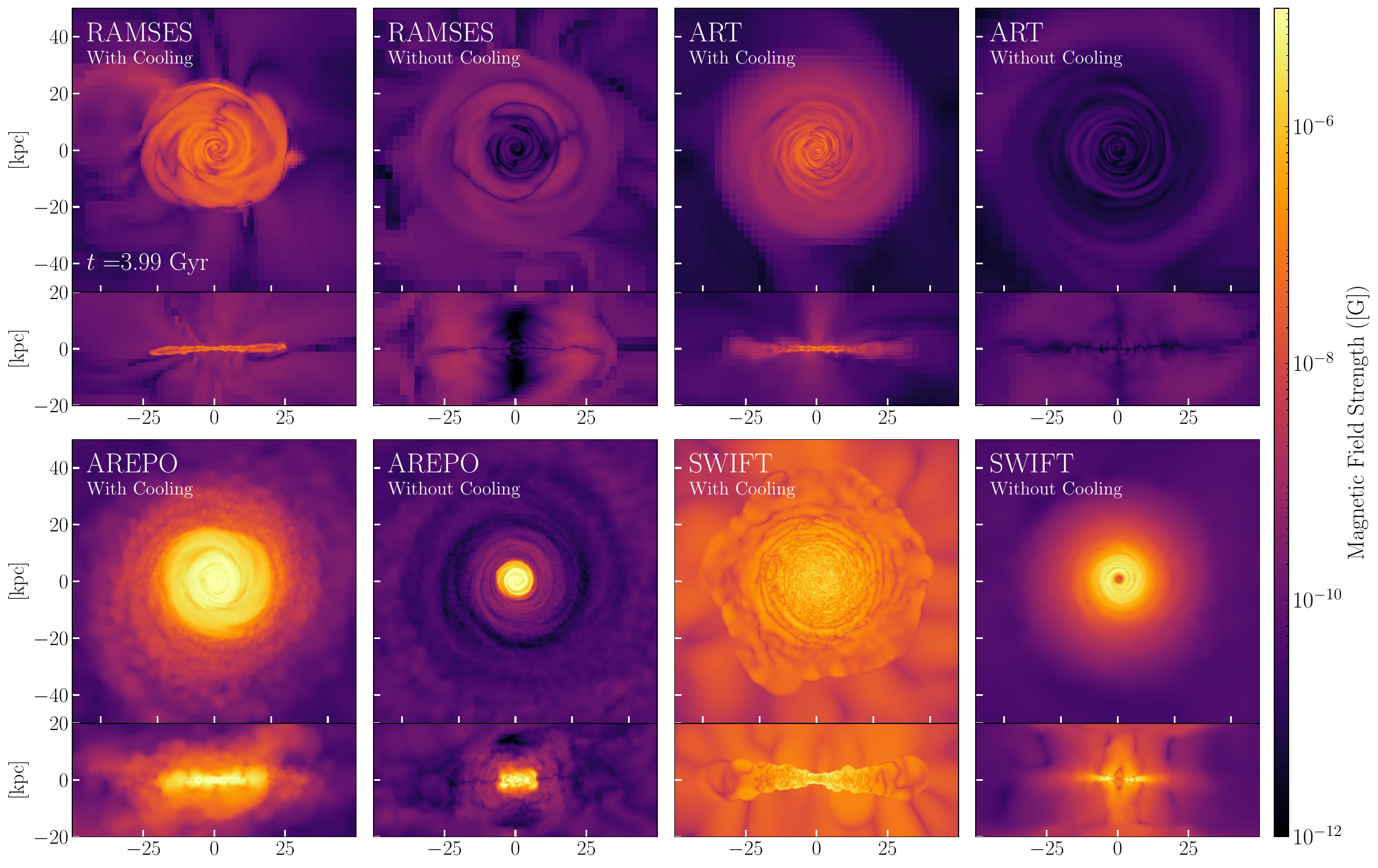}
    \caption{\textbf{Number density (top) and magnetic field strength (bottom)} slices at $t=4.0$ Gyr, run with and without cooling. The scale height of the disk in simulations without cooling increases as the disk's thermal energy build up due to adiabatic compression. As a consequence, the field along the midplane is weaker as mass is advected away from this region, and the disk becomes more extended and diffuse.}
    \label{fig:cooling_slices}
\end{figure*}
\begin{figure}[ht!]
    \centering
    \includegraphics[width=\linewidth]{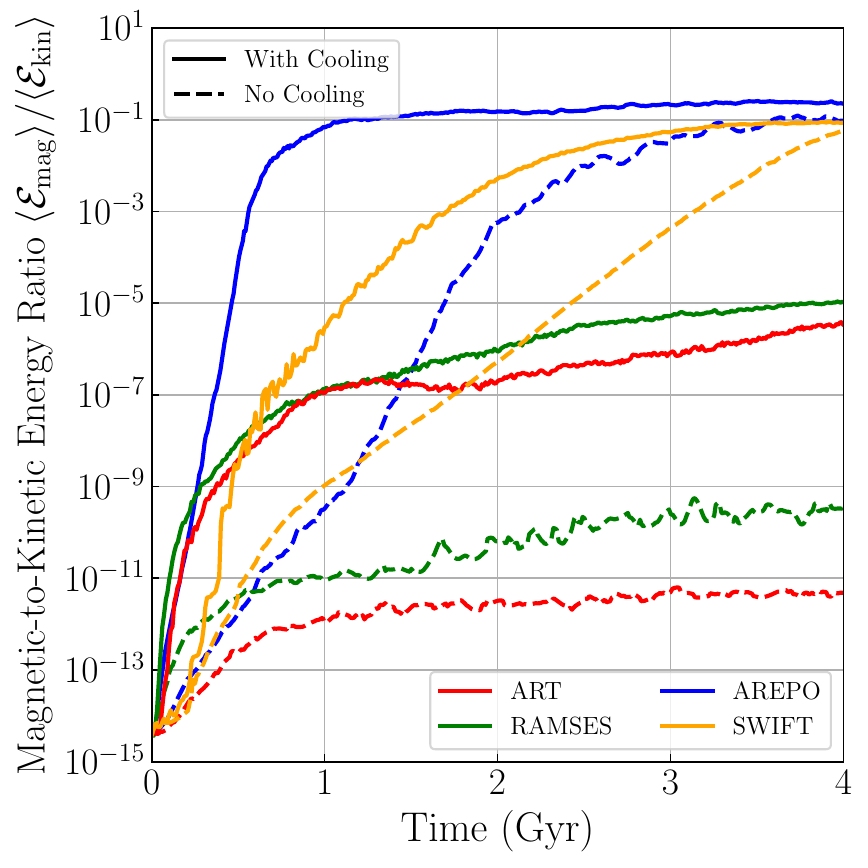}
    \caption{\textbf{Volume-averaged ratio between magnetic and kinetic energy} within a cylindrical volume of radius $r=15$ kpc and height $|z|<0.5$ kpc, for fiducial runs with and without cooling. Magnetic field amplification proceeds more slowly in the absence of cooling.}
    \label{fig:cooling_timeseries}
\end{figure}

One may observe in Figure~\ref{fig:init_cond} that the density slices in our fiducial \textsc{Arepo} and \textsc{Swift} simulations are quite noisy due to random particle sampling, whereas \textsc{Ramses} and \textsc{ART} start from much smoother density fields. Setting a uniform temperature across the simulation domain therefore implies random cell-to-cell pressure gradients in \textsc{Arepo} and \textsc{Swift} that are nearly absent in the grid-based codes. These pressure gradients could potentially drive random velocity kicks capable of seeding a mean-field dynamo throughout the disk. We tested whether this pressure noise has a measurable impact on magnetic field amplification by re-running our fiducial \textsc{Arepo} simulation with a smoother initial density field, holding all other simulation parameters fixed, and comparing the resulting field amplification to our fiducial run. This re-simulation was performed by initializing the density field with the analytic profile described in \S\ref{method}, using the \texttt{MESHRELAX\_DENSITY\_IN\_INPUT} option in \textsc{Arepo}. 

Figure~\ref{fig:ic_comparison} shows the initial number density slices for the simulation setup that samples initial conditions with randomly distributed, fixed-mass particles, compared to the setup that instead imposes the analytic density profile, along with the corresponding magnetic field amplification histories for both setups alongside the other fiducial runs.  We find no significant difference in the amplification histories between the \textsc{Arepo} simulations with the two different prescriptions for generating the initial conditions. This indicates that the random velocity perturbations introduced by our random particle sampling do not affect the robustness of our findings. As a further test, we propose performing a similar comparison but with particles forced to remain exactly on Keplerian orbits, thereby eliminating any kinematic noise associated with the initial particle sampling.

It is important to note that we did not consider other variations of the initial conditions in this code comparison, such as the initial magnetic field topology or the mass of the galaxy. 

\subsection{Sensitivity of field amplification to cooling}
Cooling plays a central role in shaping the dynamical state of interstellar and circumgalactic gas, and could therefore influence the efficiency of magnetic field amplification. In particular, the vertical structure of the disk is set by a balance between gravity, which confines gas toward the midplane, and thermal and magnetic pressure support, which opposes it. Without cooling, the thermal energy generated by adiabatic compression as gas settles into the disk could not be efficiently removed. The disk responds by puffing up to a larger pressure scale height, lowering the midplane density. Hence, the fields are less efficiently wound by shear and compression.

The differences in disk structure are evident in the slices shown in Figure~\ref{fig:cooling_slices} for runs with and without cooling, as well as in the magnetic field amplification history shown in Figure~\ref{fig:cooling_timeseries}. In the cooling runs, the stronger amplification is consistent with a greater contribution from shear-driven stretching and compression as gas settles into the disk. On the other hand, the non-cooling runs remain weakly amplified because the elevated pressure scale height of the disk makes stretching and compression more inefficient. An exception is in our \textsc{Arepo} runs, where strong amplification towards the central regions of the disk persists even in the absence of cooling. The difference in behavior across codes may also reflect how the numerical schemes capture these central flows, which requires further investigation.

\subsection{Magnetic field effects on star formation and galactic feedback}

Galaxies with different magnetic amplification histories may experience star formation and feedback at substantially different stages of their evolution. In our simulations, the magnetic-to-kinetic energy ratio differs by up to four orders of magnitude between Eulerian and Lagrangian methods at late times. Since $E_{\rm mag}\propto B^2$, this corresponds to approximately two orders of magnitude in magnetic field strength. Such differences are large enough to produce order-of-magnitude variations in mass-to-flux ratios, Alfv\'en speeds, and cosmic-ray transport timescales, suggesting that numerical differences in magnetic field amplification may propagate into predictions for star formation, feedback, and galactic outflows long before magnetic saturation is reached.

\textit{Star formation.} Magnetic support is commonly characterized by the mass-to-flux ratio, $\mu \equiv (M/\Phi)/(M/\Phi)_{\rm crit}$, where $\Phi$ is the magnetic flux threading a cloud. Since $\Phi \propto B$, a factor of $100$ difference in field strength implies a factor of $100$ difference in $\mu^{-1}$. Molecular clouds that are magnetically supercritical in one simulation may therefore remain significantly more supported against gravitational collapse in another, potentially leading to different star formation efficiencies and cloud lifetimes.

\textit{Stellar feedback.} The coupling of supernova feedback to the ISM may also depend sensitively on magnetic field amplification. The magnetic energy density is $u_B = B^2/2$. A factor of $100$ difference in field strength corresponds to a factor of $10^4$ difference in magnetic energy density. Such variations can alter the relative importance of magnetic pressure and tension forces during shell expansion, influence the structure of superbubbles, and modify the efficiency with which feedback drives galactic outflows. The dynamical importance of magnetic fields for galactic outflows can be estimated by comparing the magnetic pressure, $P_B = B^2/2$,
to the ram pressure of the outflow, $P_{\rm ram} = \rho v_{\rm out}^2$. For a characteristic CGM density of $n = 10^{-3}\,\mathrm{cm^{-3}}$ and an outflow velocity of $v_{\rm out}=300\,\mathrm{km\,s^{-1}}$, the ram pressure is $P_{\rm ram} \approx 1.5\times10^{-12}\,\mathrm{erg\,cm^{-3}}$. In comparison, magnetic field strengths of $1$, $10$, and $100\,\mu\mathrm{G}$ correspond to magnetic pressures of approximately $4\times10^{-14}$, $4\times10^{-12}$, and $4\times10^{-10}\,\mathrm{erg\,cm^{-3}}$, respectively. Thus, while $\mu$G-strength fields are dynamically subdominant ($P_B/P_{\rm ram}\sim0.03$), fields of order $10\,\mu\mathrm{G}$ are capable of exerting pressures comparable to the ram pressure of the outflow. Consequently, the large differences in magnetic field amplification found in our simulations may lead to qualitatively different predictions for the launching and propagation of galactic winds.

\textit{Cosmic ray transport and feedback.} Differences in magnetic amplification may also affect cosmic-ray transport. In the streaming limit, cosmic rays propagate at approximately the Alfv\'en speed, $v_A =B/\sqrt{4\pi\rho}$, implying an escape timescale from a galactic disk of scale height $H$,
$t_{\rm esc} \sim H/v_A \propto B^{-1}$. A factor of $100$ difference in magnetic field strength therefore implies a factor of $100$ difference in cosmic-ray streaming timescales. Because the cosmic-ray pressure gradient depends on the residence time of cosmic rays within the disk, differences in magnetic amplification may substantially alter predictions for cosmic-ray-driven winds and star formation \citep{2021ApJ...910..126S}. 

These effects extend into the circumgalactic medium (CGM), where our simulations differ substantially in the magnetization of gas above and below the disk. Because the Alfv\'en speed scales with the magnetic field strength, differences in halo magnetization would change the rate at which cosmic rays stream through the CGM. This in turn would alter both the distribution of cosmic-ray pressure and the rate at which streaming cosmic rays heat the gas, which scales as $\sim v_A |\nabla P_{\rm CR}|$, where $P_{\rm CR}$ is the cosmic-ray pressure. The degree to which cosmic rays drive feedback on the CGM therefore depends on how efficiently the magnetic field is amplified in the disk and transported into the halo.

\subsection{Observational constraints}
Our simulations do not include key physical processes in galaxies, e.g. star formation and stellar feedback, so direct comparisons with observations are difficult to make. Even so, our code comparison yields predictions for two key aspects of magnetic field amplification, i.e. the growth rate of magnetic fields and their strength within the galactic disk as it settles. We find that these quantities show substantial differences between the numerical schemes we compared. Testing these predictions therefore requires two complementary observational constraints, i.e. statistical samples spanning cosmic time to trace the evolution of magnetic fields across galaxy populations, and resolved measurements of individual galaxies that anchor magnetic field strength and structure at specific epochs. Multi-wavelength observations from the far-infrared to the radio are well suited to provide both, albeit through different techniques and in different galactic environments.

Our understanding of magnetic fields in local-volume galaxies ($z\approx0$) is built on decades of radio and far-IR polarimetry, which mapped the large-scale strength and structure of magnetic fields across dozens of nearby spirals. More recently, deep MeerKAT observations have extended magnetic field studies to fainter and more distant galaxies \citep{2023A&A...678A..56B, 2024MNRAS.528.2511T}, while LOFAR has probed CGM fields through Faraday rotation and constrained the relation between equipartition field strength and gas density from synchrotron emission \citep{2023A&A...670L..23H, 2023A&A...669A...8H}. Recent Jansky VLA (JVLA) broadband polarimetric observations have directly resolved coherent magnetic structures in nearby spirals, revealing magnetic fields indicative of large-scale coherent structure beyond the galactic disk \citep{2019A&A...632A..11M, 2020A&A...639A.112K, 2020A&A...642A.118K, 2025ApJ...978....5X}. Together, these observations have converged on a consistent picture in which ordered, large-scale spiral magnetic fields with strengths of a few to several tens of $\mu$G are ubiquitous in nearby spiral galaxies. These measurements also provide constraints on the magnetic field strength of the large-scale structure, i.e. in filaments and the intergalactic medium, limiting it to a few nG \citep{2020MNRAS.495.2607O, 2022MNRAS.512..945C, 2025A&A...693A.208C, 2025A&A...696A.203P}. Ongoing all-sky surveys such as POSSUM will provide statistical constraints on magnetic field coherence scales in both the Milky Way foreground and the intervening cosmic web by combining thousands of polarized extragalactic sources with high-angular-density coverage \citep{2024MNRAS.533.4068A, 2025PASA...42...91G,2026A&A...705A.143A}. 

However, these measurements alone do not constrain the mechanisms responsible for field amplification or how magnetic field growth couples to other galactic processes, such as star formation and stellar feedback. A complementary approach is to forward-model synthetic observables directly from galaxy simulations and confront them with the observations, e.g. FIR-radio correlation and $\gamma$-ray emission from cosmic-ray populations, rather than comparing magnetic field strengths alone. Using cosmic-ray magnetohydrodynamics (CRMHD) galaxy simulations, \citet{2021MNRAS.508.4072W}, \citet{2023MNRAS.525.4437W} and \citet{2025ApJ...980..135P} predict synchrotron emission spectra matching the observed FIR-radio correlation across several orders of magnitude in star formation rate, while \citet{2021MNRAS.505.3295W} replicate the observed FIR-$\gamma$-ray relation in galaxies by assuming a low cosmic-ray (CR) injection efficiency from supernovae paired with anisotropic CR diffusion. Similarly, \citet{2026A&A...710A.163K} use CRMHD simulations of a Milky Way-analog galaxy and reproduce the observed Fermi-LAT luminosity and spectral slope without additional parameter tuning. Synthetic Faraday rotation measures computed from the Auriga cosmological simulations also provide an independent test of the field morphology and strength, finding RM values broadly consistent with the Milky Way for observers at the solar circle and outside the galaxy \citep{2018MNRAS.481.4410P}.

Extending this picture to higher redshift, and thus constraining how rapidly fields grow in spiral galaxies, remains a much harder problem. Measurements of magnetic fields within a galaxy and its host halo are currently possible only for strongly-lensed galaxies. Using the JVLA, differential Faraday rotation between multiple lensed images of a polarized background quasar has been used to isolate the coherent magnetic field of the foreground lensing galaxy, revealing disk-dominant magnetic fields at $z\approx0.4$ similar to present-day spiral galaxies \citep{2017NatAs...1..621M, 2026A&A...705A..28K}. On the other hand, ALMA dust polarimetry has provided the first glimpse of ordered magnetic structures at $z=2.3$ \citep{2023Natur.621..483G, 2025MNRAS.540L..78D} and at $z=5.6$ \citep{2024A&A...692A..34C}. The Square Kilometre Array (SKA) is expected to increase the density of RM measurements by orders of magnitude, enabling detailed studies of magnetic field evolution over cosmic time. Surveys with the SKA will detect star-forming galaxies to higher redshifts using similar lensing techniques, and targeted follow-ups of lensed candidates will allow statistical studies of magnetic field strength as a function of galaxy mass, star formation rate, and environment \citep{2026arXiv260630025M}. Combined with optical and infrared surveys (e.g. Euclid, Rubin LSST, and Roman), these data will facilitate comparisons between magnetic field growth and galaxy assembly histories.

Future missions like the far-infrared probe mission PRIMA (The PRobe far-Infrared Mission for Astrophysics) \citep{2025JATIS..11c1628G, 2024SPIE13092E..3BB, 2025JATIS..11c1625C} will deliver unprecedented sensitivity across the far-infrared spectrum (down to $\sim 10$ pc spatial resolution within 0.5 Mpc) and is poised to revolutionize our understanding of magnetic fields in extragalactic sources and within the Milky Way \citep{2025JATIS..11c1624D, 2025JATIS..11c1620M}.

\section{Conclusions}
\label{conclusion}

We have presented the \textit{first systematic comparison of four cosmological MHD codes} --- \textsc{ART}, \textsc{Ramses}, \textsc{Arepo}, and \textsc{Swift} --- in simulating the co-evolution of an isolated Milky Way-type galaxy and its magnetic field. We summarize our main conclusions here:
\begin{enumerate}
    \item Magnetic fields in \mbox{(quasi-)Lagrangian} codes (\textsc{Arepo}, \textsc{Swift}) amplify exponentially fast, reaching near-equipartition ($\langle E_{\rm mag}\rangle/\langle E_{\rm kin}\rangle \sim 0.1$) within a few Gyr. On the other hand, magnetic fields in Eulerian codes (\textsc{ART}, \textsc{Ramses}) grow more slowly and remain several orders of magnitude below equipartition throughout the simulation.
    \item In \mbox{(quasi-)Lagrangian} codes, magnetic growth is faster in the inner disk and slows down with increasing radius. In contrast, Eulerian codes show nearly uniform growth rates across all radii, suggesting that numerical effects such as diffusion and resolution dependence may weaken the imprint of the underlying shear profile on magnetic field amplification. However, the formation of spiral arms in the galactic disk may also contribute to the differences observed.
    \item The term-by-term decomposition of the induction equation reveals systematic differences in the magnetic field amplification mechanisms in Eulerian and \mbox{(quasi-)Lagrangian} simulations. At early times, shear is the dominant driver of magnetic field growth in the Eulerian codes, whereas no single mechanism dominates in the \mbox{(quasi-)Lagrangian} codes, where shear, compression, and the magnetic divergence term contribute at comparable levels. At later times, compression becomes the primary driver of magnetic field evolution in all four codes. Throughout the simulation, the magnetic divergence term remains negligible in the Eulerian codes but is more pronounced in the \mbox{(quasi-)Lagrangian} codes, where it produces highly localized regions of both magnetic amplification and decay.
    \item Artificially reducing the velocity of mesh-generating points in \textsc{Arepo} toward the Eulerian limit progressively reduces both the growth rate and saturation level of the magnetic energy, approaching behavior observed in \textsc{ART} and \textsc{Ramses}. The Eulerian vs. Lagrangian discrepancy likely arises from differences in numerical diffusivity rather than being a physical effect --- reduced diffusion in the Lagrangian limit allows small-scale numerical noise to persist and amplify, inflating both the growth rate and saturation level.
\end{enumerate}

These results have important implications for both past and future simulation studies of galactic magnetic fields. Because the discrepancy between Eulerian and \mbox{(quasi-)Lagrangian} schemes appears to arise primarily from differences in numerical diffusivity, quantitative comparisons of magnetic field growth rates across studies using different codes should be interpreted with caution. Likewise, conclusions regarding the dominant amplification mechanism, whether shear, compressive, or turbulent driving, may be influenced by code-dependent numerical effects. Moreover, divergence-cleaning diagnostics, such as the dimensionless divergence error $(\nabla\cdot {\bf B})\,(\Delta x/B)$, should be reported alongside magnetic field results in future studies.

We leave for future work an extension of this comparison to cosmological simulations, variations in stellar feedback and subgrid prescriptions, and different initial conditions. It would also be valuable to repeat this comparison in a more controlled setup where the dominant dynamo mechanism is well understood, such as the magnetorotational instability in a local shearing box or the small-scale turbulent dynamo in homogeneous, isotropic turbulence.

\begin{acknowledgments}
M.I.U. is grateful to Philipp Girichidis, Lars Hernquist, and Siddhant Solanki for productive discussions that helped shape this work, and to Sander Somers who helped refine the manuscript. This research was supported in part by the National Science Foundation under grant No. AST 2338388. M.I.U. and J.Y.K. gratefully acknowledge support from the GRAD-MAP Summer Scholars Program. The GRAD-MAP Summer Scholars Program is financially supported by the Departments of Astronomy and Physics and the College of Computer, Mathematical, and Natural Sciences (CMNS) at the University of Maryland, College Park. C.P. acknowledges support from the European Research Council under the ERC-AdG grant PICOGAL-101019746. The simulations and analysis pipeline were run on the \textsc{Zaratan} supercomputing cluster and on the \textsc{Astra} computing server at the University of Maryland, College Park.
\end{acknowledgments}

\begin{contribution}
M.I.U. was responsible for setting up and running all simulations, analyzing the results, producing the visualizations, and writing the manuscript. B.D. implemented the ideal MHD solver in ART; B.D. and V.S. assisted with setting up the initial conditions in \textsc{ART}; R.T. assisted with \textsc{Ramses}; F.M. assisted with \textsc{Arepo}; and N.S. and M.S. assisted with \textsc{Swift}. J.Y.K. ran and analyzed early simulations using \textsc{Swift} and performed other tests through the University of Maryland GRAD-MAP Summer Scholars program. O.Z. and C.P. helped design the numerical experiments performed with \textsc{Arepo}. All authors contributed to the interpretation of the simulation results and to revising and refining the manuscript for publication.
\end{contribution}

\software{\textsc{Ramses} \citep{2002AandA...385..337T}, \textsc{ART} \citep{1997ApJS..111...73K, 2008ApJ...672...19R}, \textsc{Arepo} \citep{2010MNRAS.401..791S}, \textsc{Swift} \citep{2024MNRAS.530.2378S}, yt \citep{2011ApJS..192....9T}, \textsc{IPython} \citep{2007CSE.....9c..21P}, \textsc{numpy} \citep{2020Natur.585..357H}, \textsc{scipy} \citep{2020NaMet..17..261V}, \textsc{matplotlib} \citep{2007CSE.....9...90H}, \textsc{swiftsimio} \citep{2020JOSS....5.2430B}}

\appendix
\section{Shear-driven amplification in a differentially rotating disk ($\Omega$-effect)}
\label{sec:shear}
Here we derive the amplification rate of the magnetic field strength and energy density within a spinning disk with a radially-dependent angular velocity profile $\Omega(r)$. A similar derivation was also shown by \citet{2010A&A...523A..72D} and in other papers in the literature, but is provided here for illustration. We start from the induction equation in the ideal MHD limit ($\eta = 0$),
\begin{equation}
\frac{\partial \mathbf{B}}{\partial t} = \nabla \times (\mathbf{v} \times \mathbf{B}),
\end{equation}

\noindent and assume that the disk has a radially dependent velocity profile $\mathbf{v} = r\Omega(r)\hat{\phi}$ and an arbitrary initial magnetic field $\mathbf{B} = B_r\hat{\mathbf{r}} + B_\phi\hat{\phi} + B_z\hat{\mathbf{z}}$ near the midplane, such that 
\begin{equation}
\mathbf{v} \times \mathbf{B} = r\Omega(B_z\hat{\mathbf{r}} - B_r\hat{\mathbf{z}}).
\end{equation}

\noindent Taking the curl of the equation above in cylindrical coordinates yields a purely azimuthal vector field,
\begin{equation}
    \nabla \times (\mathbf{v} \times \mathbf{B}) = \left[r\Omega\!\left(\frac{\partial B_r}{\partial r}
    + \frac{\partial B_z}{\partial z}\right) + \Omega B_r
    + rB_r\frac{\partial\Omega}{\partial r}
    + rB_z\frac{\partial\Omega}{\partial z}\right]\hat{\phi}.
\end{equation}
If we assume azimuthal symmetry, the solenoidal condition $\nabla\cdot\mathbf{B} = 0$ in cylindrical coordinates gives $\partial_r B_r + \partial_z B_z = -B_r/r$, so the first two terms cancel,
and the expression reduces to
\begin{equation}
    \label{eq:amp}
    \frac{\partial \mathbf{B}}{\partial t} =\frac{\partial B_\phi}{\partial t}\hat{\phi}= r\!\left(B_r\frac{\partial\Omega}{\partial r} + B_z\frac{\partial\Omega}{\partial z}\right)\hat{\phi}
    = rB_r\frac{\partial\Omega}{\partial r}\hat{\phi}.
\end{equation}
where we neglect the term involving the vertical gradient of the angular velocity, $\partial_z\Omega$, for the case of thin disks.This is the classic expression for the $\Omega$-effect. If the field amplification is driven only by shear, the poloidal component grows \textbf{\textit{linearly}} over time as $(q\Omega)^{-1}$, where $q = -\partial \ln \Omega /\partial \ln r$ is the shear parameter. This calculation illustrates how differential rotation winds up the poloidal magnetic field components $(B_r, B_z)$ to generate and amplify the toroidal component $B_\phi$.

The evolution of the magnetic energy density follows directly from Equation (\ref{eq:amp}).
Since the $\Omega$-effect amplifies only the toroidal component, the total magnetic energy density
$\varepsilon_B = B^2/2$ can be written as a constant poloidal contribution plus a growing
toroidal one. Integrating Equation (\ref{eq:amp}) with $B_r$ held fixed on the
amplification timescale gives $B_\phi(t) = B_{\phi,0} - q\Omega B_r\,t$, so that
\begin{equation}
    \varepsilon_B(t)
    = \frac{1}{2}\left(B_r^2 + B_z^2 + \left(B_{\phi,0} - q\Omega B_r\,t\right)^2\right) \propto t^2.
    \label{eq:energy}
\end{equation}
While the field grows linearly in time, the energy density grows \textbf{\textit{quadratically}},
$\varepsilon_B \propto t^2$, on the same shear timescale $(q\Omega)^{-1}$.

\bibliography{references}{}
\bibliographystyle{aasjournalv7}

\end{document}